%% file: cocg.v8.tex
\documentclass[10pt, aps,prc,twocolumn,superscriptaddress,preprintnumbers,
amsmath, nofootinbib,
floatfix,
longbibliography,
showpacs 
]{revtex4-1}
\usepackage[T1]{fontenc}
\usepackage[utf8x]{inputenc} 
\usepackage{adjustbox}          
\usepackage[caption=false]{subfig}
\usepackage{url}
\usepackage{color}
\usepackage{float}
\usepackage[pdftex,colorlinks=true, linkcolor = blue, citecolor=blue,urlcolor=blue, bookmarksnumbered=true, bookmarksopen=true]{hyperref}
\usepackage{longtable}
\usepackage{amsfonts}
\usepackage{wrapfig}
\newcommand{\vvec}[1]{\boldsymbol{#1}}

\newcommand{\ddx}{\frac{\partial}{\partial x}}
\newcommand{\ddy}{\frac{\partial}{\partial y}}
\newcommand{\ddz}{\frac{\partial}{\partial z}}
\newcommand{\fm}{\;\textrm{fm}}
\newcommand{\MeV}{\;\textrm{MeV}}

\begin{document}

\title{Coordinate-space solver for superfluid many-fermion systems with shifted conjugate orthogonal conjugate gradient method}
\author{Shi Jin}
\email[]{js1421@uw.edu}
\affiliation{Department of Physics,%
  University of Washington, Seattle, Washington 98195--1560, USA}
  
  \author{Aurel Bulgac}%
\email{bulgac@uw.edu}%
\affiliation{Department of Physics,%
  University of Washington, Seattle, Washington 98195--1560, USA}
  
   \author{Kenneth Roche}%
\email{kenneth.roche@pnl.gov}%
\affiliation{Department of Physics,%
  University of Washington, Seattle, Washington 98195--1560, USA}
\affiliation{Pacific Northwest National Laboratory, 
Richland, Washington 99352, USA}

   \author{Gabriel Wlaz\l{}owski}%
\email{gabrielw@if.pw.edu.pl}%
\affiliation{Faculty of Physics, Warsaw University of Technology, %
  Ulica Koszykowa 75, 00--662 Warsaw, Poland}
\affiliation{Department of Physics,%
  University of Washington, Seattle, Washington 98195--1560, USA}

\date{\today}

\begin{abstract}

Self-consistent approaches to superfluid many-fermion systems in
three-dimensions (and their subsequent use in time-dependent studies) require a
large number of diagonalizations of very large dimension Hermitian matrices, 
which results in enormous computational costs. We present an approach based
on the shifted conjugate-orthogonal conjugate-gradient (COCG) Krylov method
for the evaluation of the Green's function, from which we subsequently
extract various densities (particle number, spin, current, kinetic
energy, anomalous, etc.)  of a nuclear system. The approach eschews the determination of the
quasiparticle wavefunctions and their corresponding quasiparticle
energies, which never explicitly appear in the construction of 
a single-particle Hamiltonian or needed for the calculation 
of various static nuclear properties, which depend only on densities.  
As benchmarks we present calculations for nuclei with
axial symmetry, including the ground state of spherical (magic or semi-magic) and
axially deformed nuclei, the saddle-point in the $^{240}$Pu
constrained fission path, and a vortex in the neutron star crust and demonstrate the superior
efficiency of the shifted COCG Krylov method over traditional approaches. 
\end{abstract}

\preprint{NT@UW-16-09}

\maketitle

\section{introduction} \label{introduction}

Density functional theory (DFT) and other self-consistent approaches
[Hartree-Fock (HF), Hartree-Fock-Bogoliubov (HFB) or Hartree-de Gennes]
have played an essential role in studying the properties of most nuclei
across the nuclear chart \cite{Bender2003}. Present phenomenological
nuclear energy density functionals (EDF) allow for accurate
descriptions of many bulk properties of nuclei such as masses, radii
and shapes, transition matrix elements, potential energy surfaces and
related inertial parameters, and even non-equilibrium properties, when
extended to time-dependent phenomena. The time-dependent extension is
straightforward~\cite{PhysRevLett.52.997,TDDFTbook,FundumentalsTDDFTbook}
and provides a unified approach for the study of both structure and
reaction dynamics.  While pairing correlations are absent in closed
shell magic nuclei, there are a lot of nuclear problems where  accurate description of the pairing correlations is
crucial. Examples of such problems include large amplitude collective motion of open shell nuclei 
and dynamics of vortices in neutron star crust~\cite{Ring2004,Bertsch:1980,Barranco:1990,Bertsch:1997,Bulgac2016,Anderson:1975,Wlazlowski2016}.

The evaluation of the nuclear DFT is numerically demanding,
particularly if one considers large fermionic systems in three dimensions and large
deformations, without any symmetry constraints. Over the years many
iterative approaches for solving the HFB equations have been proposed, including
successive diagonalizations of the HFB or HF+BCS matrices, imaginary time
evolution~\cite{Davies1980111,Ryssens2015}, and gradient
methods~\cite{Egido199570,PhysRevC.84.014312}, which typically need
significant memory requirements and operations of complexity $O(N^3)$,
where $N$ is the dimension of the HFB matrix. For a review of modern diagonalization
software, see Ref.~\cite{Golub2000}.

The most standard approach, via series of direct diagonalizations of
the HFB Hamiltonian, can be divided into two main classes. In the
first one, the HFB problem is formulated in the configuration space by
expanding the quasiparticle states of HFB on a discrete basis of
orthogonal functions, usually provided by a (deformed) harmonic
oscillator (HO) basis \cite{Stoitsov2013, Schunck2012}.
Although
typically very fast, this approach suffers from truncation errors that
typically lead to the incorrect description of the asymptotic behavior of the
system or of the large deformations of the systems, e.g. when a nucleus fissions. 
An HO-basis does
not usually provide a very effective coverage of the relevant
phase-space. Sometimes this can be improved by 
introducing wavelets~\cite{Pei:2014} and complex energy Berggren 
states~\cite{Michel:2002} to describe the continuum spectrum. 
Nuclear systems are adequately described when the
phase-space in which the dynamics occurs is properly covered by the
single-particle basis wave functions. This space is characterized typically by a volume $V=L^3$,
where $L$ is a few times the nucleus radius
and by a maximum
single-particle momentum $p_{\mathrm{cut}} = \hbar k_{\mathrm{cut}}$ proportional 
 to the
Fermi momentum $\hbar k_F$. The total number of quantum states
in such p hase space is
\begin{align}
{\mathcal N}_{PS}=4\frac{(2 p_{\mathrm{cut}}L)^3}{(2\pi\hbar)^3}\propto k_F^3 R^3,
\end{align}
where the factor 4 arises from accounting for spin and isospin
degrees of freedom, see the discussion in Ref.~\cite{Bulgac2013}. The spatial extension $L$ (that can be different in each Cartesian direction)
is chosen depending on the specific needs. These extensions in each spatial direction are different for a nucleus 
with very extended density tails, for the collision of two nuclei, for the fission of a 
heavy nucleus and its split into two fragments or 
for a nucleus in the presence or absence of a vortex 
in the neutron star crust. In an
HO-basis, by increasing the size of the basis set to cover the required spatial volume, one usually goes well
beyond $p_{\mathrm{cut}}$, leading to an inefficient coverage of the needed
phase-space.  
 Pairing correlations typically lead to a filling of all
 the momentum states up to $p_{\mathrm{cut}}$. The need to describe large
 deformations, the tails of the density distributions, and
 particularly the large amplitude dynamics of various nuclear systems
 requires large spatial simulation volumes and a large momentum cutoff, thus resulting in
 a large number of phase-space cells ${\mathcal N}_{PS}$.

Another approach is the direct HFB matrix
diagonalization in the coordinate space with a lattice spacing $dx =
\pi/k_{\mathrm{cut}}$ chosen to ensure an adequate coverage of
the phase-space~\cite{Bulgac2013,Ryssens2015,Ryssens2015a}. Thus, one
can obtain numerically accurate results for weakly bound nuclei and large
deformations \cite{Dobaczewski1984, Dobaczewski1996}. A number of
coordinate-space HFB solvers have been published over the years
\cite{Zhang:2013, Bennaceur2005, Oberacker2003, Teran2003, Blazkiewicz2005,
Pei2008, Pei:2014}, but solving the HFB equations in full
three-dimensional (3D) coordinate space is still a challenging problem
because of the large dimension of the HFB matrix discretized in a
large box.  To put this in perspective consider calculations in medium
size volume $32\times32\times48\fm^3$ with lattice resolution 
$dx=1\fm$ corresponding to $p_{\mathrm{cut}} =\hbar\pi/dx \approx 600 \MeV/c$.
Then, the $N\times N$ HFB matrix has
$N^2 =(4\times32\times32\times48)^2\approx 200~000^2$ matrix elements, and requires more
than $0.5\,\text{TB}$ memory just to store it. A typical diagonalization
(which requires $O(N^3)$ operations separately for  protons and
neutrons), takes about 40 minutes using the high-performance linear
algebra library {\footnotesize ScaLAPACK} \cite{scalapack} on the
Edison supercomputer at NERSC with 36~864 processor cores and a 
charged computational cost of 49~152 CPU hours~\cite{edison} 
(CPU, central processing unit). For self-consistent
convergence $\sim100$  diagonalizations are typically required,
generating an enormous computational cost on the order of about 10 million
CPU hours per converged calculation. After each diagonalization,
the eigenfunctions are reduced to local densities needed to construct
new quasiparticle Hamiltonian matrix coefficients before the start of
a next iteration. The manner in which new matrix coefficients are
constructed reflects the fact that, for density functional theory, the
many-body wave function contains vastly more information than is needed
in each iteration.

Here we present a new method for extracting densities directly from
the HFB Hamiltonian without calculating wavefunctions.  The method is
especially well suited for large scale calculations that inevitably
require an efficient use of supercomputers. 
It is important to compare both the computational (numerical) complexity of different algorithms that solve
the same problem, and the ease with which the methods can be 
partitioned into smaller problems that can be effectively executed in
parallel so that the entire process scales well on today's computers. 
An important indicator that characterizes this
property of an given algorithm is the \textit{strong scaling}, which describes how
the time to compute a fixed problem depends on the aggregated
scale of the computing resources used to finish the problem. 
Ideal (linear) strong scaling
for an algorithm is achieved when
the (wall) time to completion can be reduced by a constant factor $k$ while
increasing the aggregated machine scale by the same factor $k$. 
The dense, direct eigenvalue decomposition based on data decomposing
the Hamiltonian over a set of processes does not exhibit perfect strong
scaling on modern parallel computers, particularly for very large matrices, 
when communication between computational nodes starts 
dominating the computational cost. 
The factorization requires
frequent interruptions both in communicating partial results and
coordinating coarse phases
of the algorithm between computing units, and these synchronizations, 
coupled to increasingly smaller work fractions for a fixed
problem dominate the strong scaling behavior. Eventually more time is
spent managing the computation than evaluating the algorithm. 
Moreover, a typical local
energy density functional leads to a sparse HFB matrix in the
coordinate representation - a feature that is not efficiently
utilized by the eigenvalue decomposition with direct methods. The method
proposed here removes these weaknesses. The algorithm can be
constructed in a nearly communication-free manner and thus exhibits
almost prefect strong scaling over the number of points in the
coordinate space offering a near complete reduction of $O(N)$ in
complexity when evaluating the method in parallel. 
The only operation that involves the
HFB matrix is a matrix-vector (MV) multiplication, which can benefit
easily from matrix sparsity. In our implementation the MV multiplication 
is extremely efficiently implemented  using fast Fourier transform.
The programming method is straightforward in
hybrid processing models that combine traditional CPUs with hardware
accelerators such as general purpose graphics processing units
(GPUs)~\cite{cuda}. Presently, many leadership class computers are of this type.

In order to grasp the new method, consider a Hartree-Fock
equation~\cite{Ring2004}
\begin{align} \label{eq:psi_k}
H \psi_k(\vvec{r}) = \varepsilon_k \psi_k(\vvec{r}),
\end{align} 
where $\psi_k(\vvec{r})$ is the wavefunction corresponding to the energy level
$\varepsilon_k$. Without loss of generality, we assume that the
eigenvalues $\varepsilon_k$ are positive. Our goal is to compute the
particle number density 
\begin{equation}
 \rho(\vvec{r}) = \sum_{\varepsilon_k<\varepsilon_F} \psi_k (\vvec{r}) \psi_k^* (\vvec{r}),
\end{equation} 
where the summation includes only states up to a fixed Fermi energy
$\varepsilon_F$ for normal systems. The density can be obtained from the Green's function
$G(z,\vvec{r}; \vvec{r}^\prime)$, defined by the linear equation (spin degrees 
freedom are suppressed here)
\begin{align}\label{eq:gf}
(z-H) G(z,\vvec{r}; \vvec{r}^\prime) =  \delta (\vvec{r} - \vvec{r}^\prime),
\end{align}
where $z$ is a complex number. Notice that in this equation
$\vvec{r}^\prime$ can be treated as parameter and thus for a fixed
value of $\vvec{r}^\prime$ the Green's function $G(z,\vvec{r};
\vvec{r}^\prime)$ is the solution of an inhomogeneous Schr\"odinger
equation. The formal solution of this equation is
\begin{align}\label{eq:gf2}
G(z,\vvec{r}; \vvec{r}^\prime) = \sum_k \frac{\psi_k (\vvec{r}) \psi_k^* (\vvec{r}^\prime)}{z - \varepsilon_k}.
\end{align}
Once we obtain the Green's function, forcing $\vvec{r} =
\vvec{r}^\prime$, the particle density can be calculated via a contour
integral
\begin{align}\label{eq:cont}
\rho(\vvec{r}^\prime) =  \frac{1}{2\pi i} \oint_C dz\, \left . G(z,\vvec{r}; \vvec{r}^\prime ) \right |_{\vvec{r} = \vvec{r}^\prime}.
\end{align}
A contour can be chosen arbitrarily, but is required to enclose only
the eigenvalues in the the interval $[ 0,
\varepsilon_F] $ on the real axis. The problem of computing the density
for a fixed point $\vvec{r}^\prime$ is reduced to the calculation of
the Green's function $ G(z,\vvec{r}; \vvec{r}^\prime) $ for all $z$'s
on the contour, which is equivalent to solving a set of independent
linear equations~(\ref{eq:gf}). For any given local Hamiltonian the
density at any given point $\vvec{r}$ can be extracted independently
of any other point, enabling this step to be executed in parallel, with
no communications between processes. Thus by construction, the present
method exhibits perfect strong scaling up to the number of coordinate points.

For fixed $z$ and $\vvec{r}^\prime$ the algebraic
problem~(\ref{eq:gf}) can be solved using iterative methods. The
conjugate orthogonal conjugate gradient (COCG) method is a very
efficient algorithm for solving a set of linear equations $Ax=b$,
assuming that matrix $A$ is symmetric and
complex~\cite{VanderVorst1990}.  Recently, an extension of the method
called  the shifted COCG has been implemented for electron
systems~\cite{Yamamoto2008, Takayama2006}, nuclear
shell-model~\cite{Mizusaki2010}, the computation of the level density in
nuclei~\cite{Shimizu2016}, and other generalized eigenvalue problems~\cite{Sakurai2003, Sakurai2008, Ikegami2010}.
The shifted variant solves simultaneously
a family of algebraic problems $(A-\sigma)x^{\sigma}=b$ for many
shifts $\sigma\in\mathbb{C}$ simultaneously, essentially with the same speed as
standard COCG for a single shift.  In this way, the accuracy of
numerical estimation of the contour integral~(\ref{eq:cont}) can be
refined to the desired accuracy with almost no extra calculation cost.
Taking as many computing units as points in coordinate space, the
computation time will be limited only by the time needed to solve this
single algebraic problem.

The purpose of the present work is to introduce the procedure in the
context of solving the HFB equation in 3D coordinate space. The
Green's function $G(z,\vvec{r}; \vvec{r}^\prime)$ of a HFB equation
can be obtained in a similar way as solving Eq.~\eqref{eq:gf} using
the shifted COCG method, but for a generalized multi-component system
with pairing and spin-orbit coupling. The calculation of particle
density in Eq.~\eqref{eq:cont} will be generalized to all types of
local densities in the realistic nuclear energy density functional
(NEDF) \cite{Dobacz2002, Chabanat1998}. With the aim of taking
advantage of existing and future computational resources, we developed
a highly efficient parallelized {GPU} code
as the so-called engine of the shifted COCG iteration to replace the
direct parallel diagonalization procedure in the code used in
Refs. \cite{Stetcu2011, Stetcu2015, Bulgac2016}.  As a benchmark, full
self-consistent HFB calculations are performed in this work for the ground states
of spherical (magic or semi-magic) as well as axially deformed nuclei. The
constrained HFB (CHFB) calculation is also tested for the saddle point
of $^{240}\mathrm{Pu}$ in the induced fission studied in \cite{Bulgac2016}.  Finally, we
apply the method to generate nontrivial states, relevant for
astrophysical applications, containing quantized vortices and nuclear
defects immersed in a superfluid neutron matter.

We emphasize that the concept of DFT is extensively used across many fields dealing with fermionic systems, like 
quantum chemistry, solid state physics, ultracold fermionic gases and many others.
Here we focus only on the nuclear case, as typically the nuclear EDF is very complicated in comparison to functionals encountered in other fields.
The method presented here is general and can be applied to other fermionic systems as well.

\section{Theoretical Framework} \label{theory}

\subsection{Nuclear DFT within local density
  approximation} \label{subsec:nedf}

Local density approximation is one of the most successful concepts
introduced to DFT approaches. It was extended to superfluid systems, 
namely superfluid local density approximation (SLDA) \cite{Bulgac2002, Yu2003}, 
and assumes that energy density functional (EDF) 
$\mathcal{E}$ depends on various \textit{local} densities. In nuclear
systems, a generic EDF is represented as a sum of the kinetic
$\mathcal{E}_{\mathrm{kin}}$, the nuclear
$\mathcal{E}_{\mathrm{nuclear}}$, the Coulomb
$\mathcal{E}_{\mathrm{Coul}}$, and the pairing
$\mathcal{E}_{\mathrm{pair}}$ contributions
\begin{align} \label{eq:totalenergy}
\mathcal{E} = \mathcal{E}_{\mathrm{kin}} + \mathcal{E}_{\mathrm{nuclear}} + \mathcal{E}_{\mathrm{Coul}} + \mathcal{E}_{\mathrm{pair}}.
\end{align}

The kinetic part depends on the kinetic
densities and the effective nucleon masses 
\begin{align}\label{eq:kinenergy}
\mathcal{E}_{\mathrm{kin}}(\vvec{r}) = \sum_{q=n,p} \frac{\hbar^2\tau_q (\vvec{r})}{2m_q(\vvec{r})} .
\end{align}
In calculations we included the simplest center of mass correction 
by replacing the bare nucleon mass $m$ with $m/(1-1/A)$. The
total particle number is $A = N + Z$ where $N$ and $Z$ are respectively the
neutron ($n$) and proton ($p$) numbers.  The Coulomb contribution is
composed of the direct part and the exchange part

\begin{align} \label{eq:coulomb}
\begin{split}
&\mathcal{E}_{\mathrm{Coul}}(\vvec{r})  = \mathcal{E}^d_{\mathrm{Coul}}(\vvec{r}) + \mathcal{E}^e_{\mathrm{Coul}}(\vvec{r}) \\
& = \frac{e^2}{2} \int \frac{\rho_p(\vvec{r}) \rho_p(\vvec{r}^\prime)}{\lvert \vvec{r} - \vvec{r}^\prime \rvert} d^3 \vvec{r}^\prime- \frac{3e^2}{4} \left(\frac{3}{\pi} \right)^{1/3} \rho_p^{4/3}(\vvec{r}).
\end{split}
\end{align}
The pairing energy in SLDA depends on the local anomalous density:
\begin{align}\label{eq:pairing}
\mathcal{E}_{\mathrm{pair}}(\vvec{r}) = \sum_{q=n,p} g_{\mathrm{eff}}(\vvec{r}) \lvert \nu_q(\vvec{r}) \rvert^2
\end{align}
and the effective pairing coupling strength
$g_{\mathrm{eff}}(\vvec{r})$ is obtained via a renormalization
\cite{Bulgac2002, Yu2003, Borycki2006} of the bare pairing strength,
typically parametrized as
\begin{align}\label{eq:g0}
g_0 (\vvec{r}) = g_0 \left ( 1 - \alpha \frac{\rho(\vvec{r})}{\rho_0} \right ),
\end{align}
where $\rho_0 = 0.16~\mathrm{fm}^{-3}$ is the saturation
density. The parameter $\alpha = 0, 1, 1/2$ corresponds to volume, surface, and
mixed pairing respectively \cite{Dobacz2002, Bertsch2009}.

The nuclear part is the most complicated. Over the years, many forms of
the nuclear functional have been proposed, see Refs.~\cite{Bender2003,arXiv:1506.09195, Dutra2012}
for review. Typically, the functional depends on various proton and
neutron densities, including normal $\rho(\vvec{r})$, kinetic
 $\tau(\vvec{r})$, spin $\vvec{s}(\vvec{r})$ or spin
kinetic energy densities $\vvec{T}(\vvec{r})$.  In a high accuracy nuclear EDF (NEDF) 
various currents are present as well, such as the normal current
 $\vvec{j}(\vvec{r})$, and the spin current densities
$\vvec{J}(\vvec{r})$. In our previous
works~\cite{Stetcu2011,Stetcu2015,Bulgac2016} we used the popular
parametrization SLy4~\cite{Chabanat1998,Bender:2003} of the Skyrme
NEDF, that has a rather generic form,
\begin{align}\label{eq:skyrme}
\begin{split}
\mathcal{E}_{\mathrm{Skyrme}} & = \mathcal{E}_{\rho^2} + \mathcal{E}_{\rho^\gamma} + \mathcal{E}_{\rho \Delta \rho} + \mathcal{E}_{\rho \tau}  + \mathcal{E}_{\rho \nabla J} \\
& = \sum_{t=0,1} \left( C_t^{\rho} \rho_t^2   + C_t^\gamma \rho_t^{2}\rho_0^\gamma  + C_t^{\rho \Delta \rho} \rho_t \Delta \rho_t \right .\\
& \quad\quad\quad\left . + C_t^\tau \rho_t \tau_t  + C_t^{\nabla J}\rho_t \vvec{\nabla} \cdot \vvec{J}_t  \right),
\end{split}
\end{align}
where $\rho_0=\rho_n+\rho_p$ and $\rho_1=\rho_n-\rho_p$ (and similar
for $\tau_{0,1}$ and $\vvec{J}_{0,1}$) are isoscalar and isovector
densities respectively, and $C$'s are coupling constants.

Starting from the NEDF defined above, the derived HFB equation is a
4-component eigenvalue problem:
{

\begin{align}\label{eq:hfbspin}
\begin{gathered}
H 
\begin{pmatrix}
u_{k\uparrow} \\
u_{k\downarrow} \\
v_{k\uparrow} \\
v_{k\downarrow}
\end{pmatrix}
= E_k
\begin{pmatrix}
u_{k\uparrow} \\
u_{k\downarrow} \\
v_{k\uparrow} \\
v_{k\downarrow}
\end{pmatrix} \\
H =  
\begin{pmatrix}
h_{\uparrow \uparrow} -\mu  & h_{\uparrow \downarrow} & 0 & \Delta \\
h_{\downarrow \uparrow} & h_{\downarrow \downarrow} - \mu& -\Delta & 0 \\
0 & -\Delta^* &  -h^*_{\uparrow \uparrow} +\mu & -h^*_{\uparrow \downarrow} \\
\Delta^* & 0 & -h^*_{\downarrow \uparrow} & -h^*_{\downarrow \downarrow} + \mu
\end{pmatrix}
\end{gathered}
\end{align}}
where we have suppressed the spatial coordinate $\vvec{r}$ and $k$ is
the label of each quasiparticle wavefunction $[u_{k\sigma}
(\vvec{r}), v_{k\sigma}(\vvec{r})]$ where $\sigma = \uparrow,
\downarrow$. The local particle-hole Hamiltonian $h$ is obtained by
taking the appropriate functional derivatives of the energy density
functional. For the Skyrme functional~(\ref{eq:skyrme}) it takes the form
\cite{Ring2004}:
\begin{align} \label{eq:spham}
\begin{split}
h_{\sigma, \sigma'} (\vvec{r})=\left ( -\vvec{\nabla} \cdot \frac{\hbar^2}{2m^*}\vvec{\nabla} + U \right )\delta_{\sigma,\sigma'}  \\
- i \vvec{W} \cdot (\vvec{\nabla} \times \vvec{\sigma})_{\sigma,\sigma'},
\end{split}
\end{align}
where $m^*(\vvec{r})$ is the effective mass, $U(\vvec{r})$ is the
central-part of the mean-field potential, and $\vvec{W}(\vvec{r})$ is the
spin-orbit potential (for their explicit forms see Ref.~\cite{Chabanat1998}). 
The local pairing field $\Delta (\vvec{r})$ is
defined as a function of the anomalous density
\begin{align} \label{eq:pairingfield}
\Delta (\vvec{r}) = - g_{\mathrm{eff}} (\vvec{r}) \nu (\vvec{r}).
\end{align}
The HFB Hamiltonian is a functional of local densities, which are
determined by the quasiparticle wavefunctions $[u_{k\sigma}
(\vvec{r}), v_{k\sigma}(\vvec{r})]$. The explicit expressions for the
most important ones are
\begin{align}
\rho(\mathbf{r}) &= \sum_{k,\sigma}  v^*_{k\sigma}(\mathbf{r}) v_{k\sigma}(\mathbf{r}) \label{eq:rho},\\
\nu(\vvec{r}) &= \sum_k v^*_{k\uparrow}(\vvec{r})u_{k\downarrow}(\vvec{r}),  \label{eq:nu}\\
\tau(\vvec{r}) & = \sum_{k,\sigma}  \vvec{\nabla}v^*_{k\sigma}(\mathbf{r}) \cdot \vvec{\nabla}v_{k\sigma}(\mathbf{r}),   \label{eq:tau}\\
\vvec{J} (\vvec{r}) & = \left. \frac{1}{2i} (\vvec{\nabla} - \vvec{\nabla}^\prime) \times \vvec{s}(\vvec{r}, \vvec{r}^\prime) \right |_{\vvec{r}=\vvec{r}^\prime},   \label{eq:J}
\end{align}
where
\begin{align} \label{sod2}
\begin{split}
s_x (\vvec{r}, \vvec{r}^\prime) & = \sum_k \left( v_{k\uparrow}^*(\vvec{r})v_{k\downarrow}(\vvec{r}^\prime) + v_{k\downarrow}^*(\vvec{r})v_{k\uparrow}(\vvec{r}^\prime)\right), \\
s_y (\vvec{r}, \vvec{r}^\prime) & = i\sum_k \left( v_{k\uparrow}^*(\vvec{r})v_{k\downarrow}(\vvec{r}^\prime) - v_{k\downarrow}^*(\vvec{r})v_{k\uparrow}(\vvec{r}^\prime) \right), \\
s_z (\vvec{r}, \vvec{r}^\prime) & = \sum_k \left( v_{k\uparrow}^*(\vvec{r})v_{k\uparrow}(\vvec{r}^\prime) - v_{k\downarrow}^*(\vvec{r})v_{k\downarrow}(\vvec{r}^\prime)\right). \\
\end{split}
\end{align}
The summations over $k$ should be performed for quasiparticle states with
quasiparticle energies $E_k$ that satisfy $0 < E_k <
E_\mathrm{cut}$, where $E_\mathrm{cut}$ is the energy cutoff related to the momentum cutoff
$E_{\mathrm{cut}} = \hbar^2 k^2_{\mathrm{cut}} / (2m)$ ($m$ stands for the mass of nucleon).
It should be chosen sufficiently large to ensure the convergence of all observables. This
convention is applied to all summations over $k$ throughout this
paper. In the next section, we will show how to extract these densities
directly from the Green's function without explicit diagonalization of the
HFB matrix. The method can be extended to other (not listed)
densities.
 
 \subsection{Green's function and local densities}\label{sec:COCGdensities}

Denoting the  $4 \times 4$ HFB matrix in Eq.~\eqref{eq:hfbspin} as $H$, the Green's
function $G(\vvec{r}, \vvec{r}^\prime,z)$ is the solution of a matrix
equation,
\begin{align}\label{eq:gfhfbspin}
(zI_4 - H) G(z,\vvec{r}; \vvec{r}^\prime) = \delta(\vvec{r}-\vvec{r}^\prime) I_4,
\end{align}
where $I_4$ stands for a $4 \times 4$ unit matrix. Now the Green's
function is a $4 \times 4$ blocked matrix in the form:
\begin{align} \label{eq:gfspin}
G(z,\vvec{r}; \vvec{r}^\prime) = 
  \sum_k \frac{1}{z - E_k}
  \begin{pmatrix}
u_{k\uparrow}(\vvec{r}) \\
u_{k\downarrow}(\vvec{r}) \\
v_{k\uparrow}(\vvec{r}) \\
v_{k\downarrow}(\vvec{r})
\end{pmatrix}
\cdot
  \begin{pmatrix}
u_{k\uparrow}^*(\vvec{r}^\prime) \\
u_{k\downarrow}^*(\vvec{r}^\prime) \\
v_{k\uparrow}^*(\vvec{r}^\prime) \\
v_{k\downarrow}^*(\vvec{r}^\prime)
\end{pmatrix}^{T}.
\end{align}
In the above equation the summation over $k$ includes all eigenstates.  As we
discussed in Sec.~\ref{subsec:nedf}, we need to calculate the normal,
anomalous, kinetic and spin-orbit densities and to construct the HFB
matrix in the self-consistent iterations. A closer look at the explicit
expressions for the densities~(\ref{eq:rho})-(\ref{sod2}) reveals that we
need to extract only 5 of 16 entries, containing
$\{v_{k\uparrow}(\vvec{r}),v_{k\downarrow}(\vvec{r})\}\otimes\{v_{k\uparrow}^*(\vvec{r}^\prime),v_{k\downarrow}^*(\vvec{r}^\prime)\}$
and $v^*_{k\uparrow}(\vvec{r})u_{k\downarrow}(\vvec{r}^\prime)$
products. In the next subsections, we provide prescriptions for extraction
of the local densities assuming that Green's function can be
efficiently computed. For simplification of the formulas we introduce the
notation
\begin{align}
G_{\phi\rho,\psi\sigma}(z,\vvec{r}; \vvec{r}^\prime)=\sum_k \frac{\phi_{k\rho}(\vvec{r})\psi^{*}_{k\sigma}(\vvec{r}^\prime)}{z - E_k}
\end{align}
for submatrices of the Green's function, where $\phi,\psi=\{u,v\}$ are 
wavefunction coordinates and $\rho,\sigma=\{\uparrow,\downarrow\}$ are
spin coordinates.
 
\subsubsection{Normal and anomalous density}

The simplest products in the Green's function are the normal
density $\rho(\vvec{r})$ in Eq.~\eqref{eq:rho} and the anomalous density
$\nu(\vvec{r})$ in Eq.~\eqref{eq:nu}. For $\rho(\vvec{r})$, one needs
to extract the $G_{v\uparrow,v\uparrow}$ and $G_{v\downarrow,v\downarrow}$
components from the $G$ matrix and add them. Next, by performing the contour
integral, like in Eq.~\eqref{eq:cont}, we obtain the normal density for a
selected $\vvec{r}^\prime$ point
\begin{align}
\rho(\vvec{r}^\prime) =  \frac{1}{2\pi i} \sum_{\sigma}\oint_C dz\, \left. G_{v\sigma,v\sigma}(z,\vvec{r}; \vvec{r}^\prime )\right |_{\vvec{r} = \vvec{r}^\prime},
\end{align}
where the contour integral encompasses the interval
$[0,E_{\mathrm{cut}}]$ of the real axis. The expression for the anomalous
density requires only the $G_{u\downarrow,v\uparrow}$ component and by
analogy reads
\begin{align}
\nu(\vvec{r}^\prime) =  \frac{1}{2\pi i} \oint_C dz\, \left. G_{u\downarrow,v\uparrow}(z,\vvec{r}; \vvec{r}^\prime )\right |_{\vvec{r} = \vvec{r}^\prime}.
\end{align}
The extension to spin densities~(\ref{sod2}) is straightforward.

\subsubsection{Kinetic density} 

Without losing generality, we suppress the spin degrees of freedom in the equations and write the
(normal) Eq.~\eqref{eq:rho} and the kinetic energy densities
Eq.~\eqref{eq:tau} in a simplified form:
\begin{align}
\begin{split}
\rho(\vvec{r}) &= \sum_k \psi^*_k(\vvec{r})\psi_k(\vvec{r}), \\
\quad \tau(\vvec{r}) &= \sum_k \vvec{\nabla} \psi^*_k(\vvec{r}) \cdot \vvec{\nabla} \psi_k(\vvec{r}).
\end{split}
\end{align}
In order to evaluate $\tau(\vvec{r})$, we first calculate the Laplacian of the number density:
\begin{align}
\Delta \rho (\vvec{r}) = 2{\mathrm Re}[\sum_k \psi^*_k(\vvec{r})\Delta \psi_k(\vvec{r}) + \vvec{\nabla} \psi_k^*(\vvec{r}) \cdot \vvec{\nabla} \psi_k(\vvec{r})].
\end{align}
Then $\tau(\vvec{r})$ can be obtained via
\begin{align} \label{eq:tau2}
\tau(\vvec{r}) = \frac{1}{2} \Delta \rho (\vvec{r}) -  \mathrm{Re} \left( \sum_k \psi^*_k(\vvec{r})\Delta \psi_k(\vvec{r}) \right) .
\end{align}
Now the problem is reduced to the calculation of the quantity $ \sum_k
\psi^*_k(\vvec{r})\Delta \psi_k(\vvec{r}) $.  Recall that
\begin{align}
\begin{split}
\sum_k \psi_k(\vvec{r}) \psi_k^*(\vvec{r}^\prime) =  \frac{1}{2\pi i} \oint_C dz G(z,\vvec{r}; \vvec{r}^\prime)
\end{split}
\end{align}
and if we apply the Laplacian $\Delta$ on $\vvec{r}$ on both sides of the
equation and set $\vvec{r} = \vvec{r}^\prime$ we obtain
\begin{align}
\begin{split}
&\sum_k \Delta \psi_k(\vvec{r}) \left . \psi_k^*(\vvec{r}^\prime) \right |_{\vvec{r} = 
\vvec{r}^\prime} \\
& =  \frac{1}{2\pi i} \oint_c dz \left . \Delta G(z,\vvec{r}; \vvec{r}^\prime) \right |_{\vvec{r} = \vvec{r}^\prime}.
 \end{split}
\end{align}
Computation of the derivatives of the Green's
functions introduces significant numerical costs. We avoid the calculation
of the Laplacian by taking advantage of an integration by parts
\begin{align}
\begin{split}
\left . \Delta G(z,\vvec{r}; \vvec{r}^\prime) \right |_{\vvec{r} = \vvec{r}^\prime} & = \int d \vvec{r} \delta(\vvec{r} - \vvec{r}^\prime) \Delta G(z,\vvec{r}; \vvec{r}^\prime) \\
& = \int d \vvec{r} \left( \Delta \delta(\vvec{r} - \vvec{r}^\prime) \right) G(z,\vvec{r}; \vvec{r}^\prime)
\end{split}
\end{align}
where the second equality uses the boundary condition 
$\lim_{r \to\infty} G(z,\vvec{r}; \vvec{r}^\prime) = 0$. Finally
\begin{align} \label{eq:quant1}
\begin{split}
\sum_k \psi^*_k(\vvec{r}^\prime)&\Delta  \psi_k  (\vvec{r}^\prime) = \\ 
= &\frac{1}{2\pi i} \oint_C dz \int d \vvec{r} \left( \Delta \delta(\vvec{r} - \vvec{r}^\prime) \right) G(z,\vvec{r}; \vvec{r}^\prime)
\end{split}
\end{align}
which is a contour integral of a convolution. Therefore, to obtain the
total kinetic energy density Eq.~\eqref{eq:tau}, one just needs to
extract $G_{v\uparrow,v\uparrow}$ and $G_{v\downarrow,v\downarrow}$
components from the $G$ matrix, and calculate the kinetic energy
density $\tau=\tau_\uparrow+\tau_\downarrow$ based on
Eqs.~\eqref{eq:tau2} and \eqref{eq:quant1}. When implemented numerically $\Delta\delta(\vvec{r} -
\vvec{r}^\prime) $ is the numerical implementation of the Laplacian
applied to the $\delta$-function on the lattice.

\subsubsection{Spin-current density} 

The spin-current density Eq.~\eqref{eq:J} can be written in the
explicit form:
\begin{widetext}
\begin{align}\label{eq:Jxyz}
\begin{split}
J_x(\vvec{r}) & = \mathrm{Im} \sum_k  \left( v^*_{k\downarrow}(\vvec{r}) \ddy v_{k\downarrow}(\vvec{r}) - v^*_{k\uparrow}(\vvec{r}) \ddy v_{k\uparrow}(\vvec{r}) \right) 
+ \mathrm{Re} \sum_k  \left( v^*_{k\uparrow}(\vvec{r}) \ddz v_{k\downarrow}(\vvec{r}) - v^*_{k\downarrow}(\vvec{r}) \ddz v_{k\uparrow}(\vvec{r}) \right), \\
J_y(\vvec{r}) & = \mathrm{Im} \sum_k  \left( v^*_{k\uparrow}(\vvec{r}) \ddx v_{k\uparrow}(\vvec{r}) + v^*_{k\downarrow}(\vvec{r}) \ddx v_{k\downarrow}(\vvec{r}) - v^*_{k\downarrow}(\vvec{r}) \ddz v_{k\uparrow}(\vvec{r}) - v^*_{k\uparrow}(\vvec{r}) \ddz v_{k\downarrow}(\vvec{r}) \right), \\
J_z(\vvec{r}) & = \mathrm{Im} \sum_k  \left( v^*_{k\downarrow}(\vvec{r}) \ddy v_{k\uparrow}(\vvec{r}) + v^*_{k\uparrow}(\vvec{r}) \ddy v_{k\downarrow}(\vvec{r}) \right) 
+ \mathrm{Re} \sum_k  \left( v^*_{k\downarrow}(\vvec{r}) \ddx v_{k\uparrow}(\vvec{r}) - v^*_{k\uparrow}(\vvec{r}) \ddx v_{k\downarrow}(\vvec{r}) \right), 
\end{split} 
\end{align}
\end{widetext}
which are combinations of the quantities $ \sum_k \psi^*_k \vvec{\nabla}
\psi_k$ for different spin combinations. The evaluation procedure in this 
case is similar to the one we used for the density 
$\sum_k \psi^*_k \Delta \psi_k$ and we obtain
\begin{align} \label{eq:quant2}
\begin{split}
&\sum_k \psi^*_k(\vvec{r}^\prime)\vvec{\nabla}  \psi_k  (\vvec{r}^\prime) = \\ 
&= -\frac{1}{2\pi i} \oint_C dz \int d \vvec{r} \left( \vvec{\nabla} \delta(\vvec{r} - \vvec{r}^\prime) \right) G(z,\vvec{r}; \vvec{r}^\prime),
\end{split}
\end{align}
where $\vvec{\nabla}$ represents the gradient operator on
$\vvec{r}$. One needs to extract all spin components
$G_{v\rho,v\sigma}$ from the $G$ matrix and form the appropriate
combinations. 

\section{Numerical implementation} \label{numerical}

\subsection{Shifted COCG method} \label{algorithm}

Throughout our theoretical framework, the core problem is to solve
Eq.~\eqref{eq:gfhfbspin} for all $z$'s on the contour.  We solve these
equations separately for each coordinate point $\vvec{r}^{\prime}$.
When discretized, this problem is reduced to the linear equations for
a given set of contour points $z_m$ $(m = 0, 1, \cdots,
m_{\text{max}})$. These linear equations are called ``shifted'' linear
equations or shifted linear systems because the matrices $z_m - H$ are
connected by scalar shifts. We solve this problem by using the shifted
COCG method \cite{Yamamoto2008, Takayama2006}, which is an iterative
method for solving large-scale shifted linear systems with symmetric
matrices. The details of this algorithm are available in
Ref.~\cite{Yamamoto2008, Takayama2006} and here we give only a brief
review and illustrate how to apply it to the HFB matrix in
Eq.~\eqref{eq:hfbspin}.

For a given symmetric matrix $A$, we want to solve the linear
equation:
\begin{align}\label{eq:seedeq}
A\vvec{x} = \vvec{b},
\end{align}
which is called the reference system, and its shifted equations
\begin{align}\label{eq:shifteq}
(A+\sigma I) \vvec{x}^\sigma = \vvec{b},
\end{align}
where $\sigma$ is a scalar complex shift, and $\vvec{x}^\sigma$ 
is the solution of the corresponding shifted system. The reference system is
solved by the COCG method \cite{VanderVorst1990}.  We define
$\vvec{x}_n$ as the approximate solution in the $n$-th iteration,
$\vvec{r}_n$ as the corresponding residual vector $
\vvec{r}_n=\vvec{b}-A\vvec{x}_n$, search direction vector
$\vvec{p}_n$, and other coefficients $\alpha_n, \beta_n$. With the
initial condition $\vvec{x}_0 =\vvec{0}, \vvec{r}_{0}=\vvec{b},
\vvec{p}_{0} = \vvec{b}, \alpha_{-1} = 1, \beta_{-1} = 0$ we have 
to perform the following iterations:
\begin{align}
\alpha_{n-1} &= \frac{\vvec{r}_{n-1}^T\cdot\vvec{r}_{n-1}}{\vvec{p}_{n-1}^T\cdot A  \vvec{p}_{n-1}} \label{eq:COCGalpha},\\
\vvec{x}_n &= \vvec{x}_{n-1} + \alpha_{n-1} \vvec{p}_{n-1} \label{eq:COCG_xn},\\
\vvec{r}_n &= \vvec{r}_{n-1} - \alpha_{n-1} A \vvec{p}_{n-1} \label{eq:COCG_rn},\\
\beta_{n-1} &= \frac{\vvec{r}_n^T\cdot\vvec{r}_n}{\vvec{r}_{n-1}^T\cdot\vvec{r}_{n-1}} \label{eq:COCGbeta},\\
\vvec{p}_n &= \vvec{r}_n + \beta_{n-1} \vvec{p}_{n-1}\label{eq:COCG_pn}
\end{align}
where the $T$ represents transpose only, without complex conjugation and
$\cdot$ stands for the implied scalar product between left and right
vectors. Note, the evaluation of $A \vvec{p}_{n-1}$ is the most
computationally consuming part of an iteration, as it requires a
matrix-vector multiplication. The residual vector measures the
accuracy of solution in the $n$-th iteration and it is used as a breaking
condition for iterations.

Shifted systems Eq.~(\ref{eq:shifteq}) can also be solved by COCG
algorithm. For each shifted equation we introduce the corresponding
vectors $\vvec{x}_n^\sigma$, $\vvec{r}_n^\sigma$, and
$\vvec{p}_n^\sigma$ and coefficients $\alpha_n^\sigma$ and
$\beta_n^\sigma$, and initialize them with the same initial conditions
as the reference system. However, the iterations
(\ref{eq:COCGalpha})-(\ref{eq:COCG_pn}) for the shifted systems can
benefit from their collinearity, meaning that the residuals vectors for the reference
and shifted systems are connected~\cite{Yamamoto2008}
\begin{align}
\vvec{r}_n^\sigma = \frac{1}{\pi^\sigma_n} \vvec{r}_n, \label{eq:colinear}
\end{align}
where the proportionality constant in each iteration is given by
($\pi_0^\sigma = \pi_{-1}^\sigma = 1$)
\begin{align}
\begin{split}
\pi_n^\sigma = \left( 1 + \alpha_{n-1} \sigma + \alpha_{n-1} \frac{\beta_{n-2}}{\alpha_{n-2}}\right) \pi_{n-1}^\sigma \\
- \alpha_{n-1} \frac{\beta_{n-2}}{\alpha_{n-2}} \pi_{n-2}^\sigma.
\end{split}
\end{align}
Thus, for the shifted systems instead of
evaluating the time consuming Eq.~(\ref{eq:COCG_rn}) one can evaluate the
simpler Eq.~(\ref{eq:colinear}).  Moreover, the coefficients
$\alpha^\sigma$ and $\beta^\sigma$ are also connected with
corresponding coefficients of the reference system
\begin{gather}
\alpha_{n-1}^\sigma = \frac{\pi_{n-1}^\sigma}{\pi^\sigma_n} \alpha_{n-1}, \\
\beta_{n-1}^\sigma = \left( \frac{\pi_{n-1}^{\sigma}}{\pi_n^{\sigma}}\right)^2 \beta_{n-1}.
\end{gather}
These equations replace Eqs.~(\ref{eq:COCGalpha}) and
(\ref{eq:COCGbeta}).  Often the reference
system is called the seed system as it ``seeds'' data for shifted systems.
The iterations are executed simultaneously for all systems and they end when the desired
accuracy is achieved for all equations. Since when evaluating the density 
we set $\vvec{r}=\vvec{r}'$ in the Green's function in the vector Eq.~(\ref{eq:colinear}), 
one needs to evaluate only one component, and not the entire vector; see also Sec.~\ref{compcost}. 

Assume for a moment that the HFB matrix is symmetric. This is
the case when the spin-orbit term in the nuclear density functional is ignored and
the pairing potential is real. Then we can straightforwardly let $A =
z_0 I - H$, where $z_0$ is selected point from contour. Shifts are
given by $\sigma_m=z_m-z_0$.  We solve the problem on a Cartesian
mesh grid with a lattice size $N_x \times N_y \times N_z$ and lattice
spacing $dx=dy = dz$.  After discretization, the HFB matrix has dimensions
$ N \times N $ with $N = 4N_x N_y N_z$, but in this case spin-up and
spin-down components are decoupled and the matrix size is effectively
$2N_x N_y N_z$.  For a fixed spatial point $\vvec{r}^\prime$ function
$\delta(\vvec{r}-\vvec{r}^\prime)$ becomes vector, where all $N$
elements are zeros except one [$=1/(dxdydz)$], corresponding to the
selected position. We denote this vector by
$\delta_{\vvec{r}^\prime}(\vvec{r})$. Since we do not need all
elements of $G$ matrix, it is sufficient to solve problems $A
\vvec{x}_1 = \vvec{b}_1$ and $A \vvec{x}_2 = \vvec{b}_2$ (together
with shifted counterparts) where $\vvec{b}_1 =
(0,0,\delta_{\vvec{r}^\prime},0)^T$ and $\vvec{b}_2 =
(0,0,0,\delta_{\vvec{r}^\prime})^T$. Then solutions $\vvec{x}_1$ and
$\vvec{x}_2$ are third and fourth columns of Eq.~\eqref{eq:gfspin}.

In general, the HFB matrix is a hermitian but not a symmetric matrix,
which cannot be solved using the shifted COCG method directly.  
However, one can use the COCG method designed for symmetric matrices in
the case of Hermitian matrices by performing a simple matrix
transformation. Assume that $H$ is an arbitrary Hermitian matrix which
can be divided into real and imaginary parts: $ H = H_x + i H_y $,
with obvious symmetry properties $H_x = H_x^T$ and $H_y = - H_y^T$. If
we also divide the eigenvectors of $H$ into real and imaginary parts
as $\psi = x + i y$. The eigenvalue problem $H\psi = \lambda \psi$
will be converted to:
\begin{align}
(H_x+iH_y) (x+iy) = \lambda (x+iy).
\end{align}
After collecting the real and imaginary terms it will become an equivalent
eigenvalue problem:
\begin{align}
\begin{pmatrix}
H_x & -H_y \\
H_y & H_x
\end{pmatrix}
\begin{pmatrix}
x \\
y
\end{pmatrix}
 = 
 \lambda
 \begin{pmatrix}
 x \\
 y
 \end{pmatrix},
\end{align}
but now the matrix on the left hand side is real symmetric. The
Green's function $G = G_x + i G_y$ of a general HFB matrix $H$ is then
the solution of the equation [equivalent to Eq.~\eqref{eq:gfhfbspin}]
\begin{align} \label{eq:gfhfbspin2}
\left( zI - \right .
H'
\left. \right)
\begin{pmatrix}
G_x \\
G_y
\end{pmatrix}
= 
\begin{pmatrix}
\delta^{(4)} \\
0
\end{pmatrix},
\end{align}
where 
\begin{align}
H' = 
\begin{pmatrix}
H_x & -H_y \\
H_y & H_y
\end{pmatrix}
, \quad
\delta^{(4)} = \delta(\vvec{r} - \vvec{r}^\prime) \otimes I_4.
\end{align}
As before we need to solve two linear equations $A'\vvec{x}'_1 =
\vvec{b}'_1$ and $A'\vvec{x}_2 = \vvec{b}'_2$ (and their shifted
equation) with $A' = zI - H'$ in doubled space. Now, vectors of length
$2N$ need to be set for $\vvec{b}'_1 =
(0,0,\delta_{\vvec{r}^\prime},0,0,0,0,0)^T$ and $\vvec{b}'_2 =
(0,0,0,\delta_{\vvec{r}^\prime},0,0,0,0)^T$.
 
\subsection{The form of the integration contour} \label{fcontour}

\begin{figure}[t]
\includegraphics[clip, width=1.0\columnwidth]{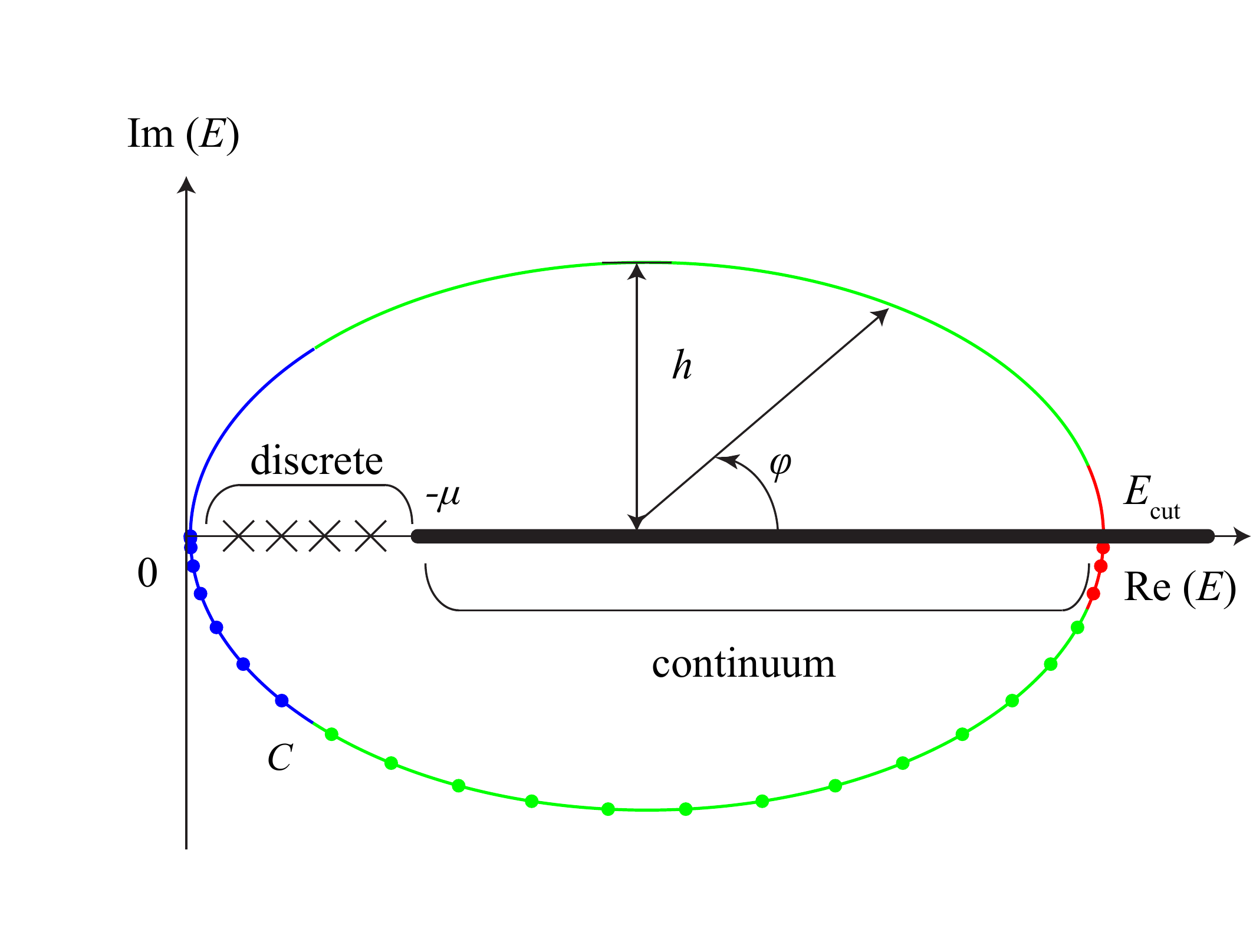}
\caption{\label{fig:contour}
The ellipse contour $C: z(\varphi)$ in Eq.~\eqref{eq:zphi} adopted to
perform the contour integral. The thick solid line represents the
positive energy continuum quasiparticle states with $E > - \mu$, while
the crosses represent the positive energy discrete bound quasiparticle
states $0<E_i < -\mu$.  Density of integration points is larger for
parts of the contour close to the real axis (depicted by blue and red
colors), and decreases as we go far away from the real axis (depicted
by green color).
}
\end{figure}

Before a further discussion of the shifted COCG method, we should provide a
clear definition of the contour, which determines the shifted systems
and is fundamental for the convergence behavior and error control of
the whole algorithm. We need to calculate an integral of the form
$\frac{1}{2\pi i} \oint_C dz f(z)$ where the contour encloses exactly
a segment of the real axis $[0, E_{\mathrm{cut}}]$. It is natural to
choose the contour to be symmetric with respect to the real axis since
the poles (the HFB spectrum) are real. We can parametrize the
contour as $z(\varphi)$ and a simple choice of its form is an ellipse:
\begin{align}\label{eq:zphi}
\begin{split}
z(\varphi) & = \frac{E_{\mathrm{cut}}}{2} + \frac{E_{\mathrm{cut}}}{2} \cos \varphi + i h \sin \varphi \\
& = E_{\mathrm{cut}} \cos^2\frac{\varphi}{2} + i h \sin \varphi 
\end{split}
\end{align}
where $0 \leq \varphi \leq 2\pi$ and the height of ellipse $h$ is the
parameter that should be chosen carefully, see Fig.~\ref{fig:contour}.
 
The integrand function behaves smoothly, only for points that are far
away from real axis.  As the contour approaches real axis, the
function $f(z)\sim\sum_{k}\frac{1}{z-E_k}$ as a function of $z$ starts
to exhibit fast oscillations, making accurate numerical integration
difficult. For this reason, we choose non-uniformly distributed
integration points along the contour.  The density of points
distribution depends on distance from the real axis with more points
closer to the real axis, and less far away from the real axis.  The
height $h$ of the eclipse is set to be significantly larger than the
expected average separation between poles, and we compute the
integrand function for angles $\varphi$ given by a  distribution function ($u$ and
the $l$ subscript corresponds to upper and lower parts of the contour
with respect to the real axis)
\begin{align} \label{phi}
\begin{split}
\varphi_u & = \frac{\pi}{2} \left \{ 1 + \tanh \left[ \alpha \tan \left( \phi - \frac{\pi}{2} \right) \right] \right \},  \\
\varphi_l &= \varphi_u +\pi, \quad 0\leq \phi \leq\pi,
\end{split}
\end{align}
where $\alpha$ is a parameter (usually $\alpha \sim$1-5) and the
integral over $\phi$ is discretized with a set of evenly distributed
points $\phi_n$ in the interval $[0, \pi]$ with step size $\Delta \phi$. 
Thus the contour integral can be converted into an weighted sum
\begin{align}
\frac{1}{2\pi i} \oint_C dz f(z)\cong\sum_{\sigma = u,l;n}  f\{z[\varphi_\sigma(\phi_n)]\} w_\sigma(\phi_n)\Delta \phi.
\end{align}
Typically, the number of integration points
does not exceed $10^4$, and further increases do not improve
the numerical result. The computation of the integrand along the contour is very
fast and represents a very small fraction of the entire calculation time,
which is dominated by the calculation of the Green's function at the
reference point.

\subsection{Convergence behavior} \label{convbh}

\begin{figure*}[t] 
\subfloat[]{\includegraphics[clip, width=0.6\columnwidth]{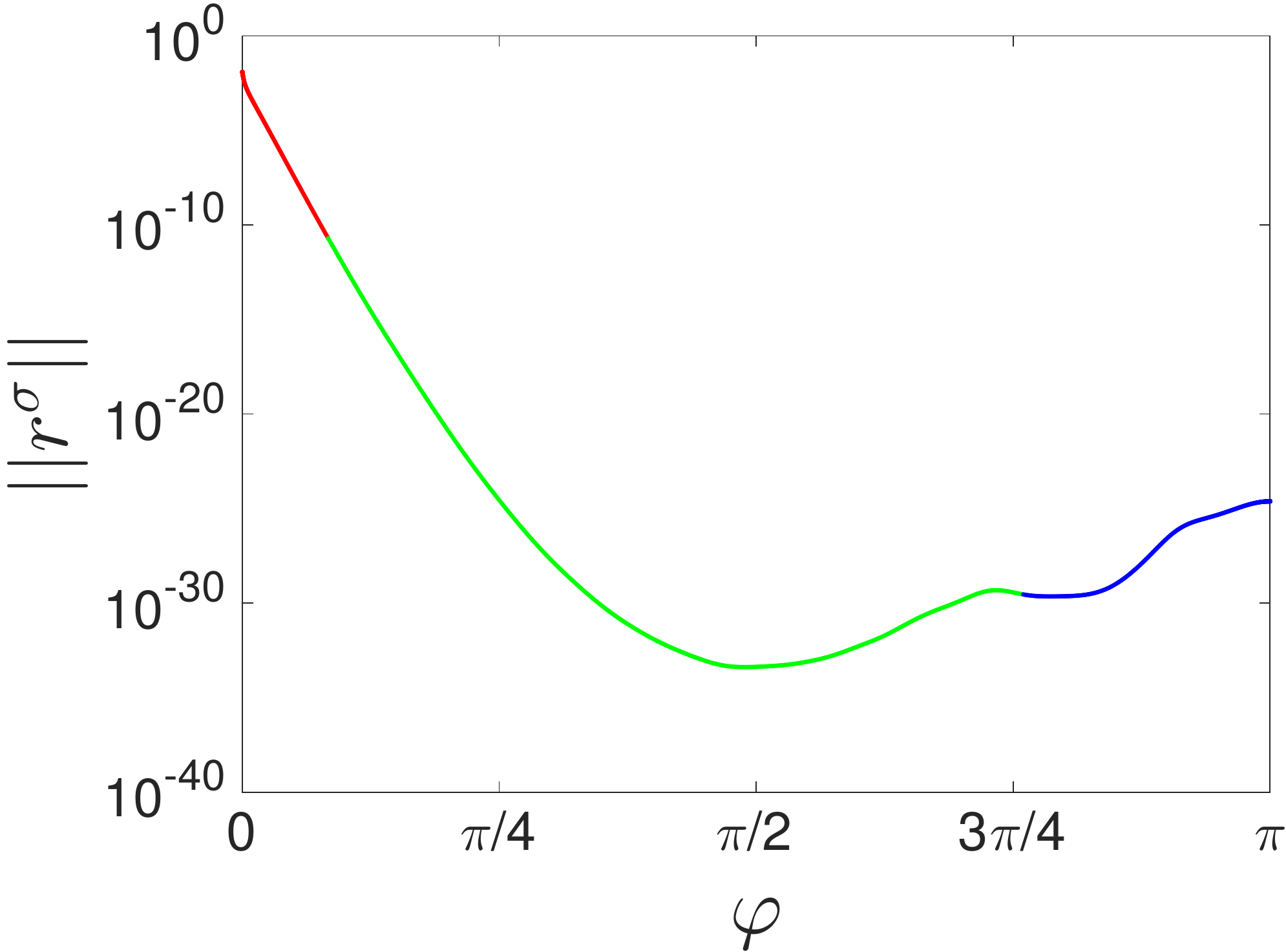}}
\subfloat[]{\includegraphics[clip, width=0.6\columnwidth]{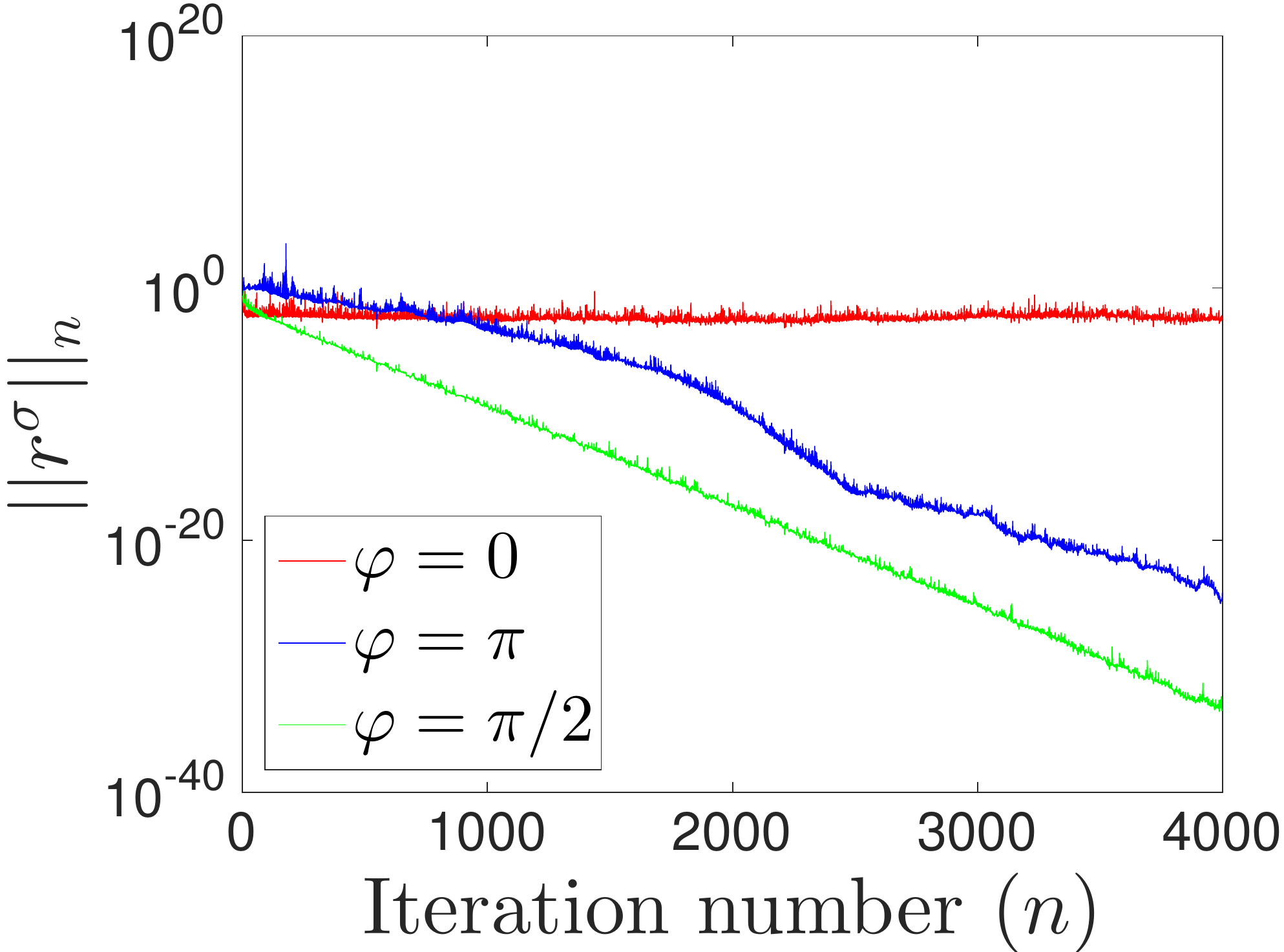}}
\subfloat[]{\includegraphics[clip, width=0.6\columnwidth]{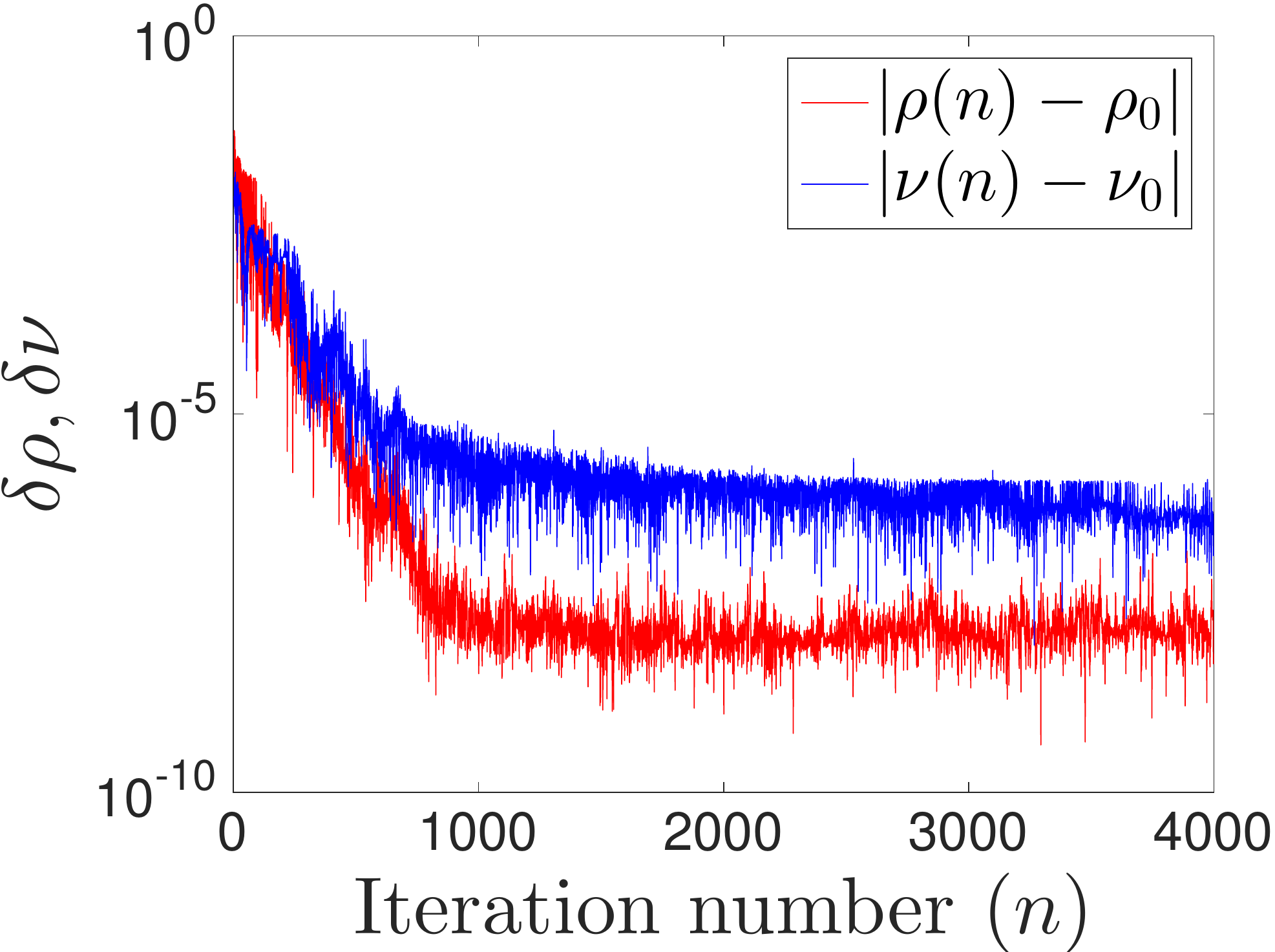}}\\
\subfloat[]{\includegraphics[clip, width=0.6\columnwidth]{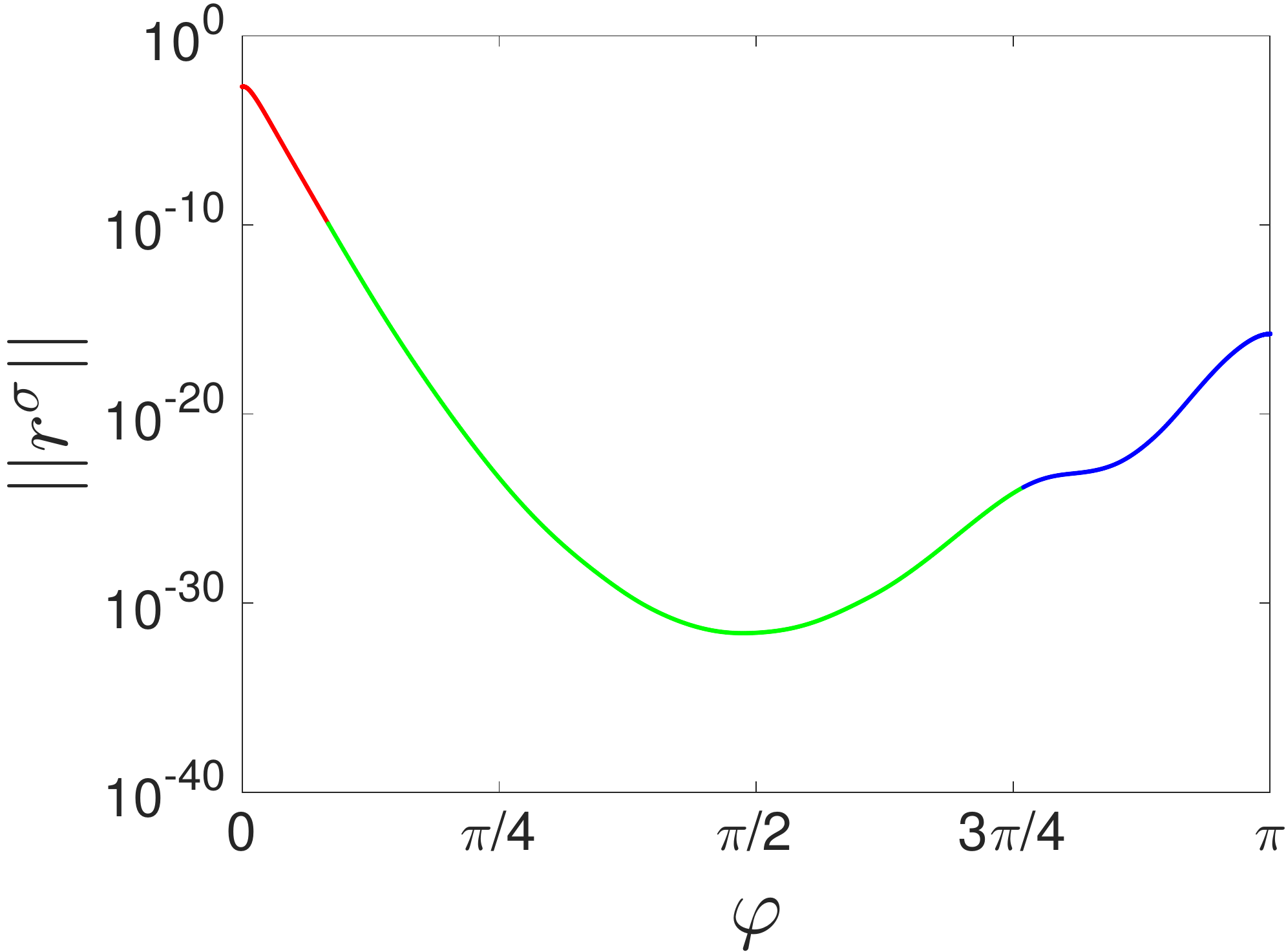}}
\subfloat[]{\includegraphics[clip, width=0.6\columnwidth]{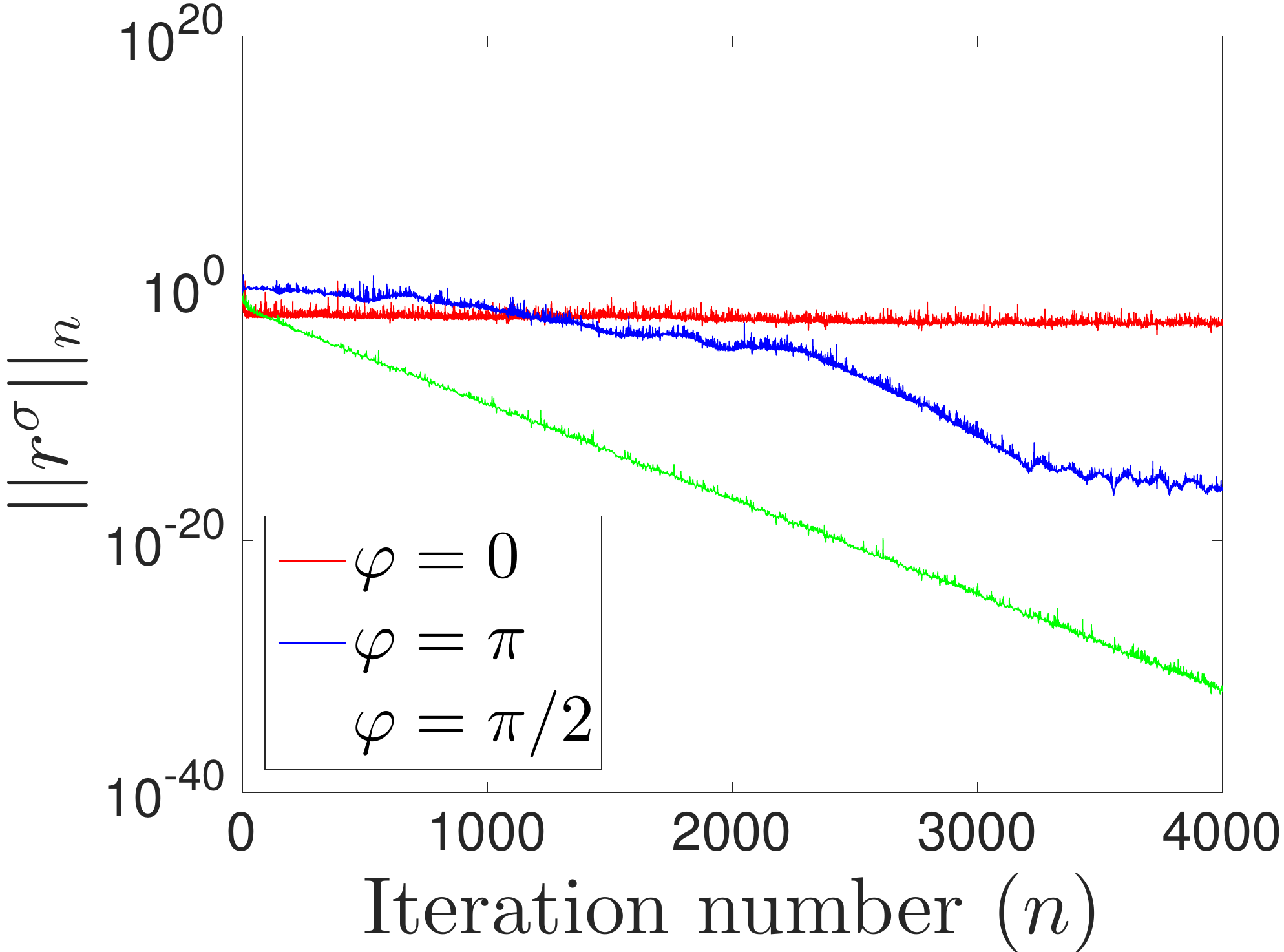}}
\subfloat[]{\includegraphics[clip, width=0.6\columnwidth]{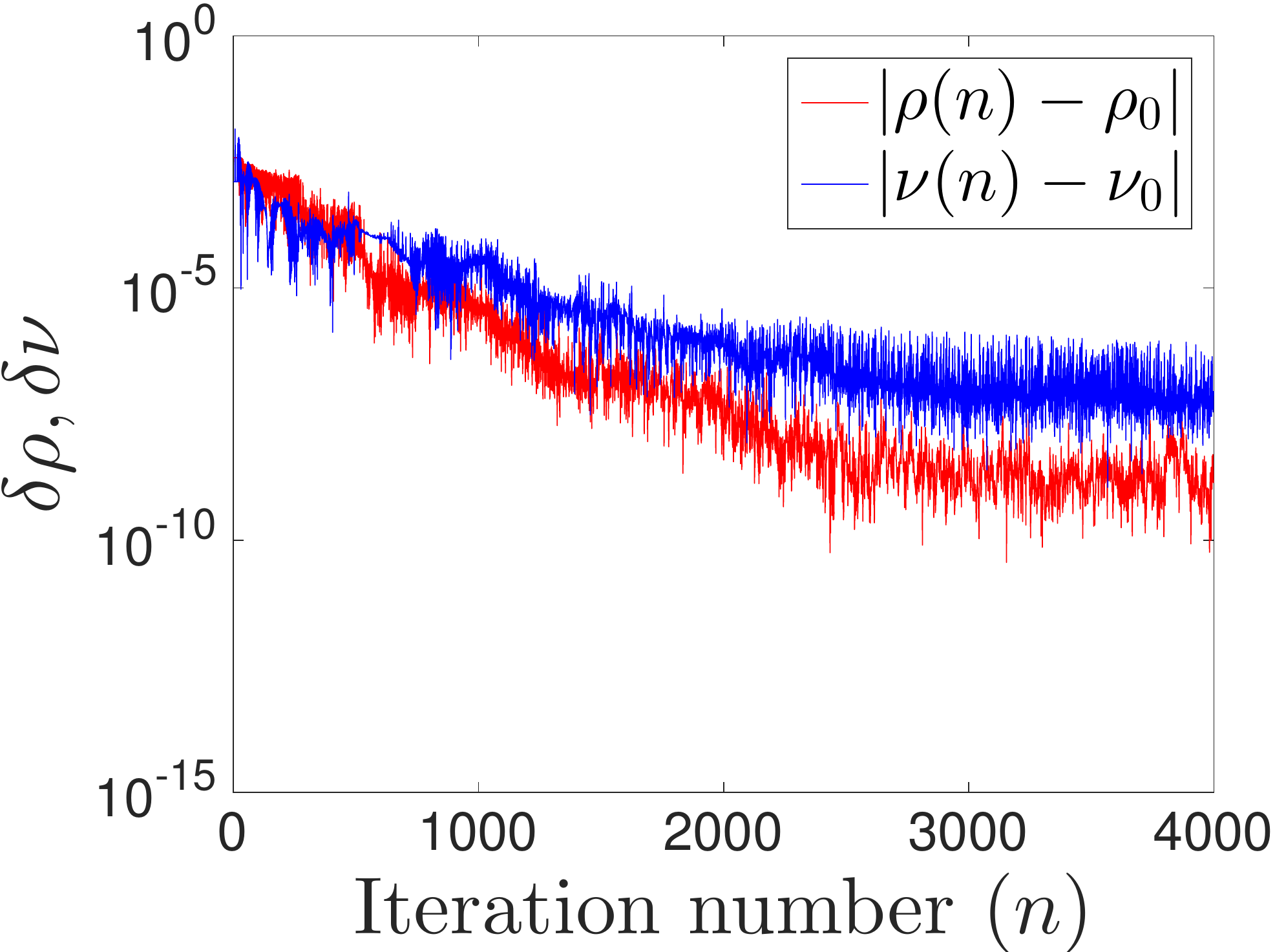}}
\caption{ \label{fig:errcont}
The convergence behavior of the shifted COCG method for the HFB equation with W-S
potential.  The top and bottom rows show results for spherically symmetric
and deformed W-S potentials respectively. The spherically symmetric
problem was solved on lattice $16^3$, while the deformed problem used a
lattice $16^2\times 20$. In both cases the lattice spacing is
$dx=1.25\fm$.  Panels~(a) and (d) show the distribution of the norms of the residual
vectors $\lVert r_n^\sigma \rVert$ on the semi-ellipse contour
$z(\varphi)$ for iteration number $n=4000$. Three colors (red, green,
blue) correspond to different parts of the contour, as depicted on
Fig.~\ref{fig:contour}. Panels~(b) and (e) show the convergence behavior of
$\lVert r_n^\sigma \rVert$ for three points $z(0)$ (red), $z(\pi/2)$
(green) and $z(\pi)$ (blue) as a function of iteration number.
Panels~(c) and (f) show  the difference between the value obtained from COCG method
in the $n$-th iteration and the exact value for the normal (red) and anomalous
(blue) densities.}
\end{figure*}

The norm of the residual vectors $\vvec{r}_n$ at the reference and of the
shifted systems $\vvec{r}_n^\sigma$ quantify the error of the approximate
solution $\vvec{x}_n$ and $\vvec{x}_n^\sigma$ in each iteration. Since
we need to calculate the Green's functions $G(z,\vvec{r};
\vvec{r}^\prime)$ for all $z$'s on the contour defined in
Sec.~\ref{fcontour}, the convergence behavior on the contour needs to
be studied.

As a simple but non-trivial test case, we consider only neutrons, and
choose a phenomenological Wood-Saxon (W-S) model
\cite{Ring2004,Bohr1998}. The central part of the single-particle
Hamiltonian Eq.~\eqref{eq:spham} is given by $U(\vvec{r}) = V_0
f(\vvec{r})$, where $f(\vvec{r})$ has the symmetrized W-S form:
\begin{align} 
f(\vvec{r}) &= \frac{1 - \exp \left( \frac{-2R_0}{a} \right) }
{\left[ 1+ \exp \left( \frac{r- R_0}{a} \right) \right] \left[ 1+ \exp \left( \frac{-r- R_0}{a} \right) \right]} 
\end{align} 
and $V_0 = -50~\mathrm{MeV}$, $r = \sqrt{x^2 + y^2 + (z/\beta)^2}$,
$R_0 = 1.12 A_0^{1/3} - 0.86 A_0^{-1/3}~\mathrm{fm}$, $A_0 = 100$, $a
= 0.54~\mathrm{fm}$. By varying the parameter $\beta$, we calculate
the densities in a W-S potential in both spherical and axially
deformed cases. The spin-orbit potential $\vvec{W}(\vvec{r})$ is
chosen as $\vvec{W} (\vvec{r} ) = \lambda \vvec{\nabla}
U(\vvec{r})$ where $\lambda = -0.5 ~\mathrm{fm}^2$, and the effective
mass is set to be the bare mass of neutron $mc^2 =
939.565~\mathrm{MeV}$.  The self-consistent pairing field
$\Delta(\vvec{r})$ in Eq.~\eqref{eq:pairingfield} should be determined
by the anomalous density $\nu(\vvec{r})$, but here we still use a
phenomenological W-S potential shape $\Delta(\vvec{r}) = 12/A_0^{1/2}
f(\vvec{r})$. In the test, the chemical potential is fixed at $\mu =
-7~\mathrm{MeV}$.  The numerical tests are performed in a cubic box of
size $L_x=L_y=L_z=20~\mathrm{fm}$ for the spherical W-S model ($\beta =
1$), and a rectangular box of size $L_x=L_y=20, L_z = 25~\mathrm{fm}$
for the deformed W-S model ($\beta = 1.5$). In both cases, the lattice
constant $dx=dy=dz = 1.25~\mathrm{fm}$. The number of lattice points is $N_x=L_x/dx$ and respectively for $y$ and $z$ direction. 
The energy cutoff is set to $E_{\mathrm{cut}} =
100~\mathrm{MeV}$.

We find that the convergence properties of the algorithm are strongly
correlated with the density of particles for a given point.  For
points with higher particle densities the method needs more iterations
to converge. This behavior can be easily understood if we recall that
the starting point for iterations is $\vvec{x}_0=0$; see
Sec.~\ref{algorithm}. For the points where the density is higher the
convergence is noticeably slower. In the following tests, we consider the
convergence properties of the method for the points requiring the highest
number of iterations. These points are located close to the center of
the simulation box.

Results for the spherically symmetric case ($\beta=1$) are presented
in the top row of Fig.~\ref{fig:errcont}.  The Figure 2(a) shows the
distribution of $\lVert {\vvec r}_n^\sigma \rVert $ on the semi-ellipse
$z(\varphi), 0 \leq \varphi \leq \pi$ on the upper half complex plane,
for iteration number $n = 4000$, and fixed position $\vvec{r}^\prime$
located close to center of the box.  In the lower half plane ( $ \pi
\leq \varphi \leq 2\pi$ ) the behavior is identical due to the
reflection symmetry of the integration contour. The shifted system converges most quickly in the
middle part of the contour (represented by color green), more slowly near
the left end of contour $z(\pi)=0$ (represented by color blue), and
very slowly near the right end of the contour $z(0) = E_\mathrm{cut}$
(represented by the color red). A closer look at the convergence behavior
at the representative points in these three parts, respectively
$z(\pi/2)$, $z(\pi)$, and $z(0)$, is shown on Figure 2 (b). Besides the
rapid convergence of the middle point $z(\pi/2)$, the residual at the
origin $z(\pi)=0$ also has a stable decrease with iterations.
However, the iterations for the right end $z(0) =
E_{\mathrm{cut}}$ fail to converge, and $\lVert r_n^\sigma \rVert$
keeps oscillating around $10^{-2}$. Consider the convergence analysis
of the conjugate gradient (CG) method \cite{Trefethen1997} in which
case the convergence ratio depends on the 2-norm condition number
$\kappa$ of the matrix $A$:
\begin{align}\label{eq:convergeratio}
\frac{\lVert r_n \rVert}{\lVert r_0 \rVert} \leq 2 \left( \frac{\sqrt{\kappa} - 1}{\sqrt{\kappa} + 1} \right)^n,
\end{align}
where the condition number $\kappa (A) = \lvert \lambda (A)
\rvert_{\mathrm{max}} / \lvert \lambda (A) \rvert_{\mathrm{min}}$ is
the ratio of maximum and minimal absolute value of the eigenvalues of
matrix $A$. From Eq.~\eqref{eq:convergeratio} it is clear that a
larger $\kappa$ leads to a slower convergence. The order of magnitude
of the condition number for matrix $A = zI -H$ can be estimated as
$\frac{\max|z-E_k|}{\min|z-E_k|}$, where $E_k$ are eigenvalues of the HFB
matrix from interval $[0,E_{\mathrm{cut}}]$. Definitely, the condition
number decreases as the imaginary part of $z$ increases, thus confirming
the finding that the convergence is fastest in middle part of the
semi-ellipse.  The largest values of $\kappa$ occur at points
close to the real axis.  We know that the spectrum of a HFB matrix is
discrete if $\lvert E_k \rvert < \lvert \mu \rvert$ and continuous
otherwise~\cite{Bulgac1999, Dobaczewski1984}.  Around $z=0$ there is a
gap in the spectrum and therefore for $z\rightarrow 0$ $\kappa
=O(E_{\mathrm{cut}}/\Delta)$, where $\Delta$ is average value of the
pairing gap in the system. When $z\rightarrow E_{\mathrm{cut}}$, $\min|z-E_k|$ can be arbitrarily small (only set by
lattice resolution), and in the limit of zero lattice spacing $dx$  the
condition number diverges.  
This ``quasi-singular'' matrix equation is hard to solve within a reasonable
iteration number. However, these continuous states near the energy
cutoff have very small occupation probabilities ($<10^{-5}$) and make
negligible contributions to the local densities. While it is possible that preconditioning
the COCG method can help~\cite{Gu:2014,Li2011,Li:2008,Wang:2009} our 
efforts in this direction did not yield any advantages.

Figure 2(c) tracks
the error of the particle density $\rho$ and anomalous density $\nu$
obtained by the calculated Green's function in each iteration. The
exact value of these two densities $\rho_0$ ($\sim 10^{-1}$) and
$\nu_0$ ($\sim 10^{-3}$) are calculated by direct
diagonalization of the same HFB matrix via {\footnotesize
ScaLAPACK}. It shows that the particle density $\rho$ can converge to
an accuracy of $10^{-8}$ within 1000 iterations, and this accuracy
will not improve as the iteration goes past that. The anomalous
density $\nu$ will converge to an accuracy of $10^{-6}$ within the
same iteration count, which is quite acceptable, since the
contribution to the energetics of a nucleus from the pairing field is
much smaller in absolute terms. The pairing field is almost two orders
of magnitude smaller than the single-particle potential. This behavior
of the anomalous density is due to the fact that the high-energy
quasiparticle continuum states contribute with significant weights as the
anomalous density diverges when the energy cutoff is increased
\cite{Bulgac1999,Bulgac2002,Borycki2006}. The same tests for deformed
W-S model are presented in Figure.~2(d)--2(f). They show a similar pattern
of the residual distribution on the contour. The local densities can reach
the same accuracy but with more iterations, because of the slower
convergence of points near $E=0$. In both the spherical and deformed
cases, the convergence behavior at $E=0$ can represent the convergence
behavior of the final local densities, and thus is chosen to be the breaking condition
for the iterations.

\begin{figure*}[t] 
\subfloat[]{\includegraphics[clip, width=0.9\columnwidth]{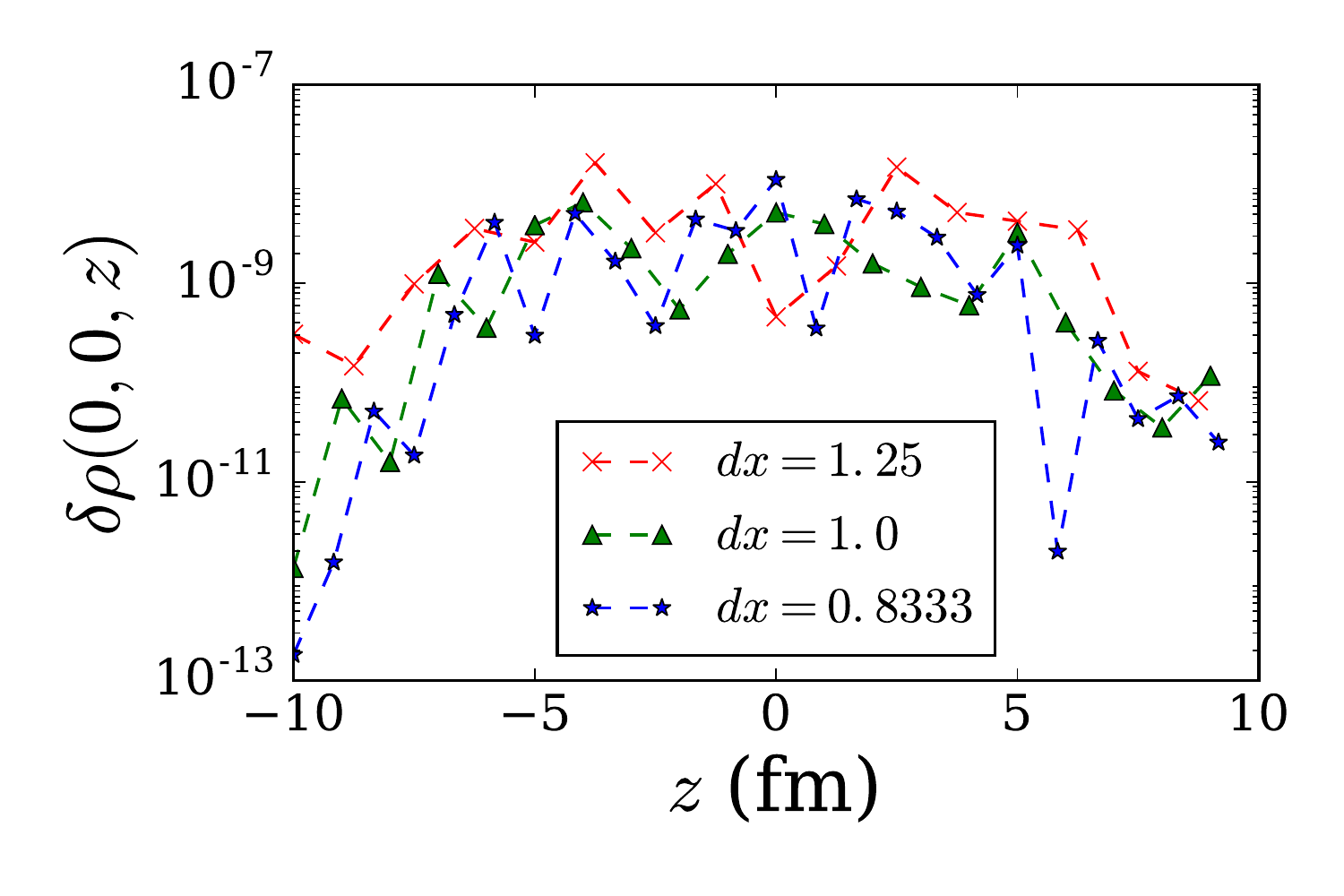}}
\subfloat[]{\includegraphics[clip, width=0.9\columnwidth]{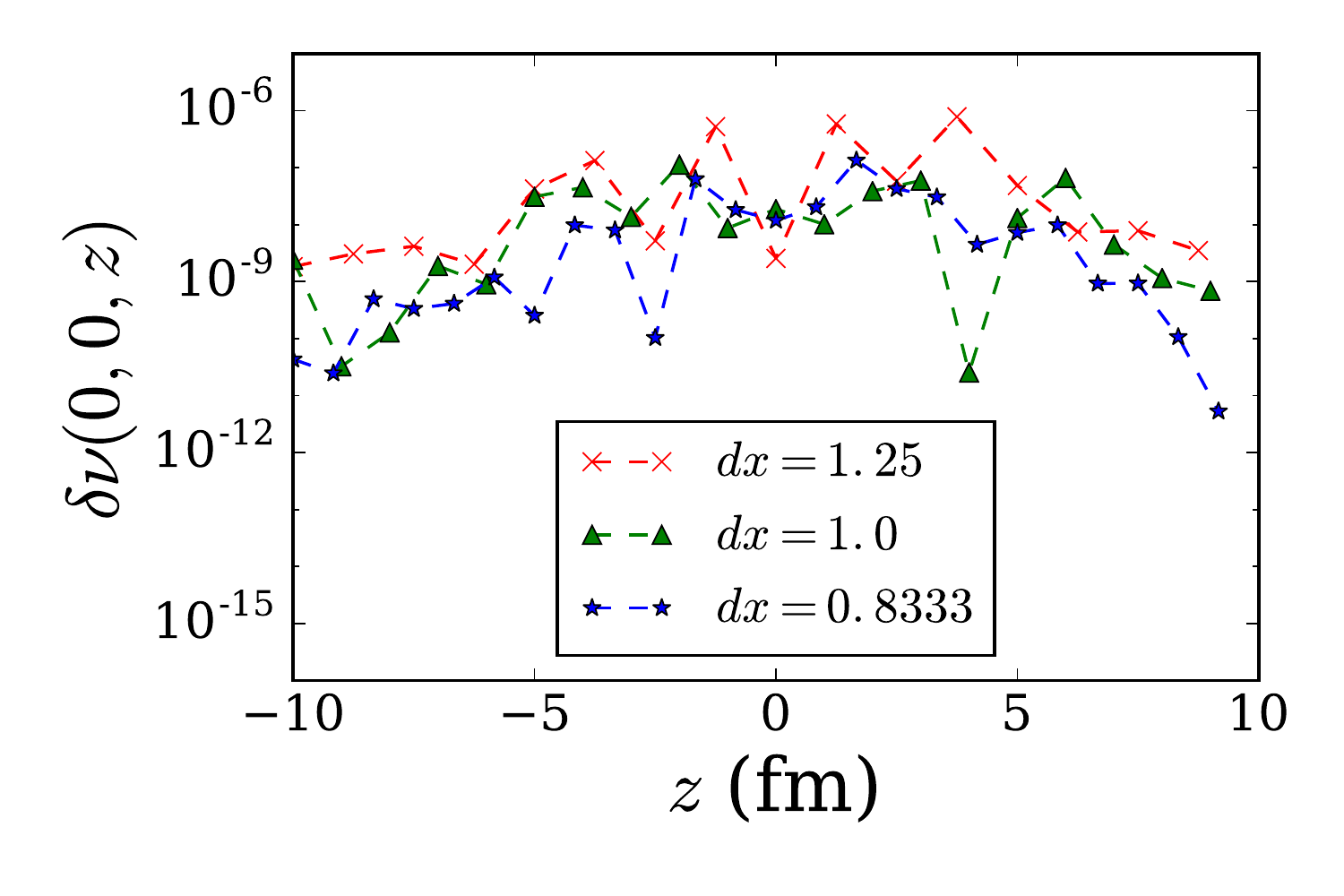}}\\
\subfloat[]{\includegraphics[clip, width=0.9\columnwidth]{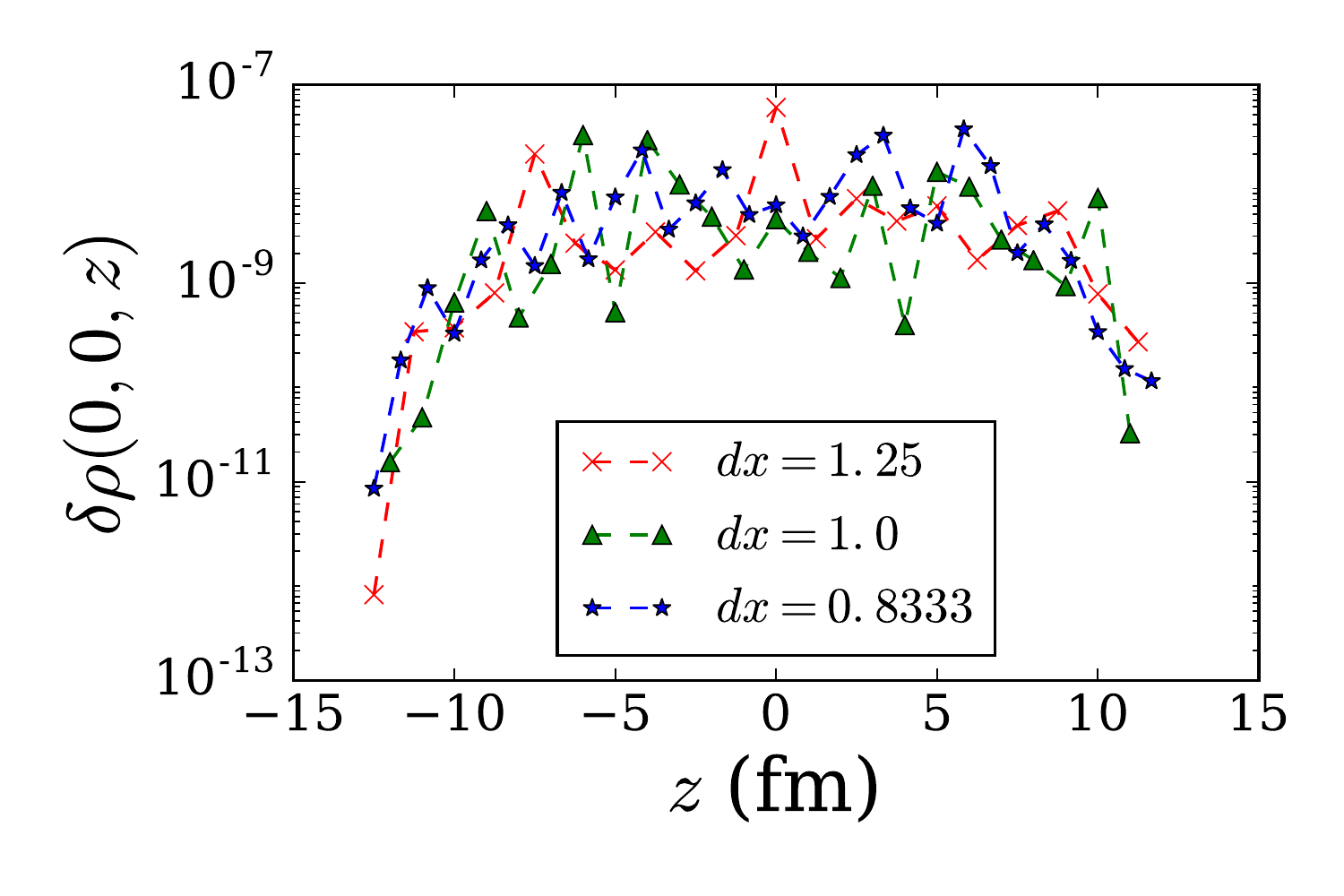}}
\subfloat[]{\includegraphics[clip, width=0.9\columnwidth]{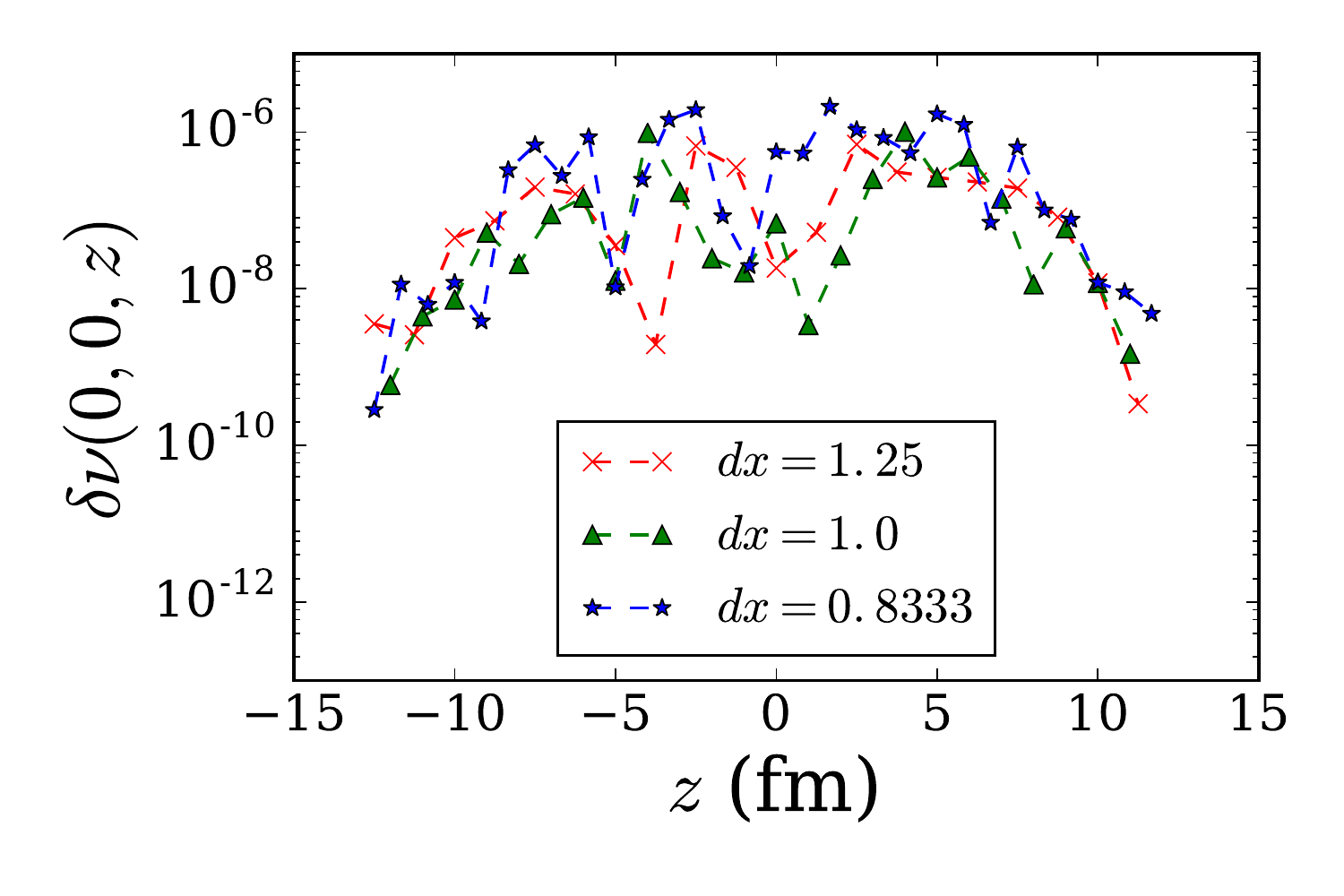}}
\caption{ \label{fig:errrhonu}
Comparisons were made between the local densities calculated by
shifted COCG iteration and the direct diagonalization approaches along
$z$ axis for fixed $x=0$ and $y=0$ coordinates with three different lattice
constants $dx$.  In the case with spherical symmetric potential (top
row) the system size is $L_x=L_y=L_z=20$~fm, and in the deformed case (bottom
row) the system size is $L_x=L_y=20$ and $L_z=25$~fm.  The left column
shows differences for the normal density
$\delta\rho=|\rho_{\textrm{cocg}}-\rho_{\textrm{diag.}}|$, and  the right
column shows them for the anomalous density
$\delta\nu=|\nu_{\textrm{cocg}}-\nu_{\textrm{diag.}}|$.}
\end{figure*}
Until now our discussion of the convergence behavior has been for a
fixed spatial point in the lattice system. To have a comprehensive
view of the accuracy of this method, we also compare the difference
between the local densities $\rho$ and $\nu$ calculated with the shifted
COCG method and with the direct diagonalization method on {\it all} lattice
spatial points. Figure.~\ref{fig:errrhonu} displays the differences for
spherical and deformed W-S model along the $z$ axis for fixed $x=0$ and
$y=0$ coordinates.  Because the local densities on this spatial line
have the largest magnitude throughout the box, we will also get the
largest errors on this line. To study the influence of the lattice
constants on the accuracy, besides the previous $dx=1.25$, we also
performed calculations with smaller lattice constants $dx=1.0$ and $dx=0.8333$. From the
figures one can see the global maximum error of $\rho$ and $\nu$ are
respectively $10^{-8}$ and $10^{-6}$, which matches the convergence
behavior of Fig.~\ref{fig:errcont}. The kinetic and spin-current
densities are calculated with similar accuracy as achieved for the
calculation of normal density.  Moreover, the accuracy is little
affected by the value of the lattice constant, a fact which can be
expected as argued in Sec.~\ref{introduction}; see also
Refs.~\cite{Bulgac2013,Ryssens2015a}.

\subsection{Exploiting symmetries} \label{subsec:axialsymmetry}

Densities on each spatial point in the system are calculated
independently.  The number of points to be processed is equal to the
number of lattice points $N_xN_yN_z$.  This number can be easily
reduced if the system exhibits symmetries, like reflection symmetries,
axial symmetry, or spherical symmetry.  For axially or spherically
symmetric systems, one can argue that it is more profitable to exploit
this symmetry directly on the level of HFB equations, i.e., assume the
correct symmetry for the wavefunctions and solve the constrained HFB
problem. However, if the solution of HFB problem will be used as
initial point for an unconstrained 3D simulation, then this approach
is not accurate enough. Typically, the solutions obtained from a solver
which explicitly uses spherical or axial symmetries once discretized on
a 3D spatial lattice are no longer orthogonal to each other and do not
represent self-consistent eigenstates (with the necessary numerical
accuracy for a stable numerical integration in time). Self-consistent
iterations in full 3D space are still required to get a high quality
state.  We demonstrate the utilization of axial symmetry for reducing
the computational cost, while the underlying HFB matrix is defined in
the full 3D coordinate space.

Consider the scalar local densities represented in cylindrical
coordinates $ ( r, \phi, z) $:
\begin{align*}
\rho(\vvec{r}) & = \rho(r,z), \\
\nu(\vvec{r}) & = \nu(r,z), \\
\tau(\vvec{r}) & = \tau(r,z).
\end{align*}  
We only need to calculate the densities on the points with different
values of $(r,z)$'s in the system. This reduces the number of points to
be calculated to approximately $1/8$ of the total number of points in
the full 3D lattice system, as only the points with $0\leq x\leq y\leq
L/2~(L=L_x=L_y) $ need to be explicitly considered. Thus the performance will improve
by a factor of 8. A reflection symmetry along the $z$-axis can add
another factor of 2, thus an overall speedup of 16.

We  re-expand the vector density $\vvec{J}(\vvec{r})$ along the 
three unit directions of the cylindrical coordinate system as $J_r$,
$J_\phi$, $J_z$, which can be related to the Cartesian components
$J_x$, $J_y$, $J_z$ by the transformation
\begin{align}\label{eq:Car2Cyl}
J_r (r,z)& = J_x \cos \phi + J_y \sin \phi, \nonumber\\
J_\phi(r,z) & = J_y \cos \phi - J_x \sin \phi, \\
J_z(r,z) & = J_z,\nonumber
\end{align}
and reversely
\begin{align}\label{eq:Cyl2Car}
J_x & = J_r \cos \phi - J_\phi \sin \phi, \nonumber\\
J_y & = J_r \sin \phi + J_\phi \cos \phi, \\
J_z & = J_z.\nonumber
\end{align}
The components $J_r$, $J_\phi$, $J_z$ have axial symmetry, i.e., all of
them depend only on $r$ and $z$, while the Cartesian components $J_x$,
$J_y$, and $J_z$ depend on all three spatial coordinates. In our
approach, we first calculate the Cartesian components $J_x$, $J_y$,
$J_z$ for the set of points with different $(r,z)$, we transform
them into the cylindrical components $J_r$, $J_\phi$, $J_z$ via
Eqs.~\eqref{eq:Car2Cyl}, and using the axial symmetry we cover all
the equivalent points in Cartesian space via reverse transformation,
using Eqs.~\eqref{eq:Cyl2Car}.

Numerically, we found that the axial symmetry is not strictly realized
due to the finite size of the lattice constant. In the tests discussed above for
a spherical or deformed W-S model, we start from a HFB Hamiltonian with
strict axial symmetry, and calculate the local densities on all points
of the lattice system without enforcing any symmetry constraints. In
theory, for example, the scalar densities ($\rho$, $\nu$, $\tau$) on the
point of Cartesian coordinates (in fm): $(0, 5, 0)$ should have the same
value ad on the point $(3, 4, 0)$, but numerically they are different
due to the cubic symmetry of the simulation box. The relative
difference between the scalar densities on these pairs of
pseudo-equivalent points varies from $10^{-2}$ to $10^{-4}$ as the
lattice constants decreases from $dx=1.25~\mathrm{fm}$ to
$dx=0.8333~\mathrm{fm}$. In order to fix this issue, we include all
points in these pairs into the calculation, with a small increase of the 
total number of lattice points needed to be calculated.

This lattice effect becomes more pronounced when one calculates the
cylindrical components of the vector density
$\vvec{J}(\vvec{r})$. The components $J_r$ and $J_z$ can be treated in the
same manner as the scalar densities, but $J_\phi$, which is
expected theoretically to vanish, suffers from a numerical noise, which
oscillates around zero. The average magnitude of this noise decreases
from $10^{-5}$ to $10^{-7}$ as $dx$ decreases from $1.25~\mathrm{fm}$
to $0.8333~\mathrm{fm}$ and for simplicity we just force them to be
zero, which will bring some discrepancy in the calculation of
$\vvec{J}(\vvec{r})$ when comparing the COCG and the direct 
diagonalization approaches. We will discuss this discrepancy in realistic
calculations for finite nuclei in Sec.~\ref{benchmark} and show that
it vanishes, as naturally expected, as the lattice constant goes to 0.

\subsection{Computational cost} \label{compcost}

The computational cost of the shifted COCG iteration is set by the
cost of solving the reference system, as it involves matrix-vector
(MV) multiplication between the matrix $A = z_0 - H$ and the vector
$\vvec{p}_n$ of size $2N$, where $N = 4N_xN_yN_z$.  The non-locality
in the HFB matrix (see the Appendix for explicit form) comes from the Laplacian or gradient operators.  When
discretized, the gradient or the Laplacian of a function (vector) can
be obtained either via a finite difference formula or through the Fourier
transform. In both cases the operation is represented by a sparse
matrix. Exploiting the sparsity accelerates the MV operations
significantly.  We compute derivatives using Fourier transforms, as it
is a more accurate method than using finite difference formulas~\cite{Ryssens2015a}.
Moreover, we advocate the direct use of fast Fourier transforms (FFT)
due to a manageable complexity $O(N\log N)$. The computation of
derivatives with FFT is expected to be faster than accurate multipoint
finite difference algorithms~\cite{Aichinger2005188}.  For shifted
systems, the collinear theorem insures that there is no need to
evaluate MV products.  The vector-scalar arithmetic in
Eqs.~(\ref{eq:COCG_xn}), (\ref{eq:COCG_pn}) and (\ref{eq:colinear})
for $\sigma \neq 0$ makes a considerable contribution to the total
computational cost if all $2Nm$ elements of $\vvec{x}^\sigma_n$,
$\vvec{p}^\sigma_n$, $\vvec{r}^\sigma_n$ are calculated, where $m$ is
the number of points on the contour.  However, in order to obtain the
local densities on a fixed spatial point $\vvec{r}^\prime$, we do not
need to know the Green's function $G(z,\vvec{r}; \vvec{r}^\prime)$ for
all spatial points $\vvec{r}$ in the system. This is because integrand
function $f(z)$ requires either $\left. G(z,\vvec{r};
\vvec{r}^\prime)\right |_{\vvec{r}= \vvec{r}^\prime}$ or overlap of
Green's function with $\vvec{\nabla} \delta(\vvec{r}-\vvec{r}^\prime)$
or $\Delta \delta(\vvec{r}-\vvec{r}^\prime)$. Derivatives of
$\delta(\vvec{r}-\vvec{r}^\prime)$ have only $l = N_x+N_y+N_z-2$
non-zero elements (which are located on three lines in $x, y, z$
directions that cross at $\vvec{r}^\prime$). The vector-scalar
arithmetic can be executed only for elements required by the integrand
functions. Finally, the size of $\vvec{x}^\sigma_n$,
$\vvec{p}^\sigma_n$, $\vvec{r}^\sigma_n$ is reduced to the order
$2lm$, which is a small number when compared with $N$ for the reference
system. The cost related to solving the  shifted systems turns out to
be negligible.  Therefore, the theoretical upper bound of the computational cost
within one full self-consistent iteration of shifted COCG method at
$N$ coordinates is determined by the 
{ FFT} operation [$O(N\log N)$] and the maximum number
of COCG Krylov iterations $M$ required to converge each point in the reference
system. The number of Krylov iterations to converge is problem 
dependent and theoretically it scales as $O(N)$, but in all of our realistic 
calculation we find that $M \approx4,000$ is sufficient to get a
high accuracy solution for nuclear problems.
Final numerical complexity scales as $O(M N^2 \log N)$ where in 
practice $M \ll N$, but in the worst case $M$ can be of the same order as $N$. 
On the other hand, the direct diagonalization scales like $O(N^3)$. It is not \textit{a
priori} clear that the COCG approach is preferred. It is at this point
that leveraging the strong scalability feature of the COCG approach
wins. Let us use a distinct parallel process for each point in the
reference system reducing the calculation to the cost $O(MN\log
N)$. Thus, within the errors of the converged results
we present, the COCG approach gains a clear advantage for
larger dimensional systems. 

We have also investigated other algorithms that can be applied to
non-symmetric shifted linear systems, the  shifted BiCG-Stab($l$)
\cite{Frommer2003} and the shifted GMRES \cite{Frommer1998}. In these
algorithms, we do not need to play the tricks described in Sec.~\ref{algorithm}
for the Hermitian matrix $H$ and the size of matrix $A$ will not be
increased by a factor of two. But, these algorithms require two
sequential matrix-vector (MV) products in each iteration; thus in
total there is no profit per iteration. Among the tested methods, the COCG
method exhibits so far the best convergence properties.

Finally, we emphasize that the presented method can efficiently
utilize heterogeneous computers. In our implementation we perform
calculations both using CPUs as well as highly efficient
multi-threaded GPUs. In our experience, the GPU implementation of
shifted COCG is more than 50$\times$ faster than its CPU counterpart.  Table \ref{table:timing}
compares the charged core hours between our GPU code implementing the shifted COCG method 
and the CPU code using the direct diagonalization method ({\footnotesize ScaLAPACK}) for 
two problems with different lattice dimensions. 
On supercomputers, this is the most relevant quantity to be compared as it essentially represents the cost that a user has to pay for calculations.  
For shifted COCG method, the axial
symmetry described in Sec.~\ref{subsec:axialsymmetry} is implemented.
These timing tests were performed at {OLCF} Titan \cite{titan}
and NERSC Edison \cite{edison} supercomputers. 
In both problems,
shifted COCG is faster than diagonalization, and this advantage becomes
more pronounced in the problem with larger dimension (no.2).

\input{table_timing.tex}

\section{Benchmark examples}  \label{benchmark}

In this section, the Green's function and the shifted COCG method,
which we denote {\footnotesize XCOCG}, is benchmarked by solving the
self-consistent HFB equation in 3D coordinate space and compared with
the codes used in Refs.~ \cite{Stetcu2011, Stetcu2015, Bulgac2016}, which we
denote {\footnotesize XDIAG}. The {\footnotesize XDIAG} code extracts wave functions
of HFB Hamiltonian~(\ref{eq:hfbspin}) in the discrete variable representation (DVR) basis~\cite{Bulgac2013}
via a direct diagonalization. \footnote{Direct methods
yield the exact solution if the precision is unlimited whereas
indirect methods may be even more accurate and are often more
efficient depending on system details. Also, iterative methods more
tend to naturally damp out roundoff errors that accumulate and become
difficult with large $N$ problems evaluated directly.} 
The diagonalization procedure is executed parallel using 
{\footnotesize ScaLAPACK} library \cite{scalapack}.
Next, densities are formed from the wave-functions 
using formulas~(\ref{eq:rho}-\ref{sod2}). These operations form a single self-consistent iteration.
We mix the intermediate solutions during the iteration process
using a linear or a Broyden mixing algorithm~\cite{Baran2008}.
The {\footnotesize XCOCG} code is a modified version of the  {\footnotesize XDIAG} code
where parallel diagonalization procedure and computation of densities
is replaced by the COCG method.
Densities are extracted according formulas provided in Sec.~\ref{sec:COCGdensities}.
The COCG part is ported to GPUs. Both codes provide the same results
up to the accuracy specified in Sec.~\ref{convbh}. 

We performed tests  on the nuclei with axial
symmetry, including spherical and axially deformed nuclei. In all of
these realistic tests, the Skyrme NEDF SLy4 \cite{Chabanat1998} is
used in p-h channel, and the SLDA treatment \cite{Bulgac2002, Yu2003}
for the pairing interaction is used in p-p channel with bare coupling strength 
$g_0 (\vvec{r}) = g_0 = -233~\mathrm{MeV}$ (volume pairing), 
and energy cutoff $E_{\mathrm{cut}} = 100~\mathrm{MeV}$. These are typical values used in realistic calculations. 
We conclude this
section by presenting states generated by the shifted COCG method  used
for studies of interaction between quantized vortices and nuclei in
neutron start crust~\cite{Wlazlowski2016}. These calculations are
not feasible for {\footnotesize XDIAG} code in a reasonable time.

\input{table_Pb208.tex}

\input{table_Ca40.tex}

\input{table_Ni62.tex}

\input{table_Zr102.tex}

\subsection{Spherical limit: Doubly magic nuclei $^{208}$Pb, $^{40}$Ca and
  semi magic nucleus $^{62}$Ni} \label{subsec:sph}
  
A common approach to testing the accuracy of a 3D coordinate solver is to compare the 
3D results for a spherical nucleus with the results obtained by a 1D spherical code that also represents the 
single-particle wave functions in coordinate space. We can choose extremely fine lattice
constants for the 1D solver and its results can be considered to be very accurate. 
Instead of a benchmark with {\footnotesize XDIAG}, we compute the double magic nucleus
$^{208}$Pb using our {\footnotesize XCOCG} code in a cubic box of size $32^3~\mathrm{fm}^3$ 
with lattice constants $dx=1.25, 1.0, 0.8~\mathrm{fm}$,
and compare them with the results obtained with the {\footnotesize HFBRAD} code~\cite{Bennaceur2005} code 
(the lattice constant $dx=0.05~\mathrm{fm}$). In Table \ref{table:Pb208} we compare various 
contributions to the total energy $E_{\mathrm{tot}}$, computed as volume integrals from 
corresponding terms in NEDF for $^{208}$Pb in these situations. From $dx=1.25$ to 
$0.8~\mathrm{fm}$, the difference of total energy between {\footnotesize XCOCG} and 
{\footnotesize HFBRAD} decreases from 0.5 MeV to 4 keV. In particular, from $dx=1.0$ to 
$0.8~\mathrm{fm}$, the values of the energy terms have a steady convergence to those 
solved by {\footnotesize HFBRAD} with the maximum difference $\leq 100~$keV. This 
numerical accuracy and convergence pattern is similar to the results in Ref.~\cite{Ryssens2015a}, 
which uses Lagrange-mesh representation \cite{Baye20151} in the calculation of spatial derivatives.
The Lagrange-mesh method is equivalent to the DVR method using {FFT} on a 3D spatial lattice.

The benchmarking with  {\footnotesize XDIAG} starts with the case of spherical nuclei. 
We solve the HFB problem for the doubly
magic nucleus $^{40}\mathrm{Ca}$ and semi-magic nucleus
$^{62}\mathrm{Ni}$ to demonstrate the accuracy of our solver in the
cases with and without pairing. In both cases, we use a cubic lattice
of size $20^3~\mathrm{fm}^3$ with different lattice constants
$\mathrm{d}x = 1.25, 1.0, 0.8333 ~ \mathrm{fm}$. In Table
\ref{table:Ca40} we compare various contributions to the total energy of $^{40}$Ca 
calculated by {\footnotesize XCOCG} and
{\footnotesize XDIAG} methods respectively. The main source of differences between the
results of the two solvers is due to the neglect of the azimuthal component
of the spin-orbit density $\vvec{J}_\phi(\vvec{r})$ when we utilize
the axial symmetry of the system. This lattice effect vanishes as the
lattice constant decreases. In particular, when $dx \leq
1~\mathrm{fm}$, the difference between the total energies is less than
10 keV. Notice also that the {\footnotesize XCOCG} energies are always lower than 
the {\footnotesize XDIAG} energies and we attribute this to the fact that 
$\vvec{J}_\phi(\vvec{r})\equiv 0$ in {\footnotesize XCOCG}. 
The same kind of calculations are performed for the semi-magic nucleus
$^{62}$Ni with non-zero neutron pairing; see
Table~\ref{table:Ni62}. Compared to $^{40}$Ca, the difference of
total energy is larger in $dx = 1.25$ case ($0.175$ MeV). This is due
to the much larger magnitude of the spin-orbit contribution, which
brings larger error in $\vvec{J}$ caused by the finite  lattice
effects. Similarly to $^{40}$Ca, as the lattice constant becomes finer,
this lattice effect vanishes and the difference of total energy drops
to values below $10$ keV when $dx \leq 1~ \mathrm{fm}$.


\subsection{Axially deformed nucleus: $^{102}$Zr}\label{subsec:deform}

The advantage of solving the HFB equation in a coordinate space basis
is that it can correctly describe the asymptotic behavior of
quasiparticle wavefunctions of nuclei with large deformations and weak
binding energies \cite{Dobaczewski1996}. Following 	Refs.~\cite{Pei2008,
Teran2003, Blazkiewicz2005}, we choose the neutron rich Zr isotope
$^{102}$Zr, which has a large prolate deformation, as the testing
ground and we choose a rectangular box of size $22.5 \times 22.5 \times 30~
\mathrm{fm}^3$ to fit its large deformation. As
in Sec.~\ref{subsec:sph}, we compare the {\footnotesize XCOCG} and
{\footnotesize XDIAG} results with different lattice sizes $dx=1.25$
and $0.9375~\mathrm{fm}$. These are shown in
Table.~\ref{table:Zr102}. The quadrupole moment $Q_{20}$ of the nucleus is
also listed in each case, where 
\begin{align} \label{eq:q20}
Q_{20} = \langle \hat{Q} \rangle = \int(2 z^2 - x^2 - y^2 ) \rho (\vvec{r})\,d^3 \vvec{r}
\end{align} 
and $\rho(\vvec{r}) = \rho_n(\vvec{r}) + \rho_p(\vvec{r})$.

The difference in the total energy is 0.3 MeV between {\footnotesize
XCOCG} and {\footnotesize XDIAG} for $dx = 1.25$, which almost equals
to the difference in the spin-orbit energy 0.27 MeV.  This error in
the spin-orbit energy contribution will affect the position of the HFB
minimum, leading to a difference in the quadrupole moment $Q_{20}$ in
the ground state of $3~\mathrm{fm}^2$ and in the energy terms
$E_{\rho^2}, E_{\rho^\gamma}$ of 1--2 MeV between {\footnotesize XCOCG}
and {\footnotesize XDIAG}. For $dx=0.9375\fm$, the errors are
significantly smaller, $\leq 1~\mathrm{keV}$ for the total energy and
$0.16~\mathrm{fm}^2$ for the quadrupole moment.

\subsection{Constrained HFB: saddle-point of $^{240}$Pu
  fission} \label{subsec:pu240}

Induced fission of $^{240}$Pu is a frequent benchmark for many
implementations based on DFT methods \cite{Younes2009, Schunck2014,
Bulgac2016, Regnier2016}. In a constrained HFB calculation in
Ref.~\cite{Bulgac2016} the nucleus is brought to a shape and an energy near
the outer saddle point of the fission barrier (at zero temperature),
used as the initial state of a time-dependent SLDA
(TDSLDA) simulation. In this work, we use our {\footnotesize XCOCG}
code to reproduce the configuration on this saddle point and compare
with that obtained via the {\footnotesize XDIAG} code.

\input{table_saddlepoint.tex}

To obtain a nucleus with a given quadrupole moment $\langle \hat{Q}
\rangle = Q_0$, we need to minimize the Routhian
\begin{align} \label{eq:routhian}
E' = E_{\mathrm{total}} + c (\langle \hat{Q} \rangle - Q_0)^2, 
\end{align}
which is equivalent to adding a Lagrange multiplier in the
single-particle Hamiltonian: $h' = h + 2c(\langle \hat{Q} \rangle -
Q_0)\hat{Q}$. In the self-consistent calculation, the constraint
strength $c$ will be updated in each iteration using the augmented
Lagrangian method~\cite{Staszczak2010}. This saddle-point
configuration of $^{240}$Pu is prepared in a rectangular box of size
$25 \times 25 \times 50~\mathrm{fm}^3$, with lattice spacing $dx =
1.25~\mathrm{fm}$. Following paper~\cite{Bulgac2016}, the quadrupole
moment constraint is set to $Q_0 = 16500~\mathrm{fm}^2$ and an
additional auxiliary external field is turned on for the formation of
the neck in this configuration.

\begin{figure}[h]
\includegraphics[clip, width=1.0\columnwidth]{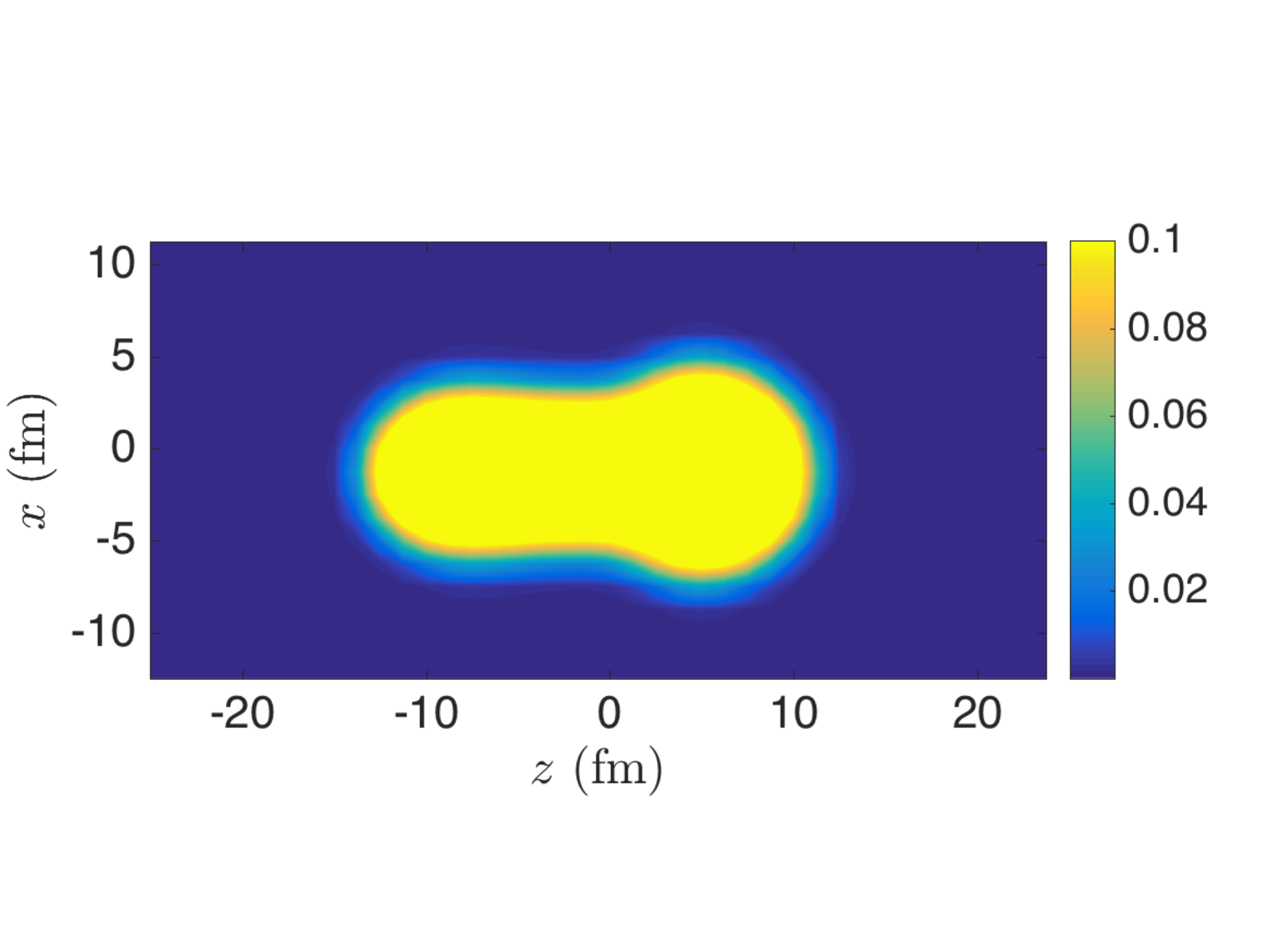}
\caption{\label{fig:saddlepoint}
Total density distribution $\rho(\vvec{r}) = \rho_n(\vvec{r}) +
\rho_p(\vvec{r})$ for the saddle-point configuration of $^{240}$Pu in
the $y=0$ plane.  }
\end{figure}

In Fig.~\ref{fig:saddlepoint} we show the density profile for the
converged solution.
Table~\ref{table:Pu240} compares the {\footnotesize XCOCG} and
{\footnotesize XDIAG} results for the saddle-point configuration of
$^{240}$Pu. As we discussed earlier, the lattice effects in the
calculation of $\vvec{J}$ cause the change of the HFB minimum, 
which results in large differences in the
energy terms $E_{\mathrm{kin}}$, $E_{\rho^2}$, $E_{\rho^{\gamma}}$, and
$E_{\rho \Delta \rho}$. In this case, because of the strong pairing, and 
extremely large deformation of the heavy nucleus, any minute change of HFB minimum will
result in larger difference in various energy terms than in the cases of the nuclei we tested
earlier. On the other hand, the larger amplitude of $\vvec{J}$,
which is reflected by the absolute value of $E_{\rho \nabla J}$, also makes the 
lattice effects more pronounced.
In principle, these differences will reduce and eventually vanish as $dx \to 0$ as we demonstrated
in the earlier sections. But the large dimension of this system and the considerable number of self-consistent 
iterations in the constrained HFB problem will make the calculation on
a finer lattice grid more expensive
with the current computational resources.

\subsection{A nucleus immersed in a neutron superfluid
  sea} \label{subsec:NScrust}

We used the shifted COCG method to generate initial states for studies
of quantum vortex dynamics in a neutron star
crust~\cite{Wlazlowski2016}.  The stationary state is a superfluid
neutron medium containing a quantum vortex and an immersed nucleus
located in the vicinity of the topological defect or vortex.  The
calculations were performed with the FaNDF\textsuperscript{0} nuclear
density functional constructed by Fayans \textit{et al.}~\cite{Fayans1998,
Fayans200049}, which is particularly well suited for these type of
studies. The bare pairing coupling constant in Eq.~(\ref{eq:pairing})
was chosen to reproduce the rescaled BCS ${}^1S_0$ pairing gap in
neutron matter; for more details, see Ref.~\cite{Wlazlowski2016}.

For these studies we used a simulation volume of size
$75 \times75 \times 60\fm^3$ with a lattice spacing $dx=1.5\fm$. The
energy cut-off was chosen to be $75\MeV$. In the box, we place a tube
(simulated by a flat-bottomed external potential) that we fill with
superfluid neutrons of density $n=0.014\fm^{-3}$ or
$0.031\fm^{-3}$. The problem has been simplified by dropping the
spin-orbit term, which is not expected to play a major role in vortex
pinning. The simplification results in an HFB matrix
(\ref{eq:hfbspin}) with a simpler block structure ($h_{\uparrow
\downarrow}=h_{ \downarrow\uparrow}=0$) and a smaller dimension
$2N_xN_yN_z$:
\begin{align}\label{eq:hfbreduced}
\begin{pmatrix}
h_{\uparrow \uparrow} -\mu  & \Delta \\
\Delta^* &  -h^*_{\downarrow \downarrow} + \mu
\end{pmatrix}
\begin{pmatrix}
u_{k\uparrow} \\
v_{k\downarrow}
\end{pmatrix}
= E_k
\begin{pmatrix}
u_{k\uparrow} \\
v_{k\downarrow}
\end{pmatrix}.
\end{align} 
After this reduction, the HFB matrix has the size $200,000\times 200,000$
and it still represents a very demanding problem for the traditional
approaches to determine the stationary states, see Section~\ref{compcost} and Table~\ref{table:timing}.   
We solved this problem
successfully with moderate computational costs to achieve
self-consistency using GPUs and the COCG approach described here
on the Titan supercomputer~\cite{titan}. Only after the final
iteration, do we use a diagonalization
to generate the wave functions on the Edison supercomputer~\cite{edison}.

\begin{figure}[th]
\includegraphics[width=1.0\columnwidth, trim=149 105 130 40, clip]{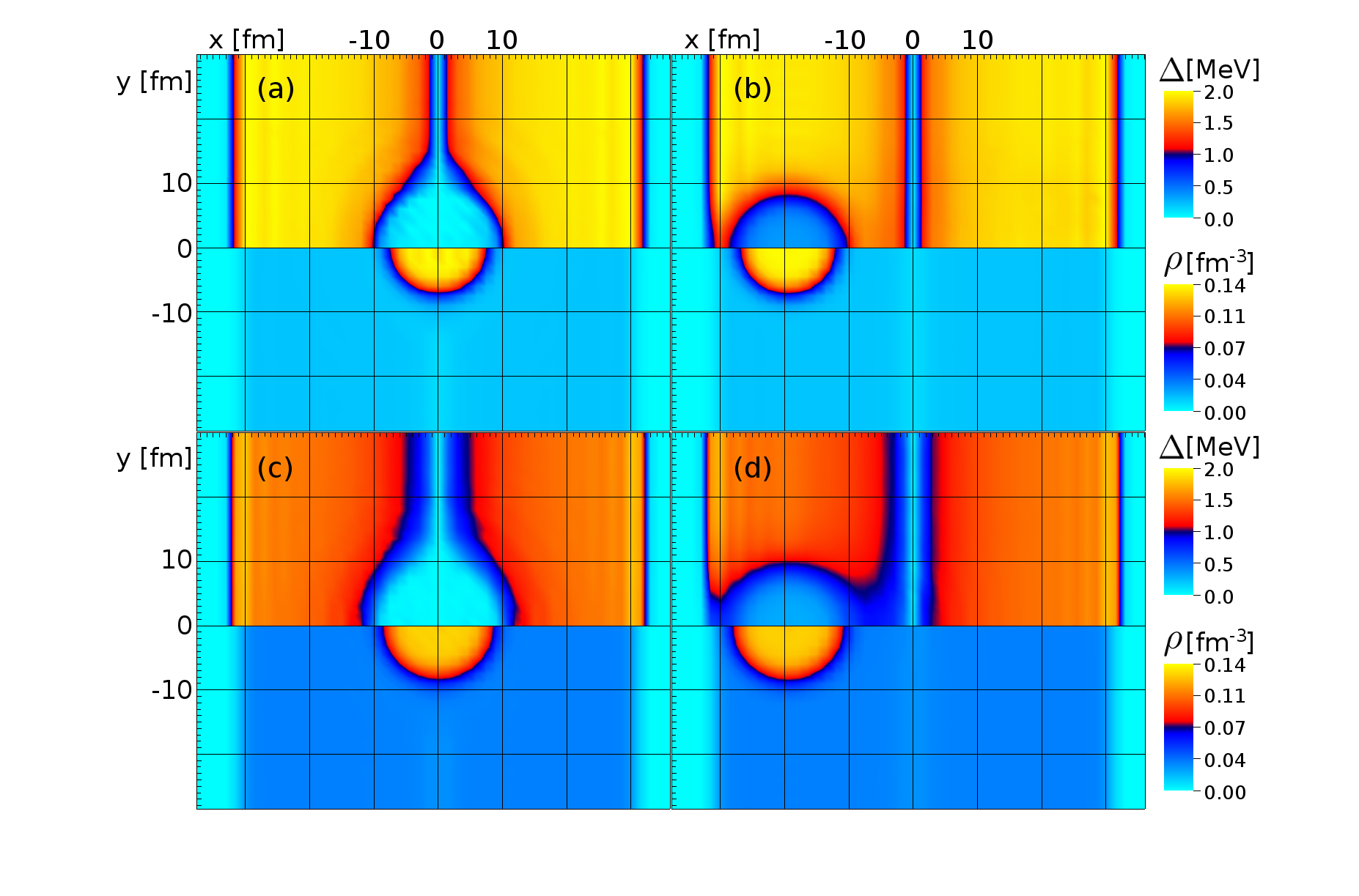}
\caption{ The lowest energy states generated by shifted COCG method for neutron 
background density $n=0.014\fm^{-3}$ [panels a and b] and $0.031\fm^{-3}$ [panels c and d]. In each 
box of size $75 \times 60\fm^2$ we show the absolute value of the neutron pairing potential $\Delta(\vvec{r})$ 
(upper half) and the total density distribution $\rho(\vvec{r})$ 
(lower half). The black lines separating blue and red regions correspond to a
value $1\MeV$ for the paring and of $0.07\fm^{-3}$ for the density respectively. \label{fig:initstates_frames}}
\end{figure}

In Fig.~\ref{fig:initstates_frames} we present stationary
configurations (with constraints) for two background neutron
densities: $n=0.014\fm^{-3}$ and $0.031\fm^{-3}$.  The ``nuclear
defect'' consists of $Z=50$ protons. Two mutual configurations were
considered: (1) a quantum vortex attached to the nucleus (pinned case);
and (2) a nucleus outside the vortex core (unpinned case). The position of
nucleus in the box was fixed by adding a constraint to the density
functional for the center of mass of the protons, in a similar fashion as
it was done for the quadrupole moment Eq.~(\ref{eq:routhian}) and for
the for saddle point in $^{240}$Pu test case. The vortex was generated
by imprinting the correct phase pattern for the pairing field $\Delta$
in the neutron channel 
$\Delta(\rho, z, \phi) = |\Delta(\rho,z)|\exp(i\phi)$, where
$\rho=\sqrt{x^2+y^2}$ is the distance from the center of the tube and
$\phi = \tan^{-1}\tfrac{y}{x}$.  In the pinned case, we took advantage of the
axial symmetry of the problem and of the reflection symmetry with respect 
to one plane in the unpinned configuration.
 
From the energetics of these systems we determine that for both
densities the configuration with the nucleus located outside the
vortex core has a lower energy per particle than the pinned configuration by about
$6$ and $4\;\textrm{keV}$ respectively for densities
$n=0.014\fm^{-3}$ and $0.031\fm^{-3}$. Thus one can expect that the
effective vortex-nucleus interaction is repulsive in nature, which was
further confirmed by studying the motion of the vortex in dynamical
simulations~\cite{Wlazlowski2016}.

\section{Further extension}
\subsection{Linear response}

The free linear response of a many-fermion system can be evaluated 
using the same COCG approach. For simplicity we will illustrate this 
procedure here only for a normal system, as the extension to the 
superfluid case is straightforward. The free polarization operator 
is defined as~\cite{Shlomo1975,Bertsch1975}
\begin{align}
\begin{split}
\Pi_0(\omega, \vvec{r},\vvec{r}')=
\sum_k \psi_k(\vvec{r}) \psi^*_k(\vvec{r}') \\
\times \left [G(\varepsilon_k+\omega, \vvec{r},\vvec{r}')  +
        G(\varepsilon_k-\omega,\vvec{r},\vvec{r}') \right ],
\end{split}
\end{align}
where $\psi_k(\vvec{r})$ and $G(\vvec{r},\vvec{r}',z)$  were defined in
Eqs. \eqref{eq:psi_k} and \eqref{eq:gf}, 
and the summation is over occupied levels. In practice the polarization 
operator is evaluated for a complex energy $\omega+i\gamma$, since 
the spreading width $\Gamma^\downarrow$ is either not accounted for
in the random phase approximation, or in order to imitate to some extent 
the imaginary part of the optical potential. The spreading width accounts for the 
fragmentation of the particle-hole transition strength due to the coupling to 
more complex states.   If the complex integration contour
in Eq.~\eqref{eq:cont} is chosen so that $|\mathrm{Im}\;z| < \gamma$ one 
can easily show that
\begin{align}
\begin{split}
\Pi_0(\omega+i\gamma, \vvec{r}, \vvec{r}')= \frac{1}{2\pi i}\oint_C 
G(z , \vvec{r},\vvec{r}') \\
\times \left [G(z+\omega +i\gamma, \vvec{r},\vvec{r}')  +
        G(z- \omega-  i\gamma, \vvec{r},\vvec{r}') \right ].
        \end{split}
\end{align}
It remains to be established if this method for the evaluation of the free
polarization operator  $\Pi_0(\omega+i\gamma, \vvec{r},\vvec{r}')$ and  
the subsequent determination of the full response is competitive and 
under what conditions with the finite-amplitude 
method~\cite{Nakatsukasa2007,Avogadro2011,Nakatsukasa2014,
Hinohara2013,Oishi2016,Mustonen2014} and/or the time-dependent
approach~\cite{Stetcu2011,Stetcu2015} for deform open-shell nuclei.

\subsection{Extraction of eigenvalues and eigenvectors using the shifted COCG Krylov method}

The shifted COCG Krylov method can be used also to determine the 
eigenvalues and the eigenvectors using the approach 
described in Refs.~\cite{Sakurai2003, Sakurai2008, Ikegami2010}. 
In this approach one has at first to evaluate the moments
\begin{align}
\mu_k = \frac{1}{2\pi i} \oint_\Gamma dz \sum_n\frac{(z-\epsilon)^k}{z-\varepsilon_n}, \;\;\; k=0,1, ... ,2N-1.
\end{align}
where $\epsilon$ is located inside the contour $\Gamma$ enclosing a segment on 
the real axis with a known number $N$ of the eigenvalues, 
and where $\varepsilon_n$ are eigenvalues. Once these moments are computed 
the eigenvalues are obtained by solving a generalized eigenvalue 
problem for two matrices of size $N \times N$.  
The number of eigenvalues in a 
given interval is not known \textit{a priori} and some eigenvalues could also be degenerate. In the 
presence of degeneracies one has to disentangle the corresponding eigenvectors.  If the degeneracy 
is due to spherical or axial symmetry one can introduce slightly different lattice constants 
$dx$, $dy$, and $dz$ or a very weak external field and lift the degeneracies
at a level that has no noticeable consequence on the physics studied.  Subsequently, the solution of the  
Schr\"odinger equation for the corresponding eigenvector $\psi_n$ 
in the case of a known non-degenerate eigenvalue $(H-\varepsilon_n)\psi_n=0$ is a trivial linear algebra problem. 
In the case of Kramers degeneracies, one can easily separate the two degenerate eigenvectors. The unknown 
number $N$ of eigenvalues in a given energy interval can be determined by evaluating 
the trace (integral over all coordinates and summation over all four components) of the
Green's function of the Hamiltonian~\eqref{eq:hfbspin}
\begin{align}
N = \mathrm{Tr} \left (\frac{1}{2\pi i}\oint_\Gamma d z \frac{1}{z-H} \right ).
\end{align} 
Thus the need to use diagonalization of very large matrices can be completely eschewed.

\section{Conclusions} \label{conclusion}

In this paper, we describe a new approach for solving the HFB type of 
equations in a coordinate representation that is different from the
traditional approaches based typically on direct
diagonalizations. In the present approach, there is no need to evaluate
the individual single-particle wave functions and their energies, but instead
we calculate the Green's function of the HFB equations, from which
we extract  various densities after evaluating a contour integral. The
Green's functions are obtained by solving a set of linear equations
with scalar shifts using the iterative shifted COCG Krylov method.  We
demonstrate the high accuracy of the iterative shifted COCG approach by
solving typical nuclear problems with and without complicated
constraints, such as the fission outer saddle-point of $^{240}$Pu and a
quantum vortex state in a neutron star crust.  A notable advantage of
this algorithm is its suitability for efficient parallelization and
effective utilization of heterogeneous computing platforms.  The
method becomes computationally superior for large spatial lattice sizes 
that are otherwise computationally very expensive 
for standard approaches, such as a direct diagonalization.

\begin{acknowledgments}

We thank G.F. Bertsch for discussions and I. Stetcu and P. Magierski
for many helpful discussions and code sharing. This work
was supported in part by U.S. DOE Office of Science Grant
DE-FG02-97ER41014. This work was also supported
in part by the Polish National Science Center (NCN) under Contract No. UMO-2014/13/D/ST3/01940.  
Calculations have been performed: at the
OLCF Titan - resources of the Oak Ridge Leadership Computing Facility,
which is a DOE Office of Science User Facility supported under
Contract DE-AC05-00OR22725; at NERSC Edison - resources of the
National Energy Research Scientific computing Center, which is
supported by the Office of Science of the U.S. Department of Energy
under Contract No. DE-AC02-05CH11231; and at Moonlight, resources of 
the Institutional Computing Program at Los Alamos National Laboratory.

\end{acknowledgments}

\appendix
\begin{widetext}
\section*{APPENDIX: Equation of Green's function for HFB equation} \label{app:gfeq}

For the HFB Hamiltonian $H$ in Eq.~\eqref{eq:hfbspin}, the
single-particle Hamiltonian in Eq.~\eqref{eq:spham} reads
\begin{align}
h= - \vvec{\nabla} \cdot \frac{\hbar^2}{2m^*(\vvec{r})}\vvec{\nabla} + U(\vvec{r})-i \vvec{W}(\vvec{r}) \cdot \left( \vvec{\nabla} \times \vvec{\sigma} \right),
\end{align}
where the kinetic energy term is represented by a real symmetric
operator in numerical implementations $T$. When discretized, besides
the local term $U(\vvec{r})_{mn} = U_n \delta_{nm}$, the nonlocal term
$T$ and spin-orbit terms require more attention. For $T$, since
{\small
\begin{align}
\begin{split}
 &- \vvec{\nabla} \cdot \frac{\hbar^2}{2m^*(\vvec{r})}\vvec{\nabla} v(\vvec{r})  =  -\frac{1}{2} \left[ \frac{\hbar^2}{2m^*(\vvec{r})} \nabla^2 v(\vvec{r})  \right .\\
& \left .+ \nabla^2 \left( \frac{\hbar^2}{2m^*(\vvec{r})}v(\vvec{r}) \right)    - \left( \nabla^2 \frac{\hbar^2}{2m^*(\vvec{r})} \right) v(\vvec{r}) \right],
\end{split}
\end{align}
}
we have
\begin{align}
\begin{split}
&T_{nm} = \left( - \vvec{\nabla} \cdot \frac{\hbar^2}{2m^*(\vvec{r})}\vvec{\nabla} \right)_{nm}  =  \\
& -\frac{1}{2} (\nabla^2)_{nm} \left(\frac{\hbar^2}{2m^*_n} + \frac{\hbar^2}{2m^*_m} \right) 
 + \frac{1}{2} \left( \nabla^2 \frac{\hbar^2}{2m^*} \right)_n \delta_{nm},
\end{split}
\end{align}
where the Laplacian operator $\nabla^2$ is a symmetric matrix in the
discrete variable representation (DVR) basis \cite{Bulgac2013} and $m^*_{n/m}=m^*(\vvec{r}_{n/m})$.  A
similar symmetrization is also performed for the spin-orbit term
\begin{align}
\begin{split}
& \vvec{W} (\vvec{r}) \cdot \left( \vvec{\nabla} \times \vvec{\sigma} \right) \\
& = \frac{1}{2} \left[ \vvec{W} (\vvec{r}) \cdot \left( \vvec{\nabla} \times \vvec{\sigma} \right) + \vvec{\nabla} \cdot \left( \vvec{\sigma} \times \vvec{W} (\vvec{r}) \right) \right],
\end{split}
\end{align}
where the gradient operator $\vvec{\nabla} =
\{ \partial_x, \partial_y, \partial_z \}$ is an anti-symmetric
operator in each spatial direction in the DVR representation. In the p-p
channel, the complex pairing field $\Delta(\vvec{r})$ is diagonal when
discretized, $\Delta(\vvec{r})_{mn} = \Delta_n \delta_{nm}$.  Finally,
after separating the real and imaginary parts of the HFB Hamiltonian
$H = A + iB$, the equation of Green's function $G = G_x + i G_y$ for a
HFB equation is
\begin{gather}
( zI - H' )
\begin{pmatrix}
G_x \\
G_y
\end{pmatrix}
 = 
\begin{pmatrix}
\delta^{(4)}\\
0
\end{pmatrix}, \quad
H' = 
\begin{pmatrix}
A & -B \\
B& A
\end{pmatrix},
\end{gather}
and where 

\begin{align}
A & =
\begin{pmatrix}
T + \tilde{U} &  \widetilde{W_x \partial_z} - \widetilde{W_z \partial_x}  & 0 & \mathrm{Re}~\Delta \\
\widetilde{W_z \partial_x} - \widetilde{W_x \partial_z}  & T + \tilde{U}  & -\mathrm{Re}~ \Delta & 0 \\
0 & -\mathrm{Re} ~ \Delta & - T - \tilde{U}  &  \widetilde{W_z \partial_x} - \widetilde{W_x \partial_z}    \\
\mathrm{Re} ~ \Delta & 0 &   \widetilde{W_x \partial_z} - \widetilde{W_z \partial_x} & -T -\widetilde{U} 
\end{pmatrix}, 
\\
B & = 
\begin{pmatrix}
 \widetilde{W_y \partial_x} - \widetilde{W_x \partial_y} & \widetilde{W_z \partial_y} - \widetilde{W_y \partial_z} & 0 & \mathrm{Im }~ \Delta \\ 
 \widetilde{W_z \partial_y} - \widetilde{W_y \partial_z} & \widetilde{W_x \partial_y} - \widetilde{W_y \partial_x}  &  -\mathrm{Im }~ \Delta & 0 \\
 0 &  \mathrm{Im }~ \Delta &  \widetilde{W_y \partial_x} - \widetilde{W_x \partial_y}  &  \widetilde{W_z \partial_y} - \widetilde{W_y \partial_z}  \\
  -\mathrm{Im }~ \Delta & 0 &  \widetilde{W_z \partial_y} - \widetilde{W_y \partial_z} & \widetilde{W_x \partial_y} - \widetilde{W_y \partial_x} 
\end{pmatrix},
\end{align}

with $\tilde{U} = U - \mu$ and $\widetilde{W_i \partial_j}$ the
anti-commutator of $W_i$ and $\partial_j$:
\begin{align}
\widetilde{W_i \partial_j} = \frac{1}{2} \left(W_i \partial_j + \partial_jW_i  \right), \quad i,j = x, y, z, i \neq j,
\end{align}
One can show that $A = A^{T}$ and $B= -B^{T}$. In the shifted
COCG method, we do not need to construct these $4N_{xyz} \times
4N_{xyz}$ matrices explicitly because we only need the MV
product between these matrices and the vectors $\vvec{x}_n$, $\vvec{p}_n$ (see
Sec.~\ref{algorithm}). Among the MV product operation, the product
between the local part of $H$ and the vectors can be regarded as a
vector-vector product. The product between the vectors and the
non-local part of $H$, due to the Laplacian and gradient operators,
can be performed using the fast Fourier transform.
\end{widetext}

\bibliography{cocg_ref,cocg_ref_gw}

\end{document}

%% file: table_timing.tex
\begin{table}[t]
  \begin{ruledtabular}
    \begin{tabular}{cccc}
    No. & $N$  & $t_\mathrm{cocg}$ (CPU hrs) & $t_\mathrm{diag}$ (CPU hrs)\\
   \hline
  1 &  $20 \times 20 \times 40 \times 4$ & 649.23 (T) & 1,547.71 (T)\\
  2 &  $50 \times 50 \times 40 \times 2$ & 9,318.4 (T) & 46,694.4 (E)
    \end{tabular}
    \end{ruledtabular}
    \caption{\label{table:timing}
   Timing comparison between the shifted COCG method and diagonalization method using ScaLAPACK for solving two problems with different dimensions $N$s in each self-consistent iteration. The first case is for finite nuclei $^{240}$Pu in the lattice of dimension $20 \times 20 \times 40$ (the extra factor of 4 is included for spin-orbit and pairing).  The second case is for the study of a nucleus immersed in a neutron superfluid sea of dimension $50 \times 50 \times 40$, where spin-orbit is ignored (the extra factor of 2 is for pairing only). In both problems, $t_{\mathrm{cocg}}$ and $t_{\mathrm{diag}}$ denote the charged core hours of the {GPU} code using shifted COCG method and the CPU code using ScaLAPACK in one self-consistent iteration. The capital letter in parentheses denotes the computing facility where the timing test is performed: ``T'' represents Titan and ``E'' represents Edison~\cite{titan,edison} and the CPU hours were determined in units of CPU hours according to the corresponding policy for these supercomputers. On Titan an hour using a single computing node, which has 16 CPUs and one GPU, is charged as 30 CPU h. On Edison an hour using a CPU is charged as 2 CPU h.
   For problem no.2 the estimated charge in case of using Titan for direct diagonalization is about 87~500 CPU hrs. 
   }
    \end{table}

%% file: table_Pb208.tex
\begin{table}[h]
  \begin{ruledtabular}
    \begin{tabular}{lrrrr}
        & \multicolumn{1}{c}{\footnotesize $dx=1.25~\mathrm{fm}$} & \multicolumn{1}{c}{\footnotesize $dx=1.0~\mathrm{fm}$ } & \multicolumn{1}{c}{$dx=0.8~\mathrm{fm}$}
        & \multicolumn{1}{c}{\footnotesize {\footnotesize HFBRAD} }\\
      \hline
$E_\mathrm{kin}$& 3868.984& 3866.098& 3866.187& 3866.163\\
$E_{\rho^2}$& -22401.965& -22383.542& -22384.385& -22384.462\\
$E_{\rho^\gamma}$& 14548.826& 14536.219& 14536.840& 14536.890\\
$E_{\rho\Delta\rho}$& 315.563& 315.236& 315.286& 315.288\\
$E_{\rho\tau}$& 1332.134& 1330.147& 1330.208& 1330.216\\
$E_{\rho\nabla J}$& -96.592& -96.451& -96.446& -96.446\\
$E_\mathrm{Coul}$& 796.848& 796.600& 796.607& 796.645\\
$E_\mathrm{tot}$& -1636.202& -1635.693& -1635.703& -1635.707\\
    \end{tabular}
  \end{ruledtabular}
  \caption{\label{table:Pb208}
Results of HFB + SLy4 calculations for $^{208}$Pb using {\footnotesize XCOCG} in a cubic box of size $32^3~\mathrm{fm}^3$ 
with lattice constants $dx=1.25, 1.0, 0.8~\mathrm{fm}$. Results obtained with the spherical 1D code {\footnotesize HFBRAD}
are presented for comparison. All energies are in MeV. 
}
\end{table}

%% file: table_Ca40.tex
\begin{table*}[t]
  \begin{ruledtabular}
    \begin{tabular}{lrrrrrrrrr}
    & \multicolumn{3}{c}{$dx = 1.25~\mathrm{fm}$} &\multicolumn{3}{c}{$dx = 1.0~\mathrm{fm}$} &\multicolumn{3}{c}{$dx = 0.8333~\mathrm{fm}$} \\
\cline{2-4}  \cline{5-7} \cline{8-10} 
& {\footnotesize XDIAG} & {\footnotesize XCOCG} &$\Delta E$ & {\footnotesize XDIAG} & {\footnotesize XCOCG} & $\Delta E$ & {\footnotesize XDIAG} & {\footnotesize XCOCG} & $\Delta E$\\
\hline
$E_\mathrm{kin}$& 623.983& 624.115& 0.132& 624.755& 624.781& 0.026& 624.836& 624.837& <0.001\\
$E_{\rho^2}$& -3714.569& -3714.702& -0.133& -3722.988& -3722.977& 0.011& -3723.483& -3723.483& <0.001\\
$E_{\rho^\gamma}$& 2396.941& 2396.967& 0.026& 2403.198& 2403.190& -0.008& 2403.555& 2403.554& <0.001\\
$E_{\rho\Delta\rho}$& 106.284& 106.368& 0.084& 106.957& 106.957& <0.001& 106.992& 106.992& <0.001\\
$E_{\rho\tau}$& 173.227& 173.122& -0.105& 173.725& 173.712& -0.013& 173.746& 173.746& <0.001\\
$E_{\rho\nabla J}$& -1.233& -1.267& -0.033& -1.279& -1.287& -0.008& -1.282& -1.282& <0.001\\
$E_\mathrm{Coul}$& 71.483& 71.484& <0.001& 71.532& 71.532& <0.001& 71.535& 71.535& <0.001\\
$E_\mathrm{tot}$& -343.885& -343.914& -0.029& -344.100& -344.093& 0.007& -344.101& -344.101& <0.001\\
    \end{tabular}
  \end{ruledtabular}
\caption{\label{table:Ca40}
Results of spherical HFB + SLy4 calculations for $^{40}$Ca using the {\footnotesize XCOCG} and {\footnotesize XDIAG} approaches 
with a cubic 3D lattice of size $L_x = 20 ~\mathrm{fm}$ with different mesh size $dx$. All energies are in MeV.
In the column of $\Delta E$, ``$<0.001$'' means $\lvert \Delta E \lvert$ is less than 1 keV and can be negligible.
}
\end{table*}

%% file: table_Ni62.tex
\begin{table*}[t]
  \begin{ruledtabular}
    \begin{tabular}{lrrrrrrrrr}
    & \multicolumn{3}{c}{$dx = 1.25~\mathrm{fm}$} &\multicolumn{3}{c}{$dx = 1.0~\mathrm{fm}$} &\multicolumn{3}{c}{$dx = 0.8333~\mathrm{fm}$} \\
\cline{2-4}  \cline{5-7} \cline{8-10} 
& {\footnotesize XDIAG} & {\footnotesize XCOCG} &$\Delta E$ & {\footnotesize XDIAG} & {\footnotesize XCOCG} & $\Delta E$ & {\footnotesize XDIAG} & {\footnotesize XCOCG} & $\Delta E$\\
\hline
$E_\mathrm{kin}$& 1092.413& 1092.737& 0.324& 1090.235& 1090.293& 0.058& 1090.107& 1090.122& 0.015\\
$E_{\rho^2}$& -6335.819& -6336.367& -0.547& -6319.916& -6320.133& -0.218& -6318.595& -6318.650& -0.055\\
$E_{\rho^\gamma}$& 4134.013& 4134.153& 0.140& 4122.721& 4122.864& 0.143& 4121.738& 4121.775& 0.037\\
$E_{\rho\Delta\rho}$& 150.018& 150.256& 0.238& 150.031& 150.055& 0.024& 149.967& 149.971& 0.004\\
$E_{\rho\tau}$& 341.670& 341.492& -0.178& 339.456& 339.460& 0.005& 339.326& 339.329& 0.003\\
$E_{\rho\nabla J}$& -52.747& -52.819& -0.072& -52.371& -52.380& -0.010& -52.367& -52.365& 0.001\\
$E_\mathrm{Coul}$& 129.928& 129.917& -0.011& 129.761& 129.762& <0.001& 129.751& 129.751& <0.001\\
$E_\mathrm{pair}$& -3.834& -3.904& -0.070& -3.989& -4.000& -0.011& -3.986& -3.994& -0.009\\
$E_\mathrm{tot}$& -544.358& -544.534& -0.175& -544.073& -544.080& -0.007& -544.058& -544.061& -0.003\\
    \end{tabular}
  \end{ruledtabular}
\caption{\label{table:Ni62}
Same as Table \ref{table:Ca40}, but for $^{62}$Ni.
}
\end{table*}

%% file: table_Zr102.tex
\begin{table*}[t]
  \begin{ruledtabular}
    \begin{tabular}{lrrrrrr}
    & \multicolumn{3}{c}{$dx = 1.25~\mathrm{fm}$} &\multicolumn{3}{c}{$dx = 0.9375~\mathrm{fm}$} \\
\cline{2-4}  \cline{5-7} 
& {\footnotesize XDIAG} & {\footnotesize XCOCG} &$\Delta E (\Delta Q)$ & {\footnotesize XDIAG} & {\footnotesize XCOCG} & $\Delta E (\Delta Q)$\\
\hline
$E_\mathrm{kin}$& 1841.034& 1841.444& 0.410& 1838.435& 1838.438& 0.003\\
$E_{\rho^2}$& -10381.846& -10383.607& -1.760& -10364.579& -10364.463& 0.115\\
$E_{\rho^\gamma}$& 6715.556& 6716.620& 1.064& 6702.965& 6702.865& -0.100\\
$E_{\rho\Delta\rho}$& 209.650& 210.055& 0.405& 208.432& 208.427& -0.005\\
$E_{\rho\tau}$& 592.475& 592.366& -0.109& 590.686& 590.669& -0.017\\
$E_{\rho\nabla J}$& -63.446& -63.716& -0.270& -62.429& -62.422& 0.008\\
$E_\mathrm{Coul}$& 230.311& 230.288& -0.023& 230.359& 230.358& -0.001\\
$E_\mathrm{pair}$& -3.067& -3.083& -0.016& -3.231& -3.233& -0.002\\
$E_\mathrm{tot}$& -859.333& -859.632& -0.299& -859.361& -859.360& 0.001\\
$Q_{20}$& 1077.44& 1080.62& 3.18& 1047.81& 1047.65& -0.16\\
    \end{tabular}
  \end{ruledtabular}
\caption{\label{table:Zr102}
Results of HFB + SLy4 calculations for $^{102}$Zr using {\footnotesize XCOCG} and {\footnotesize XDIAG} in a rectangular box of size $22.5 \times 22.5 \times 30~\mathrm{fm}^3$ with lattice constants $dx=1.25$ and  $0.9375~\mathrm{fm}$ . 
All energies are in MeV and the quadrupole moment $Q_{20}$ is in $\mathrm{fm}^2$.
}
\end{table*}

%% file: table_saddlepoint.tex
\begin{table}[htbp]
  \begin{ruledtabular}
    \begin{tabular}{lrrr}
        & \multicolumn{1}{c}{\footnotesize XDIAG} & \multicolumn{1}{c}{\footnotesize XCOCG } & \multicolumn{1}{c}{$\Delta E$}\\
      \hline
$E_\mathrm{kin}$& 4418.132& 4419.125& 0.993\\
$E_{\rho^2}$& -25104.582& -25109.048& -4.467\\
$E_{\rho^\gamma}$& 16238.995& 16241.781& 2.786\\
$E_{\rho\Delta\rho}$& 408.182& 409.032& 0.850\\
$E_{\rho\tau}$& 1465.062& 1464.900& -0.161\\
$E_{\rho\nabla J}$& -111.516& -112.088& -0.572\\
$E_\mathrm{Coul}$& 901.083& 901.062& -0.021\\
$E_\mathrm{pair}$& -8.801& -8.818& -0.017\\
$E_\mathrm{tot}$& -1793.439& -1794.054& -0.615\\
    \end{tabular}
  \end{ruledtabular}
  \caption{\label{table:Pu240}
Results of HFB + SLy4 calculations for the saddle point of $^{240}$Pu using {\footnotesize XCOCG} and {\footnotesize XDIAG} in a rectangular box of size $25 \times 25 \times 50 ~\mathrm{fm}^3$ with mesh size $dx=dy=dz=1.25~\mathrm{fm}$.
All energies are in MeV. 
}
\end{table}

%% file: cocg.v8.bbl
\begin{thebibliography}{79}%
\makeatletter
\providecommand \@ifxundefined [1]{%
 \@ifx{#1\undefined}
}%
\providecommand \@ifnum [1]{%
 \ifnum #1\expandafter \@firstoftwo
 \else \expandafter \@secondoftwo
 \fi
}%
\providecommand \@ifx [1]{%
 \ifx #1\expandafter \@firstoftwo
 \else \expandafter \@secondoftwo
 \fi
}%
\providecommand \natexlab [1]{#1}%
\providecommand \enquote  [1]{``#1''}%
\providecommand \bibnamefont  [1]{#1}%
\providecommand \bibfnamefont [1]{#1}%
\providecommand \citenamefont [1]{#1}%
\providecommand \href@noop [0]{\@secondoftwo}%
\providecommand \href [0]{\begingroup \@sanitize@url \@href}%
\providecommand \@href[1]{\@@startlink{#1}\@@href}%
\providecommand \@@href[1]{\endgroup#1\@@endlink}%
\providecommand \@sanitize@url [0]{\catcode `\\12\catcode `\$12\catcode
  `\&12\catcode `\#12\catcode `\^12\catcode `\_12\catcode `\%12\relax}%
\providecommand \@@startlink[1]{}%
\providecommand \@@endlink[0]{}%
\providecommand \url  [0]{\begingroup\@sanitize@url \@url }%
\providecommand \@url [1]{\endgroup\@href {#1}{\urlprefix }}%
\providecommand \urlprefix  [0]{URL }%
\providecommand \Eprint [0]{\href }%
\providecommand \doibase [0]{http://dx.doi.org/}%
\providecommand \selectlanguage [0]{\@gobble}%
\providecommand \bibinfo  [0]{\@secondoftwo}%
\providecommand \bibfield  [0]{\@secondoftwo}%
\providecommand \translation [1]{[#1]}%
\providecommand \BibitemOpen [0]{}%
\providecommand \bibitemStop [0]{}%
\providecommand \bibitemNoStop [0]{.\EOS\space}%
\providecommand \EOS [0]{\spacefactor3000\relax}%
\providecommand \BibitemShut  [1]{\csname bibitem#1\endcsname}%
\let\auto@bib@innerbib\@empty
\bibitem [{\citenamefont {Bender}\ \emph
  {et~al.}(2003{\natexlab{a}})\citenamefont {Bender}, \citenamefont {Heenen},\
  and\ \citenamefont {Reinhard}}]{Bender2003}%
  \BibitemOpen
  \bibfield  {author} {\bibinfo {author} {\bibfnamefont {M.}~\bibnamefont
  {Bender}}, \bibinfo {author} {\bibfnamefont {P.-H.}\ \bibnamefont {Heenen}},
  \ and\ \bibinfo {author} {\bibfnamefont {P.-G.}\ \bibnamefont {Reinhard}},\
  }\bibfield  {title} {\enquote {\bibinfo {title} {Self-consistent mean-field
  models for nuclear structure},}\ }\href {\doibase 10.1103/RevModPhys.75.121}
  {\bibfield  {journal} {\bibinfo  {journal} {Rev. Mod. Phys.}\ }\textbf
  {\bibinfo {volume} {75}},\ \bibinfo {pages} {121--180} (\bibinfo {year}
  {2003}{\natexlab{a}})}\BibitemShut {NoStop}%
\bibitem [{\citenamefont {Runge}\ and\ \citenamefont
  {Gross}(1984)}]{PhysRevLett.52.997}%
  \BibitemOpen
  \bibfield  {author} {\bibinfo {author} {\bibfnamefont {E.}~\bibnamefont
  {Runge}}\ and\ \bibinfo {author} {\bibfnamefont {E.~K.~U.}\ \bibnamefont
  {Gross}},\ }\bibfield  {title} {\enquote {\bibinfo {title}
  {Density-functional theory for time-dependent systems},}\ }\href {\doibase
  10.1103/PhysRevLett.52.997} {\bibfield  {journal} {\bibinfo  {journal} {Phys.
  Rev. Lett.}\ }\textbf {\bibinfo {volume} {52}},\ \bibinfo {pages} {997--1000}
  (\bibinfo {year} {1984})}\BibitemShut {NoStop}%
\bibitem [{\citenamefont {Marques}\ \emph {et~al.}(2006)\citenamefont
  {Marques}, \citenamefont {Ullrich}, \citenamefont {Nogueira}, \citenamefont
  {Rubio}, \citenamefont {Burke},\ and\ \citenamefont {Gross}}]{TDDFTbook}%
  \BibitemOpen
  \bibinfo {editor} {\bibfnamefont {M.A.L.}\ \bibnamefont {Marques}}, \bibinfo
  {editor} {\bibfnamefont {C.~A.}\ \bibnamefont {Ullrich}}, \bibinfo {editor}
  {\bibfnamefont {F.}~\bibnamefont {Nogueira}}, \bibinfo {editor}
  {\bibfnamefont {A.}~\bibnamefont {Rubio}}, \bibinfo {editor} {\bibfnamefont
  {K.}~\bibnamefont {Burke}}, \ and\ \bibinfo {editor} {\bibfnamefont
  {E.~U.~K.}\ \bibnamefont {Gross}},\ eds.,\ \href
  {http://dx.doi.org/10.1007/b11767107} {\emph {\bibinfo {title}
  {Time-Dependent Density Functional Theory}}},\ \bibinfo {series} {Lecture
  Notes in Physics}, Vol.\ \bibinfo {volume} {706}\ (\bibinfo  {publisher}
  {Springer},\ \bibinfo {address} {Berlin},\ \bibinfo {year}
  {2006})\BibitemShut {NoStop}%
\bibitem [{\citenamefont {Marques}\ \emph {et~al.}(2012)\citenamefont
  {Marques}, \citenamefont {Maitra}, \citenamefont {Nogueira}, \citenamefont
  {Gross},\ and\ \citenamefont {Rubio}}]{FundumentalsTDDFTbook}%
  \BibitemOpen
  \bibinfo {editor} {\bibfnamefont {A.L.~Miguel}\ \bibnamefont {Marques}},
  \bibinfo {editor} {\bibfnamefont {T.~Neepa}\ \bibnamefont {Maitra}}, \bibinfo
  {editor} {\bibfnamefont {M.S.~Fernando}\ \bibnamefont {Nogueira}}, \bibinfo
  {editor} {\bibfnamefont {E.K.U.}\ \bibnamefont {Gross}}, \ and\ \bibinfo
  {editor} {\bibfnamefont {Angel}\ \bibnamefont {Rubio}},\ eds.,\ \href
  {http://dx.doi.org/ 10.1007/978-3-642-23518-4} {\emph {\bibinfo {title}
  {Fundamentals of Time-Dependent Density Functional Theory}}},\ \bibinfo
  {series} {Lecture Notes in Physics}, Vol.\ \bibinfo {volume} {837}\ (\bibinfo
   {publisher} {Springer Berlin Heidelberg},\ \bibinfo {year}
  {2012})\BibitemShut {NoStop}%
\bibitem [{\citenamefont {Ring}\ and\ \citenamefont {Schuck}(2004)}]{Ring2004}%
  \BibitemOpen
  \bibfield  {author} {\bibinfo {author} {\bibfnamefont {P.}~\bibnamefont
  {Ring}}\ and\ \bibinfo {author} {\bibfnamefont {P.}~\bibnamefont {Schuck}},\
  }\href {https://books.google.com/books?id=PTynSM-nMA8C} {\emph {\bibinfo
  {title} {The Nuclear Many-Body Problem}}}\ (\bibinfo  {publisher}
  {Springer},\ \bibinfo {address} {Berlin},\ \bibinfo {year}
  {2004})\BibitemShut {NoStop}%
\bibitem [{\citenamefont {Bertsch}(1980)}]{Bertsch:1980}%
  \BibitemOpen
  \bibfield  {author} {\bibinfo {author} {\bibfnamefont {G.}~\bibnamefont
  {Bertsch}},\ }\bibfield  {title} {\enquote {\bibinfo {title} {The nuclear
  density of states in the space of nuclear shapes},}\ }\href {\doibase
  http://dx.doi.org/10.1016/0370-2693(80)90458-X} {\bibfield  {journal}
  {\bibinfo  {journal} {Physics Letters B}\ }\textbf {\bibinfo {volume} {95}},\
  \bibinfo {pages} {157 -- 159} (\bibinfo {year} {1980})}\BibitemShut {NoStop}%
\bibitem [{\citenamefont {Barranco}\ \emph {et~al.}(1990)\citenamefont
  {Barranco}, \citenamefont {Bertsch}, \citenamefont {Broglia},\ and\
  \citenamefont {Vigezzi}}]{Barranco:1990}%
  \BibitemOpen
  \bibfield  {author} {\bibinfo {author} {\bibfnamefont {F.}~\bibnamefont
  {Barranco}}, \bibinfo {author} {\bibfnamefont {G.F.}\ \bibnamefont
  {Bertsch}}, \bibinfo {author} {\bibfnamefont {R.A.}\ \bibnamefont {Broglia}},
  \ and\ \bibinfo {author} {\bibfnamefont {E.}~\bibnamefont {Vigezzi}},\
  }\bibfield  {title} {\enquote {\bibinfo {title} {Large-amplitude motion in
  superfluid {Fermi} droplets},}\ }\href {\doibase
  http://dx.doi.org/10.1016/0375-9474(90)93232-U} {\bibfield  {journal}
  {\bibinfo  {journal} {Nuclear Physics A}\ }\textbf {\bibinfo {volume}
  {512}},\ \bibinfo {pages} {253 -- 274} (\bibinfo {year} {1990})}\BibitemShut
  {NoStop}%
\bibitem [{\citenamefont {Bertsch}\ and\ \citenamefont
  {Bulgac}(1997)}]{Bertsch:1997}%
  \BibitemOpen
  \bibfield  {author} {\bibinfo {author} {\bibfnamefont {G.~F.}\ \bibnamefont
  {Bertsch}}\ and\ \bibinfo {author} {\bibfnamefont {A.}~\bibnamefont
  {Bulgac}},\ }\bibfield  {title} {\enquote {\bibinfo {title} {Comment on
  ``spontaneous fission: A kinetic approach''},}\ }\href {\doibase
  10.1103/PhysRevLett.79.3539} {\bibfield  {journal} {\bibinfo  {journal}
  {Phys. Rev. Lett.}\ }\textbf {\bibinfo {volume} {79}},\ \bibinfo {pages}
  {3539--3539} (\bibinfo {year} {1997})}\BibitemShut {NoStop}%
\bibitem [{\citenamefont {Bulgac}\ \emph {et~al.}(2016)\citenamefont {Bulgac},
  \citenamefont {Magierski}, \citenamefont {Roche},\ and\ \citenamefont
  {Stetcu}}]{Bulgac2016}%
  \BibitemOpen
  \bibfield  {author} {\bibinfo {author} {\bibfnamefont {A.}~\bibnamefont
  {Bulgac}}, \bibinfo {author} {\bibfnamefont {P.}~\bibnamefont {Magierski}},
  \bibinfo {author} {\bibfnamefont {K.~J.}\ \bibnamefont {Roche}}, \ and\
  \bibinfo {author} {\bibfnamefont {I.}~\bibnamefont {Stetcu}},\ }\bibfield
  {title} {\enquote {\bibinfo {title} {Induced fission of $^{240}\mathrm{Pu}$
  within a real-time microscopic framework},}\ }\href {\doibase
  10.1103/PhysRevLett.116.122504} {\bibfield  {journal} {\bibinfo  {journal}
  {Phys. Rev. Lett.}\ }\textbf {\bibinfo {volume} {116}},\ \bibinfo {pages}
  {122504} (\bibinfo {year} {2016})}\BibitemShut {NoStop}%
\bibitem [{\citenamefont {Anderson}\ and\ \citenamefont
  {Itoh}(1975)}]{Anderson:1975}%
  \BibitemOpen
  \bibfield  {author} {\bibinfo {author} {\bibfnamefont {P.~W.}\ \bibnamefont
  {Anderson}}\ and\ \bibinfo {author} {\bibfnamefont {N.}~\bibnamefont
  {Itoh}},\ }\bibfield  {title} {\enquote {\bibinfo {title} {Pulsar glitches
  and restlessness as a hard superfluid phenomenon},}\ }\href {\doibase
  10.1038/256025a0} {\bibfield  {journal} {\bibinfo  {journal} {Nature
  (London)}\ }\textbf {\bibinfo {volume} {256}},\ \bibinfo {pages} {25}
  (\bibinfo {year} {1975})}\BibitemShut {NoStop}%
\bibitem [{\citenamefont {Wlaz\l{}owski}\ \emph {et~al.}(2016)\citenamefont
  {Wlaz\l{}owski}, \citenamefont {Sekizawa}, \citenamefont {Magierski},
  \citenamefont {Bulgac},\ and\ \citenamefont {Forbes}}]{Wlazlowski2016}%
  \BibitemOpen
  \bibfield  {author} {\bibinfo {author} {\bibfnamefont {G.}~\bibnamefont
  {Wlaz\l{}owski}}, \bibinfo {author} {\bibfnamefont {K.}~\bibnamefont
  {Sekizawa}}, \bibinfo {author} {\bibfnamefont {P.}~\bibnamefont {Magierski}},
  \bibinfo {author} {\bibfnamefont {A.}~\bibnamefont {Bulgac}}, \ and\ \bibinfo
  {author} {\bibfnamefont {M.~M.}\ \bibnamefont {Forbes}},\ }\bibfield  {title}
  {\enquote {\bibinfo {title} {Vortex pinning and dynamics in the neutron star
  crust},}\ }\href {\doibase 10.1103/PhysRevLett.117.232701} {\bibfield
  {journal} {\bibinfo  {journal} {Phys. Rev. Lett.}\ }\textbf {\bibinfo
  {volume} {117}},\ \bibinfo {pages} {232701} (\bibinfo {year}
  {2016})}\BibitemShut {NoStop}%
\bibitem [{\citenamefont {Davies}\ \emph {et~al.}(1980)\citenamefont {Davies},
  \citenamefont {Flocard}, \citenamefont {Krieger},\ and\ \citenamefont
  {Weiss}}]{Davies1980111}%
  \BibitemOpen
  \bibfield  {author} {\bibinfo {author} {\bibfnamefont {K.T.R.}\ \bibnamefont
  {Davies}}, \bibinfo {author} {\bibfnamefont {H.}~\bibnamefont {Flocard}},
  \bibinfo {author} {\bibfnamefont {S.}~\bibnamefont {Krieger}}, \ and\
  \bibinfo {author} {\bibfnamefont {M.S.}\ \bibnamefont {Weiss}},\ }\bibfield
  {title} {\enquote {\bibinfo {title} {Application of the imaginary time step
  method to the solution of the static {Hartree}-{Fock} problem},}\ }\href
  {\doibase http://dx.doi.org/10.1016/0375-9474(80)90509-6} {\bibfield
  {journal} {\bibinfo  {journal} {Nucl. Phys. A}\ }\textbf {\bibinfo {volume}
  {342}},\ \bibinfo {pages} {111 -- 123} (\bibinfo {year} {1980})}\BibitemShut
  {NoStop}%
\bibitem [{\citenamefont {Ryssens}\ \emph
  {et~al.}(2015{\natexlab{a}})\citenamefont {Ryssens}, \citenamefont
  {Hellemans}, \citenamefont {Bender},\ and\ \citenamefont
  {Heenen}}]{Ryssens2015}%
  \BibitemOpen
  \bibfield  {author} {\bibinfo {author} {\bibfnamefont {W.}~\bibnamefont
  {Ryssens}}, \bibinfo {author} {\bibfnamefont {V.}~\bibnamefont {Hellemans}},
  \bibinfo {author} {\bibfnamefont {M.}~\bibnamefont {Bender}}, \ and\ \bibinfo
  {author} {\bibfnamefont {P.-H.}\ \bibnamefont {Heenen}},\ }\bibfield  {title}
  {\enquote {\bibinfo {title} {Solution of the {Skyrme}--{HF}+{BCS} equation on
  a {3D} mesh, {II}: A new version of the {Ev8} code},}\ }\href {\doibase
  http://dx.doi.org/10.1016/j.cpc.2014.10.001} {\bibfield  {journal} {\bibinfo
  {journal} {Computer Physics Communications}\ }\textbf {\bibinfo {volume}
  {187}},\ \bibinfo {pages} {175 -- 194} (\bibinfo {year}
  {2015}{\natexlab{a}})}\BibitemShut {NoStop}%
\bibitem [{\citenamefont {Egido}\ \emph {et~al.}(1995)\citenamefont {Egido},
  \citenamefont {Lessing}, \citenamefont {Martin},\ and\ \citenamefont
  {Robledo}}]{Egido199570}%
  \BibitemOpen
  \bibfield  {author} {\bibinfo {author} {\bibfnamefont {J.L.}\ \bibnamefont
  {Egido}}, \bibinfo {author} {\bibfnamefont {J.}~\bibnamefont {Lessing}},
  \bibinfo {author} {\bibfnamefont {V.}~\bibnamefont {Martin}}, \ and\ \bibinfo
  {author} {\bibfnamefont {L.M.}\ \bibnamefont {Robledo}},\ }\bibfield  {title}
  {\enquote {\bibinfo {title} {On the solution of the
  {Hartree}-{Fock}-{Bogoliubov} equations by the conjugate gradient method},}\
  }\href {\doibase http://dx.doi.org/10.1016/0375-9474(95)00370-G} {\bibfield
  {journal} {\bibinfo  {journal} {Nucl. Phys. A}\ }\textbf {\bibinfo {volume}
  {594}},\ \bibinfo {pages} {70 -- 86} (\bibinfo {year} {1995})}\BibitemShut
  {NoStop}%
\bibitem [{\citenamefont {Robledo}\ and\ \citenamefont
  {Bertsch}(2011)}]{PhysRevC.84.014312}%
  \BibitemOpen
  \bibfield  {author} {\bibinfo {author} {\bibfnamefont {L.~M.}\ \bibnamefont
  {Robledo}}\ and\ \bibinfo {author} {\bibfnamefont {G.~F.}\ \bibnamefont
  {Bertsch}},\ }\bibfield  {title} {\enquote {\bibinfo {title} {Application of
  the gradient method to {Hartree}-{Fock}-{Bogoliubov} theory},}\ }\href
  {\doibase 10.1103/PhysRevC.84.014312} {\bibfield  {journal} {\bibinfo
  {journal} {Phys. Rev. C}\ }\textbf {\bibinfo {volume} {84}},\ \bibinfo
  {pages} {014312} (\bibinfo {year} {2011})}\BibitemShut {NoStop}%
\bibitem [{\citenamefont {Golub}\ and\ \citenamefont {van~der
  Vorst}(2000)}]{Golub2000}%
  \BibitemOpen
  \bibfield  {author} {\bibinfo {author} {\bibfnamefont {G.~H.}\ \bibnamefont
  {Golub}}\ and\ \bibinfo {author} {\bibfnamefont {H.~A.}\ \bibnamefont
  {van~der Vorst}},\ }\bibfield  {title} {\enquote {\bibinfo {title}
  {Eigenvalue computation in the 20th century},}\ }\href {\doibase
  http://dx.doi.org/10.1016/S0377-0427(00)00413-1} {\bibfield  {journal}
  {\bibinfo  {journal} {Journal of Computational and Applied Mathematics}\
  }\textbf {\bibinfo {volume} {123}},\ \bibinfo {pages} {35 -- 65} (\bibinfo
  {year} {2000})}\BibitemShut {NoStop}%
\bibitem [{\citenamefont {Stoitsov}\ \emph {et~al.}(2013)\citenamefont
  {Stoitsov}, \citenamefont {Schunck}, \citenamefont {Kortelainen},
  \citenamefont {Michel}, \citenamefont {Nam}, \citenamefont {Olsen},
  \citenamefont {Sarich},\ and\ \citenamefont {Wild}}]{Stoitsov2013}%
  \BibitemOpen
  \bibfield  {author} {\bibinfo {author} {\bibfnamefont {M.V.}\ \bibnamefont
  {Stoitsov}}, \bibinfo {author} {\bibfnamefont {N.}~\bibnamefont {Schunck}},
  \bibinfo {author} {\bibfnamefont {M.}~\bibnamefont {Kortelainen}}, \bibinfo
  {author} {\bibfnamefont {N.}~\bibnamefont {Michel}}, \bibinfo {author}
  {\bibfnamefont {H.}~\bibnamefont {Nam}}, \bibinfo {author} {\bibfnamefont
  {E.}~\bibnamefont {Olsen}}, \bibinfo {author} {\bibfnamefont
  {J.}~\bibnamefont {Sarich}}, \ and\ \bibinfo {author} {\bibfnamefont
  {S.}~\bibnamefont {Wild}},\ }\bibfield  {title} {\enquote {\bibinfo {title}
  {Axially deformed solution of the {{Skyrme}-{Hartree}--{Fock}--{Bogoliubov}}
  equations using the transformed harmonic oscillator basis ({II}) {HFBTHO}
  v2.00d: A new version of the program},}\ }\href {\doibase
  http://dx.doi.org/10.1016/j.cpc.2013.01.013} {\bibfield  {journal} {\bibinfo
  {journal} {Computer Physics Communications}\ }\textbf {\bibinfo {volume}
  {184}},\ \bibinfo {pages} {1592 -- 1604} (\bibinfo {year}
  {2013})}\BibitemShut {NoStop}%
\bibitem [{\citenamefont {Schunck}\ \emph {et~al.}(2012)\citenamefont
  {Schunck}, \citenamefont {Dobaczewski}, \citenamefont {McDonnell},
  \citenamefont {Satu{\l}a}, \citenamefont {Sheikh}, \citenamefont {Staszczak},
  \citenamefont {Stoitsov},\ and\ \citenamefont {Toivanen}}]{Schunck2012}%
  \BibitemOpen
  \bibfield  {author} {\bibinfo {author} {\bibfnamefont {N.}~\bibnamefont
  {Schunck}}, \bibinfo {author} {\bibfnamefont {J.}~\bibnamefont
  {Dobaczewski}}, \bibinfo {author} {\bibfnamefont {J.}~\bibnamefont
  {McDonnell}}, \bibinfo {author} {\bibfnamefont {W.}~\bibnamefont
  {Satu{\l}a}}, \bibinfo {author} {\bibfnamefont {J.A.}\ \bibnamefont
  {Sheikh}}, \bibinfo {author} {\bibfnamefont {A.}~\bibnamefont {Staszczak}},
  \bibinfo {author} {\bibfnamefont {M.}~\bibnamefont {Stoitsov}}, \ and\
  \bibinfo {author} {\bibfnamefont {P.}~\bibnamefont {Toivanen}},\ }\bibfield
  {title} {\enquote {\bibinfo {title} {Solution of the
  {Skyrme}--{Hartree}--{Fock}--bogolyubov equations in the cartesian deformed
  harmonic-oscillator basis.: ({VII}) {HFODD} (v2.49t): A new version of the
  program},}\ }\href {\doibase http://dx.doi.org/10.1016/j.cpc.2011.08.013}
  {\bibfield  {journal} {\bibinfo  {journal} {Computer Physics Communications}\
  }\textbf {\bibinfo {volume} {183}},\ \bibinfo {pages} {166 -- 192} (\bibinfo
  {year} {2012})}\BibitemShut {NoStop}%
\bibitem [{\citenamefont {Pei}\ \emph {et~al.}(2014)\citenamefont {Pei},
  \citenamefont {Fann}, \citenamefont {Harrison}, \citenamefont {Nazarewicz},
  \citenamefont {Shi},\ and\ \citenamefont {Thornton}}]{Pei:2014}%
  \BibitemOpen
  \bibfield  {author} {\bibinfo {author} {\bibfnamefont {J.~C.}\ \bibnamefont
  {Pei}}, \bibinfo {author} {\bibfnamefont {G.~I.}\ \bibnamefont {Fann}},
  \bibinfo {author} {\bibfnamefont {R.~J.}\ \bibnamefont {Harrison}}, \bibinfo
  {author} {\bibfnamefont {W.}~\bibnamefont {Nazarewicz}}, \bibinfo {author}
  {\bibfnamefont {Yue}\ \bibnamefont {Shi}}, \ and\ \bibinfo {author}
  {\bibfnamefont {S.}~\bibnamefont {Thornton}},\ }\bibfield  {title} {\enquote
  {\bibinfo {title} {Adaptive multi-resolution 3d {Hartree}-{Fock}-{Bogoliubov}
  solver for nuclear structure},}\ }\href {\doibase 10.1103/PhysRevC.90.024317}
  {\bibfield  {journal} {\bibinfo  {journal} {Phys. Rev. C}\ }\textbf {\bibinfo
  {volume} {90}},\ \bibinfo {pages} {024317} (\bibinfo {year}
  {2014})}\BibitemShut {NoStop}%
\bibitem [{\citenamefont {Michel}\ \emph {et~al.}(2002)\citenamefont {Michel},
  \citenamefont {Nazarewicz}, \citenamefont {P\l{}oszajczak},\ and\
  \citenamefont {Bennaceur}}]{Michel:2002}%
  \BibitemOpen
  \bibfield  {author} {\bibinfo {author} {\bibfnamefont {N.}~\bibnamefont
  {Michel}}, \bibinfo {author} {\bibfnamefont {W.}~\bibnamefont {Nazarewicz}},
  \bibinfo {author} {\bibfnamefont {M.}~\bibnamefont {P\l{}oszajczak}}, \ and\
  \bibinfo {author} {\bibfnamefont {K.}~\bibnamefont {Bennaceur}},\ }\bibfield
  {title} {\enquote {\bibinfo {title} {Gamow shell model description of
  neutron-rich nuclei},}\ }\href {\doibase 10.1103/PhysRevLett.89.042502}
  {\bibfield  {journal} {\bibinfo  {journal} {Phys. Rev. Lett.}\ }\textbf
  {\bibinfo {volume} {89}},\ \bibinfo {pages} {042502} (\bibinfo {year}
  {2002})}\BibitemShut {NoStop}%
\bibitem [{\citenamefont {Bulgac}\ and\ \citenamefont
  {Forbes}(2013)}]{Bulgac2013}%
  \BibitemOpen
  \bibfield  {author} {\bibinfo {author} {\bibfnamefont {Aurel}\ \bibnamefont
  {Bulgac}}\ and\ \bibinfo {author} {\bibfnamefont {Michael~McNeil}\
  \bibnamefont {Forbes}},\ }\bibfield  {title} {\enquote {\bibinfo {title} {Use
  of the discrete variable representation basis in nuclear physics},}\ }\href
  {\doibase 10.1103/PhysRevC.87.051301} {\bibfield  {journal} {\bibinfo
  {journal} {Phys. Rev. C}\ }\textbf {\bibinfo {volume} {87}},\ \bibinfo
  {pages} {051301} (\bibinfo {year} {2013})}\BibitemShut {NoStop}%
\bibitem [{\citenamefont {Ryssens}\ \emph
  {et~al.}(2015{\natexlab{b}})\citenamefont {Ryssens}, \citenamefont {Heenen},\
  and\ \citenamefont {Bender}}]{Ryssens2015a}%
  \BibitemOpen
  \bibfield  {author} {\bibinfo {author} {\bibfnamefont {W.}~\bibnamefont
  {Ryssens}}, \bibinfo {author} {\bibfnamefont {P.-H.}\ \bibnamefont {Heenen}},
  \ and\ \bibinfo {author} {\bibfnamefont {M.}~\bibnamefont {Bender}},\
  }\bibfield  {title} {\enquote {\bibinfo {title} {Numerical accuracy of
  mean-field calculations in coordinate space},}\ }\href {\doibase
  10.1103/PhysRevC.92.064318} {\bibfield  {journal} {\bibinfo  {journal} {Phys.
  Rev. C}\ }\textbf {\bibinfo {volume} {92}},\ \bibinfo {pages} {064318}
  (\bibinfo {year} {2015}{\natexlab{b}})}\BibitemShut {NoStop}%
\bibitem [{\citenamefont {Dobaczewski}\ \emph {et~al.}(1984)\citenamefont
  {Dobaczewski}, \citenamefont {Flocard},\ and\ \citenamefont
  {Treiner}}]{Dobaczewski1984}%
  \BibitemOpen
  \bibfield  {author} {\bibinfo {author} {\bibfnamefont {J.}~\bibnamefont
  {Dobaczewski}}, \bibinfo {author} {\bibfnamefont {H.}~\bibnamefont
  {Flocard}}, \ and\ \bibinfo {author} {\bibfnamefont {J.}~\bibnamefont
  {Treiner}},\ }\bibfield  {title} {\enquote {\bibinfo {title}
  {{Hartree}-{Fock}-{Bogolyubov} description of nuclei near the neutron-drip
  line},}\ }\href {\doibase http://dx.doi.org/10.1016/0375-9474(84)90433-0}
  {\bibfield  {journal} {\bibinfo  {journal} {Nuclear Physics A}\ }\textbf
  {\bibinfo {volume} {422}},\ \bibinfo {pages} {103 -- 139} (\bibinfo {year}
  {1984})}\BibitemShut {NoStop}%
\bibitem [{\citenamefont {Dobaczewski}\ \emph {et~al.}(1996)\citenamefont
  {Dobaczewski}, \citenamefont {Nazarewicz}, \citenamefont {Werner},
  \citenamefont {Berger}, \citenamefont {Chinn},\ and\ \citenamefont
  {Decharg\'e}}]{Dobaczewski1996}%
  \BibitemOpen
  \bibfield  {author} {\bibinfo {author} {\bibfnamefont {J.}~\bibnamefont
  {Dobaczewski}}, \bibinfo {author} {\bibfnamefont {W.}~\bibnamefont
  {Nazarewicz}}, \bibinfo {author} {\bibfnamefont {T.~R.}\ \bibnamefont
  {Werner}}, \bibinfo {author} {\bibfnamefont {J.~F.}\ \bibnamefont {Berger}},
  \bibinfo {author} {\bibfnamefont {C.~R.}\ \bibnamefont {Chinn}}, \ and\
  \bibinfo {author} {\bibfnamefont {J.}~\bibnamefont {Decharg\'e}},\ }\bibfield
   {title} {\enquote {\bibinfo {title} {Mean-field description of ground-state
  properties of drip-line nuclei: Pairing and continuum effects},}\ }\href
  {\doibase 10.1103/PhysRevC.53.2809} {\bibfield  {journal} {\bibinfo
  {journal} {Phys. Rev. C}\ }\textbf {\bibinfo {volume} {53}},\ \bibinfo
  {pages} {2809--2840} (\bibinfo {year} {1996})}\BibitemShut {NoStop}%
\bibitem [{\citenamefont {Zhang}\ \emph {et~al.}(2013)\citenamefont {Zhang},
  \citenamefont {Pei},\ and\ \citenamefont {Xu}}]{Zhang:2013}%
  \BibitemOpen
  \bibfield  {author} {\bibinfo {author} {\bibfnamefont {Y.~N.}\ \bibnamefont
  {Zhang}}, \bibinfo {author} {\bibfnamefont {J.~C.}\ \bibnamefont {Pei}}, \
  and\ \bibinfo {author} {\bibfnamefont {F.~R.}\ \bibnamefont {Xu}},\
  }\bibfield  {title} {\enquote {\bibinfo {title}
  {{Hartree}-{Fock}-{Bogoliubov} descriptions of deformed weakly bound nuclei
  in large coordinate spaces},}\ }\href {\doibase 10.1103/PhysRevC.88.054305}
  {\bibfield  {journal} {\bibinfo  {journal} {Phys. Rev. C}\ }\textbf {\bibinfo
  {volume} {88}},\ \bibinfo {pages} {054305} (\bibinfo {year}
  {2013})}\BibitemShut {NoStop}%
\bibitem [{\citenamefont {Bennaceur}\ and\ \citenamefont
  {Dobaczewski}(2005)}]{Bennaceur2005}%
  \BibitemOpen
  \bibfield  {author} {\bibinfo {author} {\bibfnamefont {K.}~\bibnamefont
  {Bennaceur}}\ and\ \bibinfo {author} {\bibfnamefont {J.}~\bibnamefont
  {Dobaczewski}},\ }\bibfield  {title} {\enquote {\bibinfo {title}
  {Coordinate-space solution of the {{Skyrme}--{Hartree}--{Fock}--Bogolyubov}
  equations within spherical symmetry. {The} program {HFBRAD} (v1.00)},}\
  }\href {\doibase http://dx.doi.org/10.1016/j.cpc.2005.02.002} {\bibfield
  {journal} {\bibinfo  {journal} {Computer Physics Communications}\ }\textbf
  {\bibinfo {volume} {168}},\ \bibinfo {pages} {96 -- 122} (\bibinfo {year}
  {2005})}\BibitemShut {NoStop}%
\bibitem [{\citenamefont {Oberacker}\ \emph {et~al.}(2003)\citenamefont
  {Oberacker}, \citenamefont {Umar}, \citenamefont {Ter\'an},\ and\
  \citenamefont {Blazkiewicz}}]{Oberacker2003}%
  \BibitemOpen
  \bibfield  {author} {\bibinfo {author} {\bibfnamefont {V.~E.}\ \bibnamefont
  {Oberacker}}, \bibinfo {author} {\bibfnamefont {A.~S.}\ \bibnamefont {Umar}},
  \bibinfo {author} {\bibfnamefont {E.}~\bibnamefont {Ter\'an}}, \ and\
  \bibinfo {author} {\bibfnamefont {A.}~\bibnamefont {Blazkiewicz}},\
  }\bibfield  {title} {\enquote {\bibinfo {title}
  {{Hartree}-{Fock}-{Bogoliubov} calculations in coordinate space: Neutron-rich
  sulfur, zirconium, cerium, and samarium isotopes},}\ }\href {\doibase
  10.1103/PhysRevC.68.064302} {\bibfield  {journal} {\bibinfo  {journal} {Phys.
  Rev. C}\ }\textbf {\bibinfo {volume} {68}},\ \bibinfo {pages} {064302}
  (\bibinfo {year} {2003})}\BibitemShut {NoStop}%
\bibitem [{\citenamefont {Ter\'an}\ \emph {et~al.}(2003)\citenamefont
  {Ter\'an}, \citenamefont {Oberacker},\ and\ \citenamefont
  {Umar}}]{Teran2003}%
  \BibitemOpen
  \bibfield  {author} {\bibinfo {author} {\bibfnamefont {E.}~\bibnamefont
  {Ter\'an}}, \bibinfo {author} {\bibfnamefont {V.~E.}\ \bibnamefont
  {Oberacker}}, \ and\ \bibinfo {author} {\bibfnamefont {A.~S.}\ \bibnamefont
  {Umar}},\ }\bibfield  {title} {\enquote {\bibinfo {title} {Axially symmetric
  {Hartree}-{Fock}-{Bogoliubov} calculations for nuclei near the drip lines},}\
  }\href {\doibase 10.1103/PhysRevC.67.064314} {\bibfield  {journal} {\bibinfo
  {journal} {Phys. Rev. C}\ }\textbf {\bibinfo {volume} {67}},\ \bibinfo
  {pages} {064314} (\bibinfo {year} {2003})}\BibitemShut {NoStop}%
\bibitem [{\citenamefont {Blazkiewicz}\ \emph {et~al.}(2005)\citenamefont
  {Blazkiewicz}, \citenamefont {Oberacker}, \citenamefont {Umar},\ and\
  \citenamefont {Stoitsov}}]{Blazkiewicz2005}%
  \BibitemOpen
  \bibfield  {author} {\bibinfo {author} {\bibfnamefont {A.}~\bibnamefont
  {Blazkiewicz}}, \bibinfo {author} {\bibfnamefont {V.~E.}\ \bibnamefont
  {Oberacker}}, \bibinfo {author} {\bibfnamefont {A.~S.}\ \bibnamefont {Umar}},
  \ and\ \bibinfo {author} {\bibfnamefont {M.}~\bibnamefont {Stoitsov}},\
  }\bibfield  {title} {\enquote {\bibinfo {title} {Coordinate space
  {Hartree}-{Fock}-{Bogoliubov} calculations for the zirconium isotope chain up
  to the two-neutron drip line},}\ }\href {\doibase 10.1103/PhysRevC.71.054321}
  {\bibfield  {journal} {\bibinfo  {journal} {Phys. Rev. C}\ }\textbf {\bibinfo
  {volume} {71}},\ \bibinfo {pages} {054321} (\bibinfo {year}
  {2005})}\BibitemShut {NoStop}%
\bibitem [{\citenamefont {Pei}\ \emph {et~al.}(2008)\citenamefont {Pei},
  \citenamefont {Stoitsov}, \citenamefont {Fann}, \citenamefont {Nazarewicz},
  \citenamefont {Schunck},\ and\ \citenamefont {Xu}}]{Pei2008}%
  \BibitemOpen
  \bibfield  {author} {\bibinfo {author} {\bibfnamefont {J.~C.}\ \bibnamefont
  {Pei}}, \bibinfo {author} {\bibfnamefont {M.~V.}\ \bibnamefont {Stoitsov}},
  \bibinfo {author} {\bibfnamefont {G.~I.}\ \bibnamefont {Fann}}, \bibinfo
  {author} {\bibfnamefont {W.}~\bibnamefont {Nazarewicz}}, \bibinfo {author}
  {\bibfnamefont {N.}~\bibnamefont {Schunck}}, \ and\ \bibinfo {author}
  {\bibfnamefont {F.~R.}\ \bibnamefont {Xu}},\ }\bibfield  {title} {\enquote
  {\bibinfo {title} {Deformed coordinate-space {{Hartree}-{Fock}-{Bogoliubov}}
  approach to weakly bound nuclei and large deformations},}\ }\href {\doibase
  10.1103/PhysRevC.78.064306} {\bibfield  {journal} {\bibinfo  {journal} {Phys.
  Rev. C}\ }\textbf {\bibinfo {volume} {78}},\ \bibinfo {pages} {064306}
  (\bibinfo {year} {2008})}\BibitemShut {NoStop}%
\bibitem [{sca()}]{scalapack}%
  \BibitemOpen
  \href {http://www.netlib.org/scalapack/} {\bibinfo  {journal}
  {http://www.netlib.org/scalapack/}\ }\BibitemShut {NoStop}%
\bibitem [{edi()}]{edison}%
  \BibitemOpen
\bibfield  {journal} {  }\href
  {http://www.nersc.gov/users/computational-systems/edison/} {\bibinfo
  {journal} {http://www.nersc.gov/users/computational-systems/edison/}\
  }\BibitemShut {NoStop}%
\bibitem [{cud()}]{cuda}%
  \BibitemOpen
\bibfield  {journal} {  }\href
  {http://www.nvidia.com/object/cuda_home_new.html} {\bibinfo  {journal}
  {http://www.nvidia.com/object/cuda\_home\_new.html}\ }\BibitemShut {NoStop}%
\bibitem [{\citenamefont {van~der Vorst}\ and\ \citenamefont
  {Melissen}(1990)}]{VanderVorst1990}%
  \BibitemOpen
\bibfield  {journal} {  }\bibfield  {author} {\bibinfo {author} {\bibfnamefont
  {H.~A.}\ \bibnamefont {van~der Vorst}}\ and\ \bibinfo {author} {\bibfnamefont
  {J.~B.~M.}\ \bibnamefont {Melissen}},\ }\bibfield  {title} {\enquote
  {\bibinfo {title} {A petrov-galerkin type method for solving {$Axk=b$}, where
  {$A$} is symmetric complex},}\ }\href {\doibase 10.1109/20.106415} {\bibfield
   {journal} {\bibinfo  {journal} {IEEE Transactions on Magnetics}\ }\textbf
  {\bibinfo {volume} {26}},\ \bibinfo {pages} {706--708} (\bibinfo {year}
  {1990})}\BibitemShut {NoStop}%
\bibitem [{\citenamefont {Yamamoto}\ \emph {et~al.}(2008)\citenamefont
  {Yamamoto}, \citenamefont {Sogabe}, \citenamefont {Hoshi}, \citenamefont
  {Zhang},\ and\ \citenamefont {Fujiwara}}]{Yamamoto2008}%
  \BibitemOpen
  \bibfield  {author} {\bibinfo {author} {\bibfnamefont {Susumu}\ \bibnamefont
  {Yamamoto}}, \bibinfo {author} {\bibfnamefont {Tomohiro}\ \bibnamefont
  {Sogabe}}, \bibinfo {author} {\bibfnamefont {Takeo}\ \bibnamefont {Hoshi}},
  \bibinfo {author} {\bibfnamefont {Shao-Liang}\ \bibnamefont {Zhang}}, \ and\
  \bibinfo {author} {\bibfnamefont {Takeo}\ \bibnamefont {Fujiwara}},\
  }\bibfield  {title} {\enquote {\bibinfo {title} {Shifted
  conjugate-orthogonal--conjugate-gradient method and its application to double
  orbital extended hubbard model},}\ }\href {\doibase 10.1143/JPSJ.77.114713}
  {\bibfield  {journal} {\bibinfo  {journal} {Journal of the Physical Society
  of Japan}\ }\textbf {\bibinfo {volume} {77}},\ \bibinfo {pages} {114713}
  (\bibinfo {year} {2008})}\BibitemShut {NoStop}%
\bibitem [{\citenamefont {Takayama}\ \emph {et~al.}(2006)\citenamefont
  {Takayama}, \citenamefont {Hoshi}, \citenamefont {Sogabe}, \citenamefont
  {Zhang},\ and\ \citenamefont {Fujiwara}}]{Takayama2006}%
  \BibitemOpen
  \bibfield  {author} {\bibinfo {author} {\bibfnamefont {R.}~\bibnamefont
  {Takayama}}, \bibinfo {author} {\bibfnamefont {T.}~\bibnamefont {Hoshi}},
  \bibinfo {author} {\bibfnamefont {T.}~\bibnamefont {Sogabe}}, \bibinfo
  {author} {\bibfnamefont {S.-L.}\ \bibnamefont {Zhang}}, \ and\ \bibinfo
  {author} {\bibfnamefont {T.}~\bibnamefont {Fujiwara}},\ }\bibfield  {title}
  {\enquote {\bibinfo {title} {Linear algebraic calculation of the {Green's}
  function for large-scale electronic structure theory},}\ }\href {\doibase
  10.1103/PhysRevB.73.165108} {\bibfield  {journal} {\bibinfo  {journal} {Phys.
  Rev. B}\ }\textbf {\bibinfo {volume} {73}},\ \bibinfo {pages} {165108}
  (\bibinfo {year} {2006})}\BibitemShut {NoStop}%
\bibitem [{\citenamefont {Mizusaki}\ \emph {et~al.}(2010)\citenamefont
  {Mizusaki}, \citenamefont {Kaneko}, \citenamefont {Honma},\ and\
  \citenamefont {Sakurai}}]{Mizusaki2010}%
  \BibitemOpen
  \bibfield  {author} {\bibinfo {author} {\bibfnamefont {Takahiro}\
  \bibnamefont {Mizusaki}}, \bibinfo {author} {\bibfnamefont {Kazunari}\
  \bibnamefont {Kaneko}}, \bibinfo {author} {\bibfnamefont {Michio}\
  \bibnamefont {Honma}}, \ and\ \bibinfo {author} {\bibfnamefont {Tetsuya}\
  \bibnamefont {Sakurai}},\ }\bibfield  {title} {\enquote {\bibinfo {title}
  {Filter diagonalization of shell-model calculations},}\ }\href {\doibase
  10.1103/PhysRevC.82.024310} {\bibfield  {journal} {\bibinfo  {journal} {Phys.
  Rev. C}\ }\textbf {\bibinfo {volume} {82}},\ \bibinfo {pages} {024310}
  (\bibinfo {year} {2010})}\BibitemShut {NoStop}%
\bibitem [{\citenamefont {Shimizu}\ \emph {et~al.}(2016)\citenamefont
  {Shimizu}, \citenamefont {Utsuno}, \citenamefont {Futamura}, \citenamefont
  {Sakurai}, \citenamefont {Mizusaki},\ and\ \citenamefont
  {Otsuka}}]{Shimizu2016}%
  \BibitemOpen
  \bibfield  {author} {\bibinfo {author} {\bibfnamefont {Noritaka}\
  \bibnamefont {Shimizu}}, \bibinfo {author} {\bibfnamefont {Yutaka}\
  \bibnamefont {Utsuno}}, \bibinfo {author} {\bibfnamefont {Yasunori}\
  \bibnamefont {Futamura}}, \bibinfo {author} {\bibfnamefont {Tetsuya}\
  \bibnamefont {Sakurai}}, \bibinfo {author} {\bibfnamefont {Takahiro}\
  \bibnamefont {Mizusaki}}, \ and\ \bibinfo {author} {\bibfnamefont {Takaharu}\
  \bibnamefont {Otsuka}},\ }\bibfield  {title} {\enquote {\bibinfo {title}
  {Stochastic estimation of nuclear level density in the nuclear shell model:
  An application to parity-dependent level density in {58Ni}},}\ }\href
  {\doibase http://dx.doi.org/10.1016/j.physletb.2015.12.005} {\bibfield
  {journal} {\bibinfo  {journal} {Physics Letters B}\ }\textbf {\bibinfo
  {volume} {753}},\ \bibinfo {pages} {13 -- 17} (\bibinfo {year}
  {2016})}\BibitemShut {NoStop}%
\bibitem [{\citenamefont {Sakurai}\ and\ \citenamefont
  {Sugiura}(2003)}]{Sakurai2003}%
  \BibitemOpen
  \bibfield  {author} {\bibinfo {author} {\bibfnamefont {Tetsuya}\ \bibnamefont
  {Sakurai}}\ and\ \bibinfo {author} {\bibfnamefont {Hiroshi}\ \bibnamefont
  {Sugiura}},\ }\bibfield  {title} {\enquote {\bibinfo {title} {A projection
  method for generalized eigenvalue problems using numerical integration},}\
  }\href {\doibase 10.1016/S0377-0427(03)00565-X} {\bibfield  {journal}
  {\bibinfo  {journal} {Journal of Computational and Applied Mathematics}\
  }\textbf {\bibinfo {volume} {159}},\ \bibinfo {pages} {119 -- 128} (\bibinfo
  {year} {2003})}\BibitemShut {NoStop}%
\bibitem [{\citenamefont {Sakurai}\ \emph {et~al.}(2008)\citenamefont
  {Sakurai}, \citenamefont {Kodaki}, \citenamefont {Tadano}, \citenamefont
  {Takahashi}, \citenamefont {Sato},\ and\ \citenamefont
  {Nagashima}}]{Sakurai2008}%
  \BibitemOpen
  \bibfield  {author} {\bibinfo {author} {\bibfnamefont {Tetsuya}\ \bibnamefont
  {Sakurai}}, \bibinfo {author} {\bibfnamefont {Yoshihisa}\ \bibnamefont
  {Kodaki}}, \bibinfo {author} {\bibfnamefont {Hiroto}\ \bibnamefont {Tadano}},
  \bibinfo {author} {\bibfnamefont {Daisuke}\ \bibnamefont {Takahashi}},
  \bibinfo {author} {\bibfnamefont {Mitsuhisa}\ \bibnamefont {Sato}}, \ and\
  \bibinfo {author} {\bibfnamefont {Umpei}\ \bibnamefont {Nagashima}},\
  }\bibfield  {title} {\enquote {\bibinfo {title} {A parallel method for large
  sparse generalized eigenvalue problems using a gridrpc system},}\ }\href
  {\doibase 10.1016/j.future.2008.01.002} {\bibfield  {journal} {\bibinfo
  {journal} {Future Generation Computer Systems}\ }\textbf {\bibinfo {volume}
  {24}},\ \bibinfo {pages} {613 -- 619} (\bibinfo {year} {2008})}\BibitemShut
  {NoStop}%
\bibitem [{\citenamefont {Ikegami}\ \emph {et~al.}(2010)\citenamefont
  {Ikegami}, \citenamefont {Sakurai},\ and\ \citenamefont
  {Nagashima}}]{Ikegami2010}%
  \BibitemOpen
  \bibfield  {author} {\bibinfo {author} {\bibfnamefont {Tsutomu}\ \bibnamefont
  {Ikegami}}, \bibinfo {author} {\bibfnamefont {Tetsuya}\ \bibnamefont
  {Sakurai}}, \ and\ \bibinfo {author} {\bibfnamefont {Umpei}\ \bibnamefont
  {Nagashima}},\ }\bibfield  {title} {\enquote {\bibinfo {title} {A filter
  diagonalization for generalized eigenvalue problems based on the
  sakurai--sugiura projection method},}\ }\href {\doibase
  10.1016/j.cam.2009.09.029} {\bibfield  {journal} {\bibinfo  {journal}
  {Journal of Computational and Applied Mathematics}\ }\textbf {\bibinfo
  {volume} {233}},\ \bibinfo {pages} {1927 -- 1936} (\bibinfo {year}
  {2010})}\BibitemShut {NoStop}%
\bibitem [{\citenamefont {{J. Dobaczewski}}\ \emph {et~al.}(2002)\citenamefont
  {{J. Dobaczewski}}, \citenamefont {{W. Nazarewicz}},\ and\ \citenamefont
  {{M.V. Stoitsov}}}]{Dobacz2002}%
  \BibitemOpen
  \bibfield  {author} {\bibinfo {author} {\bibnamefont {{J. Dobaczewski}}},
  \bibinfo {author} {\bibnamefont {{W. Nazarewicz}}}, \ and\ \bibinfo {author}
  {\bibnamefont {{M.V. Stoitsov}}},\ }\bibfield  {title} {\enquote {\bibinfo
  {title} {Nuclear ground-state properties from mean-field calculations},}\
  }\href {\doibase 10.1140/epja/i2001-10218-8} {\bibfield  {journal} {\bibinfo
  {journal} {Eur. Phys. J. A}\ }\textbf {\bibinfo {volume} {15}},\ \bibinfo
  {pages} {21--26} (\bibinfo {year} {2002})}\BibitemShut {NoStop}%
\bibitem [{\citenamefont {Chabanat}\ \emph {et~al.}(1998)\citenamefont
  {Chabanat}, \citenamefont {Bonche}, \citenamefont {Haensel}, \citenamefont
  {Meyer},\ and\ \citenamefont {Schaeffer}}]{Chabanat1998}%
  \BibitemOpen
  \bibfield  {author} {\bibinfo {author} {\bibfnamefont {E.}~\bibnamefont
  {Chabanat}}, \bibinfo {author} {\bibfnamefont {P.}~\bibnamefont {Bonche}},
  \bibinfo {author} {\bibfnamefont {P.}~\bibnamefont {Haensel}}, \bibinfo
  {author} {\bibfnamefont {J.}~\bibnamefont {Meyer}}, \ and\ \bibinfo {author}
  {\bibfnamefont {R.}~\bibnamefont {Schaeffer}},\ }\bibfield  {title} {\enquote
  {\bibinfo {title} {A {Skyrme} parametrization from subnuclear to neutron star
  densities {Part II}. {Nuclei far from stabilities}},}\ }\href {\doibase
  http://dx.doi.org/10.1016/S0375-9474(98)00180-8} {\bibfield  {journal}
  {\bibinfo  {journal} {Nuclear Physics A}\ }\textbf {\bibinfo {volume}
  {635}},\ \bibinfo {pages} {231 -- 256} (\bibinfo {year} {1998})}\BibitemShut
  {NoStop}%
\bibitem [{\citenamefont {Stetcu}\ \emph {et~al.}(2011)\citenamefont {Stetcu},
  \citenamefont {Bulgac}, \citenamefont {Magierski},\ and\ \citenamefont
  {Roche}}]{Stetcu2011}%
  \BibitemOpen
  \bibfield  {author} {\bibinfo {author} {\bibfnamefont {I.}~\bibnamefont
  {Stetcu}}, \bibinfo {author} {\bibfnamefont {A.}~\bibnamefont {Bulgac}},
  \bibinfo {author} {\bibfnamefont {P.}~\bibnamefont {Magierski}}, \ and\
  \bibinfo {author} {\bibfnamefont {K.~J.}\ \bibnamefont {Roche}},\ }\bibfield
  {title} {\enquote {\bibinfo {title} {Isovector giant dipole resonance from
  the {3D} time-dependent density functional theory for superfluid nuclei},}\
  }\href {\doibase 10.1103/PhysRevC.84.051309} {\bibfield  {journal} {\bibinfo
  {journal} {Phys. Rev. C}\ }\textbf {\bibinfo {volume} {84}},\ \bibinfo
  {pages} {051309} (\bibinfo {year} {2011})}\BibitemShut {NoStop}%
\bibitem [{\citenamefont {Stetcu}\ \emph {et~al.}(2015)\citenamefont {Stetcu},
  \citenamefont {Bertulani}, \citenamefont {Bulgac}, \citenamefont
  {Magierski},\ and\ \citenamefont {Roche}}]{Stetcu2015}%
  \BibitemOpen
  \bibfield  {author} {\bibinfo {author} {\bibfnamefont {I.}~\bibnamefont
  {Stetcu}}, \bibinfo {author} {\bibfnamefont {C.~A.}\ \bibnamefont
  {Bertulani}}, \bibinfo {author} {\bibfnamefont {A.}~\bibnamefont {Bulgac}},
  \bibinfo {author} {\bibfnamefont {P.}~\bibnamefont {Magierski}}, \ and\
  \bibinfo {author} {\bibfnamefont {K.~J.}\ \bibnamefont {Roche}},\ }\bibfield
  {title} {\enquote {\bibinfo {title} {{Relativistic Coulomb Excitation within
  the Time Dependent Superfluid Local Density Approximation}},}\ }\href
  {\doibase 10.1103/PhysRevLett.114.012701} {\bibfield  {journal} {\bibinfo
  {journal} {Phys. Rev. Lett.}\ }\textbf {\bibinfo {volume} {114}},\ \bibinfo
  {pages} {012701} (\bibinfo {year} {2015})}\BibitemShut {NoStop}%
\bibitem [{\citenamefont {Bulgac}\ and\ \citenamefont {Yu}(2002)}]{Bulgac2002}%
  \BibitemOpen
  \bibfield  {author} {\bibinfo {author} {\bibfnamefont {Aurel}\ \bibnamefont
  {Bulgac}}\ and\ \bibinfo {author} {\bibfnamefont {Yongle}\ \bibnamefont
  {Yu}},\ }\bibfield  {title} {\enquote {\bibinfo {title} {Renormalization of
  the {Hartree}-{Fock}-{Bogoliubov} equations in the case of a zero range
  pairing interaction},}\ }\href {\doibase 10.1103/PhysRevLett.88.042504}
  {\bibfield  {journal} {\bibinfo  {journal} {Phys. Rev. Lett.}\ }\textbf
  {\bibinfo {volume} {88}},\ \bibinfo {pages} {042504} (\bibinfo {year}
  {2002})}\BibitemShut {NoStop}%
\bibitem [{\citenamefont {Yu}\ and\ \citenamefont {Bulgac}(2003)}]{Yu2003}%
  \BibitemOpen
  \bibfield  {author} {\bibinfo {author} {\bibfnamefont {Yongle}\ \bibnamefont
  {Yu}}\ and\ \bibinfo {author} {\bibfnamefont {Aurel}\ \bibnamefont
  {Bulgac}},\ }\bibfield  {title} {\enquote {\bibinfo {title} {Energy density
  functional approach to superfluid nuclei},}\ }\href {\doibase
  10.1103/PhysRevLett.90.222501} {\bibfield  {journal} {\bibinfo  {journal}
  {Phys. Rev. Lett.}\ }\textbf {\bibinfo {volume} {90}},\ \bibinfo {pages}
  {222501} (\bibinfo {year} {2003})}\BibitemShut {NoStop}%
\bibitem [{\citenamefont {Borycki}\ \emph {et~al.}(2006)\citenamefont
  {Borycki}, \citenamefont {Dobaczewski}, \citenamefont {Nazarewicz},\ and\
  \citenamefont {Stoitsov}}]{Borycki2006}%
  \BibitemOpen
  \bibfield  {author} {\bibinfo {author} {\bibfnamefont {P.~J.}\ \bibnamefont
  {Borycki}}, \bibinfo {author} {\bibfnamefont {J.}~\bibnamefont
  {Dobaczewski}}, \bibinfo {author} {\bibfnamefont {W.}~\bibnamefont
  {Nazarewicz}}, \ and\ \bibinfo {author} {\bibfnamefont {M.~V.}\ \bibnamefont
  {Stoitsov}},\ }\bibfield  {title} {\enquote {\bibinfo {title} {Pairing
  renormalization and regularization within the local density approximation},}\
  }\href {\doibase 10.1103/PhysRevC.73.044319} {\bibfield  {journal} {\bibinfo
  {journal} {Phys. Rev. C}\ }\textbf {\bibinfo {volume} {73}},\ \bibinfo
  {pages} {044319} (\bibinfo {year} {2006})}\BibitemShut {NoStop}%
\bibitem [{\citenamefont {Bertsch}\ \emph {et~al.}(2009)\citenamefont
  {Bertsch}, \citenamefont {Bertulani}, \citenamefont {Nazarewicz},
  \citenamefont {Schunck},\ and\ \citenamefont {Stoitsov}}]{Bertsch2009}%
  \BibitemOpen
  \bibfield  {author} {\bibinfo {author} {\bibfnamefont {G.~F.}\ \bibnamefont
  {Bertsch}}, \bibinfo {author} {\bibfnamefont {C.~A.}\ \bibnamefont
  {Bertulani}}, \bibinfo {author} {\bibfnamefont {W.}~\bibnamefont
  {Nazarewicz}}, \bibinfo {author} {\bibfnamefont {N.}~\bibnamefont {Schunck}},
  \ and\ \bibinfo {author} {\bibfnamefont {M.~V.}\ \bibnamefont {Stoitsov}},\
  }\bibfield  {title} {\enquote {\bibinfo {title} {Odd-even mass differences
  from self-consistent mean field theory},}\ }\href {\doibase
  10.1103/PhysRevC.79.034306} {\bibfield  {journal} {\bibinfo  {journal} {Phys.
  Rev. C}\ }\textbf {\bibinfo {volume} {79}},\ \bibinfo {pages} {034306}
  (\bibinfo {year} {2009})}\BibitemShut {NoStop}%
\bibitem [{\citenamefont {Bulgac}\ \emph {et~al.}(2015)\citenamefont {Bulgac},
  \citenamefont {Forbes},\ and\ \citenamefont {Jin}}]{arXiv:1506.09195}%
  \BibitemOpen
  \bibfield  {author} {\bibinfo {author} {\bibfnamefont {Aurel}\ \bibnamefont
  {Bulgac}}, \bibinfo {author} {\bibfnamefont {Michael~McNeil}\ \bibnamefont
  {Forbes}}, \ and\ \bibinfo {author} {\bibfnamefont {Shi}\ \bibnamefont
  {Jin}},\ }\bibfield  {title} {\enquote {\bibinfo {title} {Nuclear energy
  density functionals: What do we really know?}}\ }\href
  {http://arxiv.org/abs/1506.09195} {\bibfield  {journal} {\bibinfo  {journal}
  {arXiv:1506.09195}\ } (\bibinfo {year} {2015})}\BibitemShut {NoStop}%
\bibitem [{\citenamefont {Dutra}\ \emph {et~al.}(2012)\citenamefont {Dutra},
  \citenamefont {Louren\ifmmode~\mbox{\c{c}}\else \c{c}\fi{}o}, \citenamefont
  {S\'a~Martins}, \citenamefont {Delfino}, \citenamefont {Stone},\ and\
  \citenamefont {Stevenson}}]{Dutra2012}%
  \BibitemOpen
  \bibfield  {author} {\bibinfo {author} {\bibfnamefont {M.}~\bibnamefont
  {Dutra}}, \bibinfo {author} {\bibfnamefont {O.}~\bibnamefont
  {Louren\ifmmode~\mbox{\c{c}}\else \c{c}\fi{}o}}, \bibinfo {author}
  {\bibfnamefont {J.~S.}\ \bibnamefont {S\'a~Martins}}, \bibinfo {author}
  {\bibfnamefont {A.}~\bibnamefont {Delfino}}, \bibinfo {author} {\bibfnamefont
  {J.~R.}\ \bibnamefont {Stone}}, \ and\ \bibinfo {author} {\bibfnamefont
  {P.~D.}\ \bibnamefont {Stevenson}},\ }\bibfield  {title} {\enquote {\bibinfo
  {title} {Skyrme interaction and nuclear matter constraints},}\ }\href
  {\doibase 10.1103/PhysRevC.85.035201} {\bibfield  {journal} {\bibinfo
  {journal} {Phys. Rev. C}\ }\textbf {\bibinfo {volume} {85}},\ \bibinfo
  {pages} {035201} (\bibinfo {year} {2012})}\BibitemShut {NoStop}%
\bibitem [{\citenamefont {Bender}\ \emph
  {et~al.}(2003{\natexlab{b}})\citenamefont {Bender}, \citenamefont {Heenen},\
  and\ \citenamefont {Reinhard}}]{Bender:2003}%
  \BibitemOpen
  \bibfield  {author} {\bibinfo {author} {\bibfnamefont {Michael}\ \bibnamefont
  {Bender}}, \bibinfo {author} {\bibfnamefont {Paul-Henri}\ \bibnamefont
  {Heenen}}, \ and\ \bibinfo {author} {\bibfnamefont {Paul-Gerhard}\
  \bibnamefont {Reinhard}},\ }\bibfield  {title} {\enquote {\bibinfo {title}
  {Self-consistent mean-field models for nuclear structure},}\ }\href {\doibase
  10.1103/RevModPhys.75.121} {\bibfield  {journal} {\bibinfo  {journal} {Rev.
  Mod. Phys.}\ }\textbf {\bibinfo {volume} {75}},\ \bibinfo {pages} {121--180}
  (\bibinfo {year} {2003}{\natexlab{b}})}\BibitemShut {NoStop}%
\bibitem [{\citenamefont {Bohr}\ and\ \citenamefont
  {Mottelson}(1998)}]{Bohr1998}%
  \BibitemOpen
  \bibfield  {author} {\bibinfo {author} {\bibfnamefont {A.}~\bibnamefont
  {Bohr}}\ and\ \bibinfo {author} {\bibfnamefont {B.R.}\ \bibnamefont
  {Mottelson}},\ }\href {https://books.google.com/books?id=bDXgCO3Z4bIC} {\emph
  {\bibinfo {title} {Nuclear Structure}}},\ Vol.~\bibinfo {volume} {1}\
  (\bibinfo  {publisher} {World Scientific},\ \bibinfo {address} {Singapore},\
  \bibinfo {year} {1998})\BibitemShut {NoStop}%
\bibitem [{\citenamefont {Trefethen}\ and\ \citenamefont
  {Bau}(1997)}]{Trefethen1997}%
  \BibitemOpen
  \bibfield  {author} {\bibinfo {author} {\bibfnamefont {L.N.}\ \bibnamefont
  {Trefethen}}\ and\ \bibinfo {author} {\bibfnamefont {D.}~\bibnamefont
  {Bau}},\ }\href {https://books.google.com/books?id=JaPtxOytY7kC} {\emph
  {\bibinfo {title} {Numerical Linear Algebra}}}\ (\bibinfo  {publisher}
  {Society for Industrial and Applied Mathematics},\ \bibinfo {address}
  {Philadelphia},\ \bibinfo {year} {1997})\BibitemShut {NoStop}%
\bibitem [{\citenamefont {Bulgac}()}]{Bulgac1999}%
  \BibitemOpen
  \bibfield  {author} {\bibinfo {author} {\bibfnamefont {Aurel}\ \bibnamefont
  {Bulgac}},\ }\bibfield  {title} {\enquote {\bibinfo {title}
  {{Hartree}-{Fock}-{Bogoliubov} approximation for finite systems},}\ }\href
  {http://arxiv.org/abs/nucl-th/9907088} {\bibinfo  {journal} {CIP, Bucharest,
  Report No. FT-194-1980; nucl-th/9907088}\ }\BibitemShut {NoStop}%
\bibitem [{\citenamefont {Gu}\ \emph {et~al.}(2014)\citenamefont {Gu},
  \citenamefont {Huang}, \citenamefont {Li}, \citenamefont {Li}, \citenamefont
  {Sogabe},\ and\ \citenamefont {Clemens}}]{Gu:2014}%
  \BibitemOpen
\bibfield  {journal} {  }\bibfield  {author} {\bibinfo {author} {\bibfnamefont
  {X.~M.}\ \bibnamefont {Gu}}, \bibinfo {author} {\bibfnamefont {T.~Z.}\
  \bibnamefont {Huang}}, \bibinfo {author} {\bibfnamefont {L.}~\bibnamefont
  {Li}}, \bibinfo {author} {\bibfnamefont {H.~B.}\ \bibnamefont {Li}}, \bibinfo
  {author} {\bibfnamefont {T.}~\bibnamefont {Sogabe}}, \ and\ \bibinfo {author}
  {\bibfnamefont {M.}~\bibnamefont {Clemens}},\ }\bibfield  {title} {\enquote
  {\bibinfo {title} {Quasi-minimal residual variants of the {COCG} and {COCR}
  methods for complex symmetric linear systems in electromagnetic
  simulations},}\ }\href {\doibase 10.1109/TMTT.2014.2365472} {\bibfield
  {journal} {\bibinfo  {journal} {IEEE Transactions on Microwave Theory and
  Techniques}\ }\textbf {\bibinfo {volume} {62}},\ \bibinfo {pages}
  {2859--2867} (\bibinfo {year} {2014})}\BibitemShut {NoStop}%
\bibitem [{\citenamefont {Li}\ \emph {et~al.}(2011)\citenamefont {Li},
  \citenamefont {Nie}, \citenamefont {Meng}, \citenamefont {Zhang},\ and\
  \citenamefont {Sun}}]{Li2011}%
  \BibitemOpen
  \bibfield  {author} {\bibinfo {author} {\bibfnamefont {Yue~Hui}\ \bibnamefont
  {Li}}, \bibinfo {author} {\bibfnamefont {Zai~Ping}\ \bibnamefont {Nie}},
  \bibinfo {author} {\bibfnamefont {Min}\ \bibnamefont {Meng}}, \bibinfo
  {author} {\bibfnamefont {Xiang~Qian}\ \bibnamefont {Zhang}}, \ and\ \bibinfo
  {author} {\bibfnamefont {Xiang~Yang}\ \bibnamefont {Sun}},\ }\bibfield
  {title} {\enquote {\bibinfo {title} {An efficient {MAINV} preconditioned
  {COCG} method for {FEM} analysis of millimeter wave filters},}\ }\href
  {\doibase 10.1007/s10762-010-9761-6} {\bibfield  {journal} {\bibinfo
  {journal} {Journal of Infrared, Millimeter, and Terahertz Waves}\ }\textbf
  {\bibinfo {volume} {32}},\ \bibinfo {pages} {216--224} (\bibinfo {year}
  {2011})}\BibitemShut {NoStop}%
\bibitem [{\citenamefont {Li}\ \emph {et~al.}(2008)\citenamefont {Li},
  \citenamefont {Huang},\ and\ \citenamefont {Ren}}]{Li:2008}%
  \BibitemOpen
  \bibfield  {author} {\bibinfo {author} {\bibfnamefont {L.}~\bibnamefont
  {Li}}, \bibinfo {author} {\bibfnamefont {T.-Z.}\ \bibnamefont {Huang}}, \
  and\ \bibinfo {author} {\bibfnamefont {Z.-G.}\ \bibnamefont {Ren}},\
  }\bibfield  {title} {\enquote {\bibinfo {title} {A preconditioned {COCG}
  method for solving complex symmetric linear systems arising from scattering
  problems},}\ }\href {\doibase 10.1163/156939308787537793} {\bibfield
  {journal} {\bibinfo  {journal} {Journal of Electromagnetic Waves and
  Applications}\ }\textbf {\bibinfo {volume} {22}},\ \bibinfo {pages}
  {2023--2034} (\bibinfo {year} {2008})}\BibitemShut {NoStop}%
\bibitem [{\citenamefont {Wang}\ \emph {et~al.}(2009)\citenamefont {Wang},
  \citenamefont {Du},\ and\ \citenamefont {Sun}}]{Wang:2009}%
  \BibitemOpen
  \bibfield  {author} {\bibinfo {author} {\bibfnamefont {Yingxi}\ \bibnamefont
  {Wang}}, \bibinfo {author} {\bibfnamefont {Kui}\ \bibnamefont {Du}}, \ and\
  \bibinfo {author} {\bibfnamefont {Weiwei}\ \bibnamefont {Sun}},\ }\bibfield
  {title} {\enquote {\bibinfo {title} {Preconditioning iterative algorithm for
  the electromagnetic scattering from a large cavity},}\ }\href {\doibase
  10.1002/nla.615} {\bibfield  {journal} {\bibinfo  {journal} {Numerical Linear
  Algebra with Applications}\ }\textbf {\bibinfo {volume} {16}},\ \bibinfo
  {pages} {345--363} (\bibinfo {year} {2009})}\BibitemShut {NoStop}%
\bibitem [{\citenamefont {Aichinger}\ and\ \citenamefont
  {Krotscheck}(2005)}]{Aichinger2005188}%
  \BibitemOpen
  \bibfield  {author} {\bibinfo {author} {\bibfnamefont {M.}~\bibnamefont
  {Aichinger}}\ and\ \bibinfo {author} {\bibfnamefont {E.}~\bibnamefont
  {Krotscheck}},\ }\bibfield  {title} {\enquote {\bibinfo {title} {A fast
  configuration space method for solving local {Kohn--Sham} equations},}\
  }\href {\doibase http://dx.doi.org/10.1016/j.commatsci.2004.11.002}
  {\bibfield  {journal} {\bibinfo  {journal} {Computational Materials Science}\
  }\textbf {\bibinfo {volume} {34}},\ \bibinfo {pages} {188 -- 212} (\bibinfo
  {year} {2005})}\BibitemShut {NoStop}%
\bibitem [{\citenamefont {Frommer}(2003)}]{Frommer2003}%
  \BibitemOpen
  \bibfield  {author} {\bibinfo {author} {\bibfnamefont {A.}~\bibnamefont
  {Frommer}},\ }\bibfield  {title} {\enquote {\bibinfo {title} {Bicgstab(ℓ)
  for families of shifted linear systems},}\ }\href {\doibase
  10.1007/s00607-003-1472-6} {\bibfield  {journal} {\bibinfo  {journal}
  {Computing}\ }\textbf {\bibinfo {volume} {70}},\ \bibinfo {pages} {87--109}
  (\bibinfo {year} {2003})}\BibitemShut {NoStop}%
\bibitem [{\citenamefont {Frommer}\ and\ \citenamefont
  {Gl{\"a}ssner}(1998)}]{Frommer1998}%
  \BibitemOpen
  \bibfield  {author} {\bibinfo {author} {\bibfnamefont {Andreas}\ \bibnamefont
  {Frommer}}\ and\ \bibinfo {author} {\bibfnamefont {Uwe}\ \bibnamefont
  {Gl{\"a}ssner}},\ }\bibfield  {title} {\enquote {\bibinfo {title} {Restarted
  {GMRES} for shifted linear systems},}\ }\href {\doibase
  10.1137/S1064827596304563} {\bibfield  {journal} {\bibinfo  {journal} {SIAM
  Journal on Scientific Computing}\ }\textbf {\bibinfo {volume} {19}},\
  \bibinfo {pages} {15--26} (\bibinfo {year} {1998})}\BibitemShut {NoStop}%
\bibitem [{tit()}]{titan}%
  \BibitemOpen
  \href {https://www.olcf.ornl.gov/computing-resources/titan-cray-xk7/}
  {\bibinfo  {journal}
  {https://www.olcf.ornl.gov/computing-resources/titan-cray-xk7/}\
  }\BibitemShut {NoStop}%
\bibitem [{\citenamefont {Baran}\ \emph {et~al.}(2008)\citenamefont {Baran},
  \citenamefont {Bulgac}, \citenamefont {Forbes}, \citenamefont {Hagen},
  \citenamefont {Nazarewicz}, \citenamefont {Schunck},\ and\ \citenamefont
  {Stoitsov}}]{Baran2008}%
  \BibitemOpen
\bibfield  {journal} {  }\bibfield  {author} {\bibinfo {author} {\bibfnamefont
  {Andrzej}\ \bibnamefont {Baran}}, \bibinfo {author} {\bibfnamefont {Aurel}\
  \bibnamefont {Bulgac}}, \bibinfo {author} {\bibfnamefont {Michael~McNeil}\
  \bibnamefont {Forbes}}, \bibinfo {author} {\bibfnamefont {Gaute}\
  \bibnamefont {Hagen}}, \bibinfo {author} {\bibfnamefont {Witold}\
  \bibnamefont {Nazarewicz}}, \bibinfo {author} {\bibfnamefont {Nicolas}\
  \bibnamefont {Schunck}}, \ and\ \bibinfo {author} {\bibfnamefont {Mario~V.}\
  \bibnamefont {Stoitsov}},\ }\bibfield  {title} {\enquote {\bibinfo {title}
  {Broyden's method in nuclear structure calculations},}\ }\href {\doibase
  10.1103/PhysRevC.78.014318} {\bibfield  {journal} {\bibinfo  {journal} {Phys.
  Rev. C}\ }\textbf {\bibinfo {volume} {78}},\ \bibinfo {pages} {014318}
  (\bibinfo {year} {2008})}\BibitemShut {NoStop}%
\bibitem [{\citenamefont {Baye}(2015)}]{Baye20151}%
  \BibitemOpen
  \bibfield  {author} {\bibinfo {author} {\bibfnamefont {Daniel}\ \bibnamefont
  {Baye}},\ }\bibfield  {title} {\enquote {\bibinfo {title} {The lagrange-mesh
  method},}\ }\href {\doibase http://dx.doi.org/10.1016/j.physrep.2014.11.006}
  {\bibfield  {journal} {\bibinfo  {journal} {Physics Reports}\ }\textbf
  {\bibinfo {volume} {565}},\ \bibinfo {pages} {1 -- 107} (\bibinfo {year}
  {2015})}\BibitemShut {NoStop}%
\bibitem [{\citenamefont {Younes}\ and\ \citenamefont
  {Gogny}(2009)}]{Younes2009}%
  \BibitemOpen
  \bibfield  {author} {\bibinfo {author} {\bibfnamefont {W.}~\bibnamefont
  {Younes}}\ and\ \bibinfo {author} {\bibfnamefont {D.}~\bibnamefont {Gogny}},\
  }\bibfield  {title} {\enquote {\bibinfo {title} {Microscopic calculation of
  $^{240}\mathrm{Pu}$ scission with a finite-range effective force},}\ }\href
  {\doibase 10.1103/PhysRevC.80.054313} {\bibfield  {journal} {\bibinfo
  {journal} {Phys. Rev. C}\ }\textbf {\bibinfo {volume} {80}},\ \bibinfo
  {pages} {054313} (\bibinfo {year} {2009})}\BibitemShut {NoStop}%
\bibitem [{\citenamefont {Schunck}\ \emph {et~al.}(2014)\citenamefont
  {Schunck}, \citenamefont {Duke}, \citenamefont {Carr},\ and\ \citenamefont
  {Knoll}}]{Schunck2014}%
  \BibitemOpen
  \bibfield  {author} {\bibinfo {author} {\bibfnamefont {N.}~\bibnamefont
  {Schunck}}, \bibinfo {author} {\bibfnamefont {D.}~\bibnamefont {Duke}},
  \bibinfo {author} {\bibfnamefont {H.}~\bibnamefont {Carr}}, \ and\ \bibinfo
  {author} {\bibfnamefont {A.}~\bibnamefont {Knoll}},\ }\bibfield  {title}
  {\enquote {\bibinfo {title} {Description of induced nuclear fission with
  {Skyrme} energy functionals: Static potential energy surfaces and fission
  fragment properties},}\ }\href {\doibase 10.1103/PhysRevC.90.054305}
  {\bibfield  {journal} {\bibinfo  {journal} {Phys. Rev. C}\ }\textbf {\bibinfo
  {volume} {90}},\ \bibinfo {pages} {054305} (\bibinfo {year}
  {2014})}\BibitemShut {NoStop}%
\bibitem [{\citenamefont {Regnier}\ \emph {et~al.}(2016)\citenamefont
  {Regnier}, \citenamefont {Dubray}, \citenamefont {Schunck},\ and\
  \citenamefont {Verri\`ere}}]{Regnier2016}%
  \BibitemOpen
  \bibfield  {author} {\bibinfo {author} {\bibfnamefont {D.}~\bibnamefont
  {Regnier}}, \bibinfo {author} {\bibfnamefont {N.}~\bibnamefont {Dubray}},
  \bibinfo {author} {\bibfnamefont {N.}~\bibnamefont {Schunck}}, \ and\
  \bibinfo {author} {\bibfnamefont {M.}~\bibnamefont {Verri\`ere}},\ }\bibfield
   {title} {\enquote {\bibinfo {title} {Fission fragment charge and mass
  distributions in $^{239}\mathrm{Pu}(n,f)$ in the adiabatic nuclear energy
  density functional theory},}\ }\href {\doibase 10.1103/PhysRevC.93.054611}
  {\bibfield  {journal} {\bibinfo  {journal} {Phys. Rev. C}\ }\textbf {\bibinfo
  {volume} {93}},\ \bibinfo {pages} {054611} (\bibinfo {year}
  {2016})}\BibitemShut {NoStop}%
\bibitem [{\citenamefont {Staszczak}\ \emph {et~al.}(2010)\citenamefont
  {Staszczak}, \citenamefont {Stoitsov}, \citenamefont {Baran},\ and\
  \citenamefont {Nazarewicz}}]{Staszczak2010}%
  \BibitemOpen
  \bibfield  {author} {\bibinfo {author} {\bibfnamefont {A.}~\bibnamefont
  {Staszczak}}, \bibinfo {author} {\bibfnamefont {M.}~\bibnamefont {Stoitsov}},
  \bibinfo {author} {\bibfnamefont {A.}~\bibnamefont {Baran}}, \ and\ \bibinfo
  {author} {\bibfnamefont {W.}~\bibnamefont {Nazarewicz}},\ }\bibfield  {title}
  {\enquote {\bibinfo {title} {Augmented lagrangian method for constrained
  nuclear density functional theory},}\ }\href {\doibase
  10.1140/epja/i2010-11018-9} {\bibfield  {journal} {\bibinfo  {journal} {Eur.
  Phys. J. A}\ }\textbf {\bibinfo {volume} {46}},\ \bibinfo {pages} {85--90}
  (\bibinfo {year} {2010})}\BibitemShut {NoStop}%
\bibitem [{\citenamefont {Fayans}(1998)}]{Fayans1998}%
  \BibitemOpen
  \bibfield  {author} {\bibinfo {author} {\bibfnamefont {S.~A.}\ \bibnamefont
  {Fayans}},\ }\bibfield  {title} {\enquote {\bibinfo {title} {Towards a
  universal nuclear density functional},}\ }\href {\doibase 10.1134/1.567841}
  {\bibfield  {journal} {\bibinfo  {journal} {Journal of Experimental and
  Theoretical Physics Letters}\ }\textbf {\bibinfo {volume} {68}},\ \bibinfo
  {pages} {169--174} (\bibinfo {year} {1998})}\BibitemShut {NoStop}%
\bibitem [{\citenamefont {Fayans}\ \emph {et~al.}(2000)\citenamefont {Fayans},
  \citenamefont {Tolokonnikov}, \citenamefont {Trykov},\ and\ \citenamefont
  {Zawischa}}]{Fayans200049}%
  \BibitemOpen
  \bibfield  {author} {\bibinfo {author} {\bibfnamefont {S.A.}\ \bibnamefont
  {Fayans}}, \bibinfo {author} {\bibfnamefont {S.V.}\ \bibnamefont
  {Tolokonnikov}}, \bibinfo {author} {\bibfnamefont {E.L.}\ \bibnamefont
  {Trykov}}, \ and\ \bibinfo {author} {\bibfnamefont {D.}~\bibnamefont
  {Zawischa}},\ }\bibfield  {title} {\enquote {\bibinfo {title} {Nuclear
  isotope shifts within the local energy-density functional approach},}\ }\href
  {\doibase http://dx.doi.org/10.1016/S0375-9474(00)00192-5} {\bibfield
  {journal} {\bibinfo  {journal} {Nuclear Physics A}\ }\textbf {\bibinfo
  {volume} {676}},\ \bibinfo {pages} {49 -- 119} (\bibinfo {year}
  {2000})}\BibitemShut {NoStop}%
\bibitem [{\citenamefont {Shlomo}\ and\ \citenamefont
  {Bertsch}(1975)}]{Shlomo1975}%
  \BibitemOpen
  \bibfield  {author} {\bibinfo {author} {\bibfnamefont {S.}~\bibnamefont
  {Shlomo}}\ and\ \bibinfo {author} {\bibfnamefont {G.}~\bibnamefont
  {Bertsch}},\ }\bibfield  {title} {\enquote {\bibinfo {title} {Nuclear
  response in the continuum},}\ }\href {\doibase
  http://dx.doi.org/10.1016/0375-9474(75)90292-4} {\bibfield  {journal}
  {\bibinfo  {journal} {Nuclear Physics A}\ }\textbf {\bibinfo {volume}
  {243}},\ \bibinfo {pages} {507 -- 518} (\bibinfo {year} {1975})}\BibitemShut
  {NoStop}%
\bibitem [{\citenamefont {Bertsch}\ and\ \citenamefont
  {Tsai}(1975)}]{Bertsch1975}%
  \BibitemOpen
  \bibfield  {author} {\bibinfo {author} {\bibfnamefont {G.F.}\ \bibnamefont
  {Bertsch}}\ and\ \bibinfo {author} {\bibfnamefont {S.F.}\ \bibnamefont
  {Tsai}},\ }\bibfield  {title} {\enquote {\bibinfo {title} {A study of the
  nuclear response function},}\ }\href {\doibase
  http://dx.doi.org/10.1016/0370-1573(75)90003-4} {\bibfield  {journal}
  {\bibinfo  {journal} {Physics Reports}\ }\textbf {\bibinfo {volume} {18}},\
  \bibinfo {pages} {125 -- 158} (\bibinfo {year} {1975})}\BibitemShut {NoStop}%
\bibitem [{\citenamefont {Nakatsukasa}\ \emph {et~al.}(2007)\citenamefont
  {Nakatsukasa}, \citenamefont {Inakura},\ and\ \citenamefont
  {Yabana}}]{Nakatsukasa2007}%
  \BibitemOpen
  \bibfield  {author} {\bibinfo {author} {\bibfnamefont {Takashi}\ \bibnamefont
  {Nakatsukasa}}, \bibinfo {author} {\bibfnamefont {Tsunenori}\ \bibnamefont
  {Inakura}}, \ and\ \bibinfo {author} {\bibfnamefont {Kazuhiro}\ \bibnamefont
  {Yabana}},\ }\bibfield  {title} {\enquote {\bibinfo {title} {Finite amplitude
  method for the solution of the random-phase approximation},}\ }\href
  {\doibase 10.1103/PhysRevC.76.024318} {\bibfield  {journal} {\bibinfo
  {journal} {Phys. Rev. C}\ }\textbf {\bibinfo {volume} {76}},\ \bibinfo
  {pages} {024318} (\bibinfo {year} {2007})}\BibitemShut {NoStop}%
\bibitem [{\citenamefont {Avogadro}\ and\ \citenamefont
  {Nakatsukasa}(2011)}]{Avogadro2011}%
  \BibitemOpen
  \bibfield  {author} {\bibinfo {author} {\bibfnamefont {Paolo}\ \bibnamefont
  {Avogadro}}\ and\ \bibinfo {author} {\bibfnamefont {Takashi}\ \bibnamefont
  {Nakatsukasa}},\ }\bibfield  {title} {\enquote {\bibinfo {title} {Finite
  amplitude method for the quasiparticle random-phase approximation},}\ }\href
  {\doibase 10.1103/PhysRevC.84.014314} {\bibfield  {journal} {\bibinfo
  {journal} {Phys. Rev. C}\ }\textbf {\bibinfo {volume} {84}},\ \bibinfo
  {pages} {014314} (\bibinfo {year} {2011})}\BibitemShut {NoStop}%
\bibitem [{\citenamefont {Nakatsukasa}(2014)}]{Nakatsukasa2014}%
  \BibitemOpen
  \bibfield  {author} {\bibinfo {author} {\bibfnamefont {Takashi}\ \bibnamefont
  {Nakatsukasa}},\ }\bibfield  {title} {\enquote {\bibinfo {title} {Finite
  amplitude method in linear response {TDDFT} calculations},}\ }\href
  {http://stacks.iop.org/1742-6596/533/i=1/a=012054} {\bibfield  {journal}
  {\bibinfo  {journal} {Journal of Physics: Conference Series}\ }\textbf
  {\bibinfo {volume} {533}},\ \bibinfo {pages} {012054} (\bibinfo {year}
  {2014})}\BibitemShut {NoStop}%
\bibitem [{\citenamefont {Hinohara}\ \emph {et~al.}(2013)\citenamefont
  {Hinohara}, \citenamefont {Kortelainen},\ and\ \citenamefont
  {Nazarewicz}}]{Hinohara2013}%
  \BibitemOpen
  \bibfield  {author} {\bibinfo {author} {\bibfnamefont {Nobuo}\ \bibnamefont
  {Hinohara}}, \bibinfo {author} {\bibfnamefont {Markus}\ \bibnamefont
  {Kortelainen}}, \ and\ \bibinfo {author} {\bibfnamefont {Witold}\
  \bibnamefont {Nazarewicz}},\ }\bibfield  {title} {\enquote {\bibinfo {title}
  {Low-energy collective modes of deformed superfluid nuclei within the
  finite-amplitude method},}\ }\href {\doibase 10.1103/PhysRevC.87.064309}
  {\bibfield  {journal} {\bibinfo  {journal} {Phys. Rev. C}\ }\textbf {\bibinfo
  {volume} {87}},\ \bibinfo {pages} {064309} (\bibinfo {year}
  {2013})}\BibitemShut {NoStop}%
\bibitem [{\citenamefont {Oishi}\ \emph {et~al.}(2016)\citenamefont {Oishi},
  \citenamefont {Kortelainen},\ and\ \citenamefont {Hinohara}}]{Oishi2016}%
  \BibitemOpen
  \bibfield  {author} {\bibinfo {author} {\bibfnamefont {Tomohiro}\
  \bibnamefont {Oishi}}, \bibinfo {author} {\bibfnamefont {Markus}\
  \bibnamefont {Kortelainen}}, \ and\ \bibinfo {author} {\bibfnamefont {Nobuo}\
  \bibnamefont {Hinohara}},\ }\bibfield  {title} {\enquote {\bibinfo {title}
  {Finite amplitude method applied to the giant dipole resonance in heavy
  rare-earth nuclei},}\ }\href {\doibase 10.1103/PhysRevC.93.034329} {\bibfield
   {journal} {\bibinfo  {journal} {Phys. Rev. C}\ }\textbf {\bibinfo {volume}
  {93}},\ \bibinfo {pages} {034329} (\bibinfo {year} {2016})}\BibitemShut
  {NoStop}%
\bibitem [{\citenamefont {Mustonen}\ \emph {et~al.}(2014)\citenamefont
  {Mustonen}, \citenamefont {Shafer}, \citenamefont {Zenginerler},\ and\
  \citenamefont {Engel}}]{Mustonen2014}%
  \BibitemOpen
  \bibfield  {author} {\bibinfo {author} {\bibfnamefont {M.~T.}\ \bibnamefont
  {Mustonen}}, \bibinfo {author} {\bibfnamefont {T.}~\bibnamefont {Shafer}},
  \bibinfo {author} {\bibfnamefont {Z.}~\bibnamefont {Zenginerler}}, \ and\
  \bibinfo {author} {\bibfnamefont {J.}~\bibnamefont {Engel}},\ }\bibfield
  {title} {\enquote {\bibinfo {title} {Finite-amplitude method for
  charge-changing transitions in axially deformed nuclei},}\ }\href {\doibase
  10.1103/PhysRevC.90.024308} {\bibfield  {journal} {\bibinfo  {journal} {Phys.
  Rev. C}\ }\textbf {\bibinfo {volume} {90}},\ \bibinfo {pages} {024308}
  (\bibinfo {year} {2014})}\BibitemShut {NoStop}%
\end{thebibliography}%
